\documentclass{aa}

\usepackage{graphicx}
\usepackage[dvipsnames]{xcolor}
\usepackage[normalem]{ulem} 

\usepackage{txfonts}

\usepackage[colorlinks=true
,urlcolor=blue
,anchorcolor=blue
,citecolor=blue
,filecolor=blue
,linkcolor=blue
,menucolor=blue
,linktocpage=true
,pdfproducer=medialab
,pdfa=true]{hyperref}

\usepackage{multicol}
\usepackage{appendix}
\usepackage{cuted}

\begin{document}

   \title{On the shape of pancakes:}

   \subtitle{Catastrophe theory and Gaussian statistics in 2D}

   \author{Abineet Parichha
          \inst{1}\fnmsep\thanks{corresponding author}
          ,
          Stephane Colombi\inst{1}
          ,
          Shohei Saga\inst{1,2,3}
          \and
          Atsushi Taruya\inst{4,5}
          }

   \institute{$^1$ Sorbonne Universit\'e, CNRS, UMR7095, Institut d'Astrophysique de Paris, 98bis boulevard Arago, F-75014 Paris, France\\
   $^2$ Institute for Advanced Research, Nagoya University, Furo-cho Chikusa-ku, Nagoya 464-8601, Japan\\
   $^3$ Kobayashi-Maskawa Institute for the Origin of Particles and the Universe, Nagoya University, Chikusa-ku, Nagoya, 464-8602, Japan\\
   $^4$ Center for Gravitational Physics and Quantum Information, Yukawa Institute for Theoretical Physics, Kyoto University, Kyoto
   606-8502, Japan\\
   $^5$ Kavli Institute for the Physics and Mathematics of the Universe (WPI), Todai institute for Advanced Study, University of Tokyo,
   Kashiwa, Chiba 277-8568, Japan\\
              \email{abineet.parichha@iap.fr}
             \thanks{author email}
             }

   \date{\today}

  \abstract
   {Cold dark matter (CDM) can be thought of as a 2D (or 3D) sheet of particles in 4D (or 6D) phase-space due to its negligible velocity dispersion. The large scale structure, also called the cosmic web, is thus a result of the topology of the CDM manifold. Initial crossing of particle trajectories occurs at the critical points of this manifold, forming singularities that seed most of the collapsed structures. The cosmic web can thus be characterized using the points of singularities. In this context, we employ catastrophe theory in 2D to study the motion around such singularities and analytically model the shape of the emerging structures, particularly the pancakes, which later evolve into halos and filaments - the building blocks of the 2D web. We compute higher order corrections to the shape of the pancakes, including properties such as the curvature and the scale of transition from their C to S shape. Using Gaussian statistics (with the assumption of Zeldovich flow) for our model parameters, we also compute the distributions of observable features related to the shape of pancakes and their variation across halo and filament populations in 2D cosmologies. We find that a larger fraction of pancakes evolve into filaments, they are more curved if they are to evolve into halos, are dominantly C-shaped, and the nature of shell-crossing is highly anisotropic. Extending this work to 3D will allow testing of predictions against actual observations of the cosmic web and searching for signatures of non-Gaussianity at corresponding scales.}


   \maketitle

\section{Introduction}
\label{sec:intro}

The large-scale structure of the Universe exhibits a striking and complex pattern known as the cosmic web, a vast network of halos, walls, filaments, and voids that emerged from tiny primordial density perturbations \citep{Zeldovich_1970,Peebles_1980,Bond_1996,Springel_2005}. In the standard $\Lambda$CDM cosmological framework, approximately 84\% of the Universe’s matter content \citep[see, e.g.,][]{Planck_2018} is constituted of cold dark matter (CDM) --- a non-relativistic, collisionless component that drives structure formation. As the Universe expanded and evolved under gravity, overdense regions grew, giving rise to a hierarchical structure in which CDM clustered into halos, connected by elongated filaments and walls, and surrounded by underdense voids. The baryons having decoupled from the early hot plasma condensed into the collapsed halos, which became the primary sites of galaxy formation. This web-like arrangement of galaxies has been revealed in a series of pioneering redshift surveys, from the early work of \citet{Joeveer_1978} and \citet{deLapparent_1986} to the large-scale surveys such as 2dF \citep{Colless_2003}, SDSS \citep{Gott_2005}, and 2MASS \citep{Huchra_2012}. Beyond its role in structuring the Universe, the cosmic web serves as a sensitive probe of fundamental physics: its morphology and connectivity carry imprints of the nature of dark matter, the dynamics of structure formation, and the statistical properties of the primordial density field, including possible non-Gaussianities \citep{Libeskind_2018,Codis_2012,Cautun_2014,Kitaura_2019}. As such, the study of the cosmic web is of growing importance in the era of precision cosmology and large-scale surveys such as DESI \citep{DESI_2016} and Euclid \citep{Euclid_2025}.

As CDM is the dominant driver of structure formation, analyses of the cosmic web are naturally framed in terms of dark matter dynamics. CDM is modeled as a self-gravitating and collisionless fluid obeying the Vlasov–Poisson equations:
\begin{align}
    &\frac{\partial f}{\partial t} + \mathbf{u}\cdot \nabla_{\mathbf{r}}f - \nabla_{\mathbf{r}} \Phi \cdot \nabla_{\mathbf{u}}f = 0,\\
    &\Delta_{\mathbf{r}}\Phi = 4\pi G\, \rho = 4\pi G \int f(\mathbf{r}, \mathbf{u}, t)\, {\rm d}^3\mathbf{u},
\end{align}
where $f(\mathbf{r}, \mathbf{u}, t)$ represents the phase-space density at position $\mathbf{r}$, velocity $\mathbf{u}$ and time $t$; and $\Phi$ is the gravitational potential. Owing to negligible velocity dispersion, the CDM distribution can be thought of as a 2D (or 3D) sheet \citep{Abel_2012,Shandarin_2012} in 4D (or 6D) phase space. The system of equations are usually numerically resolved with an $N$-body approach, that is, representing the CDM sheet by an ensemble of particles that follow the standard Lagrangian equations of motion in an expanding Universe:
\begin{equation}
    \label{eq:lagrange_eq_motion} \frac{{\rm d}^2\mathbf{x}}{{\rm d}t^2} + 2H\frac{{\rm d}\mathbf{x}}{{\rm d}t} = -\frac{1}{a^2} \mathbf{\nabla}_{\mathbf{x}}\Phi(\mathbf{x}) \hspace{0.5 cm} ; \hspace{0.2 cm} \Delta_{\mathbf{x}} \Phi(\mathbf{x}) = 4\pi G \,\bar{\rho}\, a^2\, \delta(\mathbf{x}),
\end{equation}
where $\mathbf{x}$ is the comoving position, $a$ is the cosmological expansion factor corresponding to time $t$ and $\delta$ is the matter density contrast, defined as $\delta = \rho/\bar{\rho} - 1$ with $\bar{\rho}$ being the background matter density.

On the analytical front, the single-stream evolution from the primordial Gaussian random field to the onset of shell-crossing can be accurately described using Lagrangian perturbation theory (LPT) \citep[see, e.g.,][and references therein]{Bernardeau_2002}. The pioneering Zel’dovich approximation (ZA, also 1-LPT) \citep{Zeldovich_1970}, was the first to capture the anisotropic nature of collapse and the emergence of pancake-like structures that evolve into a hierarchy of halos, filaments, and walls in a cosmological setting. Building on this, \citet{Arnold_1982_ASZ} framed the emerging structures in terms of the topology of singularities in the CDM manifold in 1D and 2D. \citet{Arnold_1982_Petrovskogo,Arnold_1983_Rays} presented the normal forms of all generic singularities in 3D, laying the foundation of catastrophe theory \citep{Arnold_1984}. Under this formalism, the structural hierarchy—halos, filaments, walls, and voids—as well as the singularities that seed them, are characterized through the eigenvalues and eigenvectors of the deformation tensor. On the statistical side, \citet{Doroshkevich_1970, Doroshkevich_1978} derived the distribution of these eigenvalues and applied it to make statistical predictions about the pancake area and curvature, while \citet{Bardeen_1986} developed the statistics of peaks in Gaussian random fields. Later, \citet{Bond_1996} coined the term “cosmic web,” linking the statistics of eigenvalues to predictions about the filamentary network connecting halos. 

Considerable effort has been devoted to tracing the cosmic web by reducing it to the underlying skeleton or spine of the large-scale structure. Early descriptions include the adhesion model \citep{Gurbatov_1989}, while non-local ridge-tracing techniques using the density field in 2D and 3D were developed by \citet{Novikov_2006} and \citet{Sousbie_2008}, respectively. Incorporating the information from the deformation tensor, more recent approaches \citep{Hidding_2014,Feldbrugge_2018} use catastrophe theory to define the web as the locus of singularities. Extending this framework by combining Gaussian statistics, \citet{Feldbrugge_2023_feb, Feldbrugge_2023_sept, Feldbrugge_2025} probe the statistical properties of the progenitors of the hierarchical structures in 2D, such as the correlation function of singular points, filament formation times, and multistream properties.

In this work, we focus on the 2D case, which is analytically more tractable yet retains the essential features of the hierarchical structures present in 3D. In section \ref{sec:catastrophe_theory}, we draw on catastrophe theory \citep{Arnold_1982_ASZ} to derive the 2D normal form (valid up to $O(q^4)$, where $\vec{q}$ is the Lagrangian coordinate) for the subset of singularities that seed pancakes, which we then use to characterize their shape and related properties. Section~\ref{sec:gaussian_stats} extends the distribution of deformation-tensor eigenvalues \citep{Doroshkevich_1970} to include $O(q^4)$ derivatives of the Gaussian field, treated as parameters in the normal form. Adopting the peak statistics formalism of \citet{Bardeen_1986}, we compute their conditional distribution at singular points where pancakes form. Together, these ingredients yield the distributions of several observable pancake features. Our framework is in line with recent advances \citep{Feldbrugge_2025} and their precursors, but places particular emphasis on pancake geometry—especially curvature and higher-order corrections that produce C- and S-type morphologies. Finally, section~\ref{sec:conclusions} summarizes the main results of our 2D analysis, outlines extensions to 3D, and discusses the observational ramifications. We have relegated the extended formulae appearing in the derivation of our analytical results in sections \ref{sec:catastrophe_theory} and \ref{sec:gaussian_stats} to appendices \ref{sec:appendix_A} and \ref{sec:appendix_B}, respectively. In appendix \ref{sec:numerical_verification}, we present corroboration of our analytical and statistical predictions with measurements from numerical simulations of Zeldovich flow initialized with Gaussian random fields.

\section{Analytical model using catastrophe theory}
\label{sec:catastrophe_theory}

\subsection{Lagrangian formalism, shell-crossing and singularities}
\begin{figure*}
    \centering
    \includegraphics[width=\linewidth]{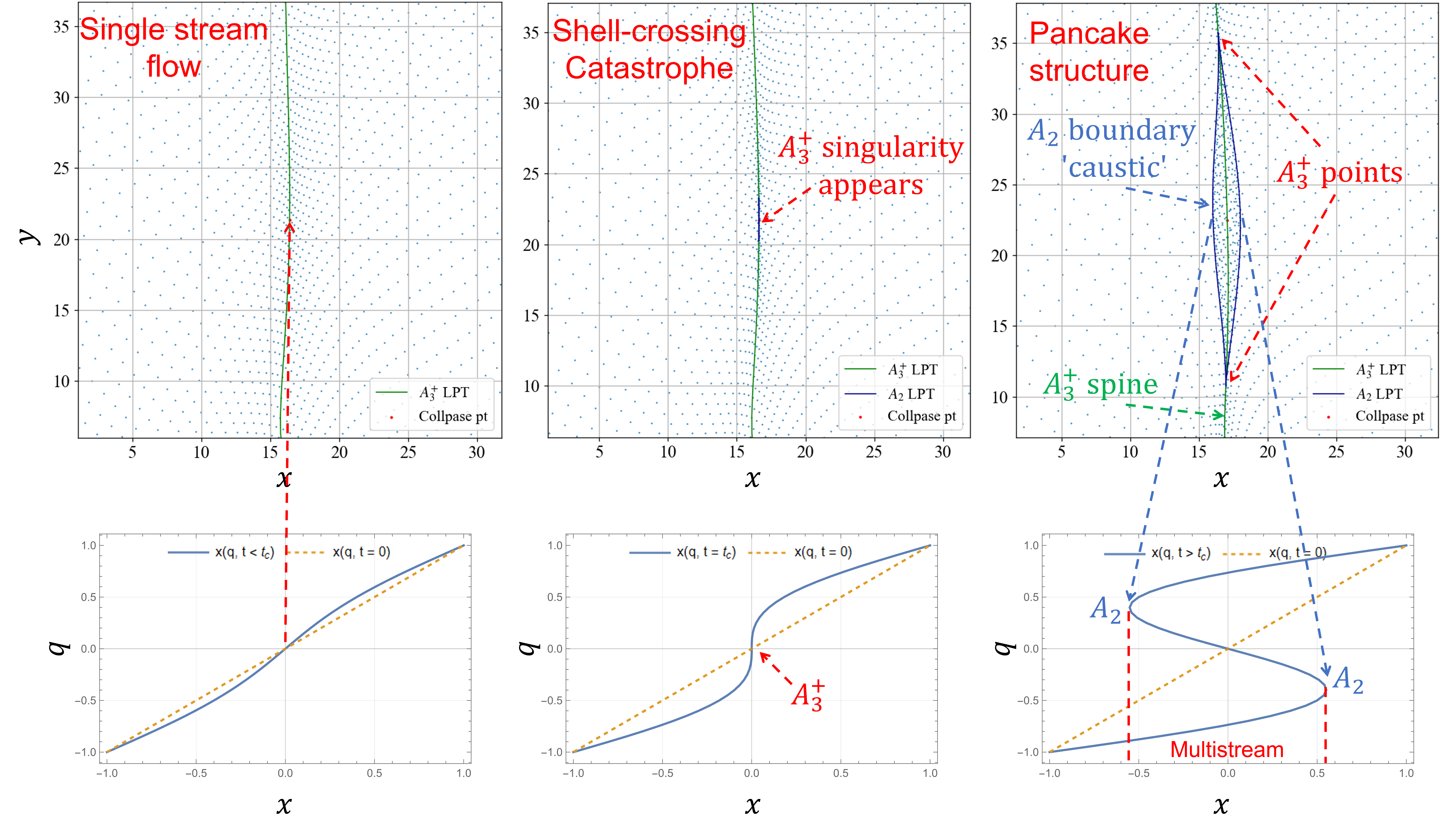}
    \caption{Illustration of 2D shell-crossing in Eulerian $x-y$ space (top row) along with an $q_x$-axis slice through the shell-crossing point in $q-x$ space (bottom row) to help visualize the folding of the CDM sheet. The left column depicts a snapshot during the single stream flow prior to shell-crossing. The middle column shows the moment of shell-crossing. It is an event of catastrophe marked by the appearance of an $A_3^+$ singularity (red dot) that seeds the 2D pancake-like structure, shown in the right column. The boundary of the pancake encompassing the multi-streaming region is defined by $A_2$ singularities (blue), which grows along the $A_3^+$ spine (green).}
    \label{fig:shell_crossing}
\end{figure*}

In Lagrangian formalism, we track the motion of particles $\mathbf{x}(\vec{q}, t)$ in terms of their displacement $\Psi(\vec{q},t)$ from initial positions $\vec{q}$:
\begin{equation}
    \label{eq:x_q_psi}
    \mathbf{x}(\vec{q},t) = \vec{q} + \Psi(\vec{q},t).
\end{equation}
In Lagrangian $\vec{q}$-space, the density is uniform and equal to the background density in our expanding universe, that is the Eulerian $\vec{x}$-space. The Eulerian density is given by the Jacobian of the transformation between Lagrangian and Eulerian spaces:
\begin{align}
    \label{eq:J}
    J_{ij}(\vec{q},t) &= \frac{\partial x_i}{\partial q_j} (\vec{q},t) = \delta^D_{ij} + \frac{\partial \Psi_i}{\partial q_j} (\vec{q},t)\\
    \label{eq:rho_J}
    \rho(\mathbf{x}, t) &= \bar{\rho}(t) (1 + \delta(\mathbf{x}, t)) = \frac{\bar{\rho}(t)}{J(\vec{q},t)},
\end{align}
where $J$ is the determinant of the Jacobian and $\delta^D$ is the Dirac-delta function. We define the deformation matrix $d_{ij}(\vec{q},t) = \delta^D_{ij} - J_{ij}(\vec{q},t) = -\partial \Psi_i \: / \: \partial q_j (\vec{q},t)$ with eigenvalues $\alpha > \beta$ and corresponding eigenvectors $\hat{n}^{\alpha}, \hat{n}^{\beta}$, which together encode the complete dynamical information. In terms of eigenvalues,
\begin{equation}
    \label{eq:rho_alpha}
    \rho(\mathbf{x}, t) \propto J(\vec{q},t)^{-1} = [(1 - \alpha(\vec{q},t))(1 - \beta(\vec{q},t))]^{-1}.
\end{equation}
Within the framework of catastrophe theory, the particle motion can be viewed as the evolving shape of the CDM sheet in $\vec{x} - \vec{q}$ space. Refer to fig. \ref{fig:shell_crossing}. The topology of the sheet is determined by the set of critical points: $J = \| \partial x_i / \partial q_j \| = 0$. The tangent to the sheet at critical points is perpendicular to the Eulerian space and hence, the density at these points projected onto the Eulerian space is singular, as implied by eq. \eqref{eq:rho_J}. From eq. \eqref{eq:rho_alpha}, the density being singular implies
\begin{equation}
    \label{eq:A2_cond}
    \alpha(\vec{q},t) = 1
\end{equation}
for points undergoing shell-crossing for the first time (refer to section 5(a) in \citet{Doroshkevich_1978} and eq. (10) in \citet{Arnold_1982_ASZ}). This condition identifies the $A_2$ class of singularities in 2D. 

When shell-crossing occurs for the first time in a local neighborhood, the sheet starts folding onto itself, turning perpendicular to the Eulerian space and forming a cusp. The first critical point (or singularity) appears, indicating a change in topology (or catastrophe). We shall refer to this point as the shell-crossing point. It is the first point to satisfy the condition $\alpha(\vec{q}, t) = 1$ in its local neighborhood and is therefore a local maximum of the eigenvalue field, that is $\vec{\nabla}_{\vec{q}}\alpha(\vec{q}, t) = 0$. In 2D, it is a singularity of class $A_3^+$, also called cusp singularity. Shell-crossing typically happens along one axis; it is nearly impossible to have shell-crossings along multiple axes simultaneously \citep[refer to eq. (12) in][]{Doroshkevich_1970}. Therefore, as the CDM sheet folds further, its projection onto the Eulerian space is shaped like a pancake, bulging towards the center. In 2D, it is shaped like an eye. The cusps at either end of the fold in CDM sheet branch out in opposite directions ($\hat{n}_{\beta}$) transverse to the shell-crossing direction ($\hat{n}_{\alpha}$) and follow the locus of $A_3^+$ singularities (refer to section 5(b) in \citet{Doroshkevich_1978} and eq. (13) in \citet{Arnold_1982_ASZ}), given by :
\begin{equation}
    \label{eq:A3_cond}
    \bar{\nabla}_{\vec{q}} \alpha(\vec{q},t) \cdot \hat{n}^{\alpha}(\vec{q},t) = 0.
\end{equation}
Thus, the $A_3^+$ locus serves as the spine along which the pancake grows, with its boundary (lateral edges of the fold in CDM sheet) being defined by the $A_2$ locus eq. \eqref{eq:A2_cond}. The $A_2$ boundary divides the multi-stream and single-stream regions in Eulerian space. As more particles enter the multi-stream region, the size of the pancake ($A_2$ boundary) grows, with the two $A_3^+$ singularities at either tips branching out farther.

In fact, the locus of $A_3^+$ singularities passes through all the critical - local maxima, minima, and saddle - points of the eigenvalue field, since $\bar{\nabla}_{\vec{q}}\alpha(\vec{q},t)=0$ automatically satisfies eq. \eqref{eq:A3_cond}. These are the points around which collapses occur and structures emerge. Therefore, the $A_3^+$ locus traces not only the spine of pancakes, but also the entire the skeletal network of structures in our 2D setup, akin to the cosmic web of the universe in 3D. Put differently, the study of the cosmic web can be reduced to analyzing the critical points of the eigenvalue field $\alpha(\vec{q}, t)$ within the framework of Morse-Smale theory \citep{Smale_1960}.

Growing pancakes form only around points of local maxima with positive eigenvalue $\alpha$. These points, which we refer to as shell-crossing points, are the focus of our formalism. If there is no shell-crossing along the transverse direction(s), the pancake grows into a filament in 2D (wall in 3D). If there is shell-crossing along the secondary transverse direction, the pancake turns into a cluster in 2D (filament in 3D). Further, in 3D, shell-crossing along the tertiary transverse direction turns it into a cluster. Clusters, followed by violent relaxation in the multistream regime \citep{Lynden_1967}, lead to the formation of primordial halos.
 
On the other hand, the local minima and saddle points with positive eigenvalue $\alpha$ undergo shell-crossing to form disappearing pancakes and tube-shaped structures, respectively. However, these are not the first points to do so in their neighborhood, which is already deep in the multistream regime by the time these points turn singular. Although the same formalism, with minor tweaks discussed in the next subsection, holds true for these points asymptotically, its applicability in studying the large-scale structure around these points is limited.

Eqs. \eqref{eq:lagrange_eq_motion}, \eqref{eq:x_q_psi}, \eqref{eq:J}, and \eqref{eq:rho_J} form a complete set of equations that can be solved for a given set of initial conditions. For cosmological setups, the initial conditions are generally provided in terms of the matter power-spectrum $P_{\delta}(k)$ of a Gaussian random field for the overdensity $\delta(\vec{q})$, defined as $\langle \delta(\bar{k}) \delta^*(\bar{k}') \rangle = (2 \pi)^2 \delta_D^2(\bar{k} - \bar{k}') P_{\delta}(k)$, where $k$ is the Fourier mode and $\delta_D$ is the Dirac-delta function.

If there is no vorticity in Lagrangian space in our setup, then the displacement field derives from a potential $\phi$
\begin{equation}
    \label{eq:vort_free_cond}
    \Psi(\vec{q},t) = -\bar{\nabla} \phi(\vec{q},t),
\end{equation}
which is related to the overdensity $\delta(\vec{q})$ in the linear regime: $\Delta_{\vec{q}} \phi(\vec{q}) = \delta(\vec{q})$. The displacement potential $\phi(\vec{q})$ is therefore Gaussian distributed as well, with the power spectrum
\begin{equation}
    \label{eq:P_phi_delta_k} P_{\phi}(k) = k^{-4} P_{\delta} (k).
\end{equation}
Eq. \eqref{eq:rho_J} is valid as long as the flow is single-stream and one can obtain recursive solutions using LPT. After shell-crossing, the flow is multistream inside the boundary of the pancake defined by $A_2$ singularities, and LPT breaks down. Eq. \eqref{eq:rho_J} must be modified into 
\begin{equation}
    \rho(\mathbf{x}, t) = \bar{\rho}(t) \sum_{\vec{q} \: : \: \mathbf{x}(\vec{q}, t) = \mathbf{x}} J(\vec{q}, t) ^{-1}
\end{equation}
to take into account the contribution from the overlapping folds of the CDM sheet. This is the basis of post-collapse perturbation theory (PCPT) \citep{Colombi_2015,Taruya_2017, Rampf_2021, Shohei_2023} in which the corrections to the force field from the overlapping folds are computed that allows the post-collapse motion to be tracked deeper into the multi-stream regime. 

For our purpose of studying the structure of an emerging pancake in 2D, we derive the normal form of the first singularity that appears in a local neighborhood at shell-crossing. We start by expanding the motion around the shell-crossing point and time $(\vec{q}_c, t_c)$ into a Taylor series up to $O(q^3, t)$:
\begin{equation}
    \label{eq:taylor_exp}
    \mathbf{x}( \vec{q}, t) = \sum_{0 \le k \le 1} \sum_{0 \le i+j \le 3} \frac{\partial^{i+j+k} \: \mathbf{x}}{\partial q_x^i \partial q_y^j \partial t^k} \Bigg|_{\vec{q}_c, t_c} ( q_x - q_{x,c} )^i ( q_y - q_{y,c} )^j ( t - t_{c} )^k.
\end{equation}
The first assumption made is that of ballistic approximation, that is, expansion in $\mathbf{x}$ up to linear order in time. Higher-order expansion in time would require computation of the force field as is done in PCPT. We shall show in a following subsection (refer to eq. \eqref{eq:A2_eq_Eul} and the subsequent discussion) that in order to be self-consistent under this assumption, the expansion in velocity must be linear order in space: $d\mathbf{x}/dt \sim O(q)$.
The second assumption is that the motion of particles is vorticity-free in Lagrangian space: \begin{equation}
    \label{eq:x_vort_free}
    \mathbf{x}( \vec{q}, t) = \bar{\nabla}_q \left( q^2/2 - \phi( \vec{q}, t) \right),
\end{equation}
which is obtained by substituting eq. \eqref{eq:vort_free_cond} in eq. \eqref{eq:x_q_psi}. Thus, the coefficients of our expansion comprise derivatives of $\phi$ up to 4th order in $\vec{q}$ and 1st order in $t$. We use the following notation for the derivatives of $\phi$:
\begin{equation}
    \phi^{i, j, k} ({\vec{q}, t}) = \frac{\partial^{i+j+k} \: \phi}{\partial q_x^i \partial q_y^j \partial t^k} ({\vec{q}, t}).
\end{equation}
In equations where the function input $({\vec{q}, t})$ has been dropped to make them concise, it is to be assumed that the function is evaluated at the shell-crossing point $ ({\vec{q}_c, t_c})$.

\subsection{Coefficient constraints}
Based on the geometry of the CDM sheet in the neighborhood of the shell-crossing point $(\vec{q}_c, t_c)$, we impose constraints on the set of coefficients of our Taylor expansion eq. \eqref{eq:taylor_exp}.

\begin{itemize}
    \item $O(q^0)$ derivatives: $\phi^{0,0,0},\phi^{0,0,1}$\\
    These neither appear in eq. \eqref{eq:taylor_exp} nor affect the dynamics and hence, are immaterial.\\

    \item $O(q^1)$ derivatives: $\phi^{1,0,0},\phi^{0,1,0},\phi^{1,0,1},\phi^{0,1,1}$\\
    Upon translating on to the shell-crossing point in Lagrangian space $\vec{q}_c = 0$ and in Eulerian space $\mathbf{x}(\vec{q}_c, t_c)  = 0$, the $O(q^1,t^0)$ derivatives vanish: $\phi^{1,0,0}, \phi^{0,1,0}$ = 0. Similarly, upon subtracting the velocity at the shell-crossing point globally from velocity field such that $d\mathbf{x}/dt \: (\vec{q}_c, t_c) = 0$, the $O(q^1,t^1)$ derivatives vanish as well: $\phi^{1,0,1},\phi^{0,1,1} = 0$.\\
    
    \item $O(q^2)$ derivatives: $\phi^{2,0,0},\phi^{1,1,0},\phi^{0,2,0},\phi^{2,0,1},\phi^{1,1,1},\phi^{0,2,1}$\\
    In the absence of vorticity in Lagrangian space, deformation matrix $d_{ij} = \delta_{ij}^D - J_{ij}$ can be rewritten by substituting eq. \eqref{eq:x_vort_free} in eq. \eqref{eq:J}:
    \begin{equation}
        d_{ij} (\vec{q}, t) = 
        \begin{bmatrix}
            \phi^{2,0}(\vec{q}, t) & \phi^{1,1}(\vec{q}, t) \\
            \phi^{1,1}(\vec{q}, t) & \phi^{0,2}(\vec{q}, t)
        \end{bmatrix},
    \end{equation}
    which is simply the Hessian of the potential $\phi$ and has the eigenvalues $\alpha > \beta$. Since $d_{ij}$ is symmetrical, its eigenvectors $\hat{n}^{\alpha}, \hat{n}^{\beta}$ are orthogonal. Refer to the eqs. \eqref{eq:alpha_LPT} and \eqref{eq:n_alpha_LPT} in the appendix for the full expressions of the eigenvalues and eigenvectors.

    By performing rotation, both in Lagrangian and Eulerian spaces, to align the reference axis $\hat{q}_x$ along the shell-crossing axis $\hat{n}^{\alpha}(\vec{q}_c, t_c)$, we diagonalize $d_{ij}(\vec{q}_c, t_c)$, which implies $\phi^{1,1,0} = 0$ and $\phi^{2,0,0} > \phi^{0,2,0}$. The eigenvalues and eigenvectors evaluated at $(\vec{q}_c, t_c)$ are
    \begin{align}
        &\alpha(\vec{q}_c, t_c) = \phi^{2,0,0} \hspace{2.3 cm} \beta(\vec{q}_c, t_c) = \phi^{0,2,0},\\
        \nonumber
        &\vec{n}^{\alpha}(\vec{q}_c, t_c) =
        \begin{bmatrix}
            2 (\phi^{2,0,0} - \phi^{0,2,0})\\
            0
        \end{bmatrix}
        \hspace{0.2 cm} 
        \vec{n}^{\beta}(\vec{q}_c, t_c) =
        \begin{bmatrix}
            0\\
            2 (\phi^{0,2,0} - \phi^{2,0,0})
        \end{bmatrix}.
    \end{align}

    The density $\rho(\mathbf{x}_c, t_c) \propto [(1 - \alpha(\vec{q}_c, t_c))(1 - \beta(\vec{q}_c, t_c))]^{-1}$, eq. \eqref{eq:rho_alpha}. The shell-crossing point corresponds to a singularity in the density field, which implies $\alpha(\vec{q}_c, t_c) =  \phi ^{2,0,0} = 1$.
    
    If we assume Zeldovich flow prior to shell-crossing, then $\phi(\vec{q},t) = D_+(t) \phi(\vec{q})$ implies $\phi^{i,j,1} \propto \phi^{i,j,0}$. This suggests that the constraints on the signs of $\phi^{i,j,1}$ are the same as that of $\phi^{i,j,0}$, which are $\phi^{2,0,1} > 0, \phi^{1,1,1} = 0, \phi^{2,0,1} > \phi^{0,2,1}$. Even for higher-order LPT motion, the constraints on the signs should remain the same, as the magnitude of the corrections would typically be less than that of the leading order. Therefore, the constraints on the signs of $\phi^{i,j,1}$ are not just limited to Zeldovich flow.\\

    \item $O(q^3)$ derivatives: $\phi^{3,0,0},\phi^{2,1,0},\phi^{1,2,0},\phi^{0,3,0}$ \\
    The expression for $\bar{\nabla}_{\vec{q}}\alpha(\vec{q}, t)$ is provided in the appendix, eq. \eqref{eq:grad_alpha_LPT}. Evaluating it at $(\vec{q}_c,t_c)$ using the previously obtained constraints $\phi^{1,1,0} = 0, \phi^{2,0,0} = 1, \phi^{2,0,0} > \phi^{0,2,0}$ \citep[refer to eq. (109) in][]{Feldbrugge_2023_sept}:
    \begin{equation}
        \bar{\nabla}_{\vec{q}}\alpha(\vec{q}_c, t_c) =
        \begin{bmatrix}
            \phi^{3,0,0} \\
            \phi^{2,1,0}
        \end{bmatrix},
    \end{equation}
    Since the shell-crossing point is a local maximum of the eigenvalue $\alpha(\vec{q},t)$, $\bar{\nabla}_{\vec{q}}\alpha(\vec{q}_c, t_c)$ must vanish, which implies $\phi^{3,0,0} = \phi^{2,1,0} = 0$.\\
    The other two derivatives $\phi^{1,2,0}, \phi^{0,3,0}$ are unconstrained.\\
    Since we restrict $d\mathbf{x}/dt$ up to $O(q)$ (refer to eq. \eqref{eq:A2_eq_Eul} and the subsequent discussion), the $O(q^{\ge 3}, t^1)$ derivatives of $\phi$ are not considered.\\

    \item $O(q^4)$ derivatives: $\phi^{4,0,0},\phi^{3,1,0},\phi^{2,2,0},\phi^{1,3,0}, \phi^{0,4,0}$\\
     The expression for the Hessian of the higher eigenvalue field, defined as $H_{ij}(\alpha(\vec{q}, t)) = \partial^2 \alpha(\vec{q}, t) / \partial q_i \partial q_j$, is provided in the appendix, eq. \eqref{eq:H_alpha_LPT}.  Evaluating it at $(\vec{q}_c,t_c)$ using the previously obtained constraints: $\phi^{1,1,0} = 0, \phi^{2,0,0} = 1, \phi^{2,0,0} > \phi^{0,2,0}, \phi^{3,0,0} = \phi^{2,1,0} = 0$ \citep[refer to eq. (6.5) in][]{Feldbrugge_2023_feb}:
    \begin{equation}
        \label{eq:H}
        H_{ij}(\alpha(\vec{q}_c, t_c)) = 
        \begin{bmatrix}
            \phi^{4,0,0} & \phi^{3,1,0} \\
            \phi^{3,1,0} &  \phi^{2,2,0} + 2 (\phi^{1,2,0})^2 / (1 - \phi^{0,2,0}) 
        \end{bmatrix}.
    \end{equation}
    It is symmetric but not diagonal, unless $\phi^{3,1,0} = 0$. Consequently, its eigenvectors $\hat{n}_H^{\alpha}(\vec{q}_c,t_c)$ and $\hat{n}_H^{\beta}(\vec{q}_c,t_c)$ are orthogonal, though not necessarily aligned with the eigenvectors of $d_{ij}(\vec{q}_c, t_c)$, namely $\hat{n}^{\alpha}(\vec{q}_c,t_c)$ and $\hat{n}^{\beta}(\vec{q}_c,t_c)$, which serve as our reference axes. The expressions for the eigenvalues $\alpha_H > \beta_H$ and corresponding eigenvectors at $(\vec{q}_c,t_c)$ are provided in the appendix, eqs. \eqref{eq:alpha_H_LPT} and \eqref{eq:n_alpha_H_LPT}. For the shell-crossing point to be a non-degenerate local maximum of $\alpha(\vec{q},t)$, both the eigenvalues of its Hessian, which are also the curvatures at that point, must be negative. This is possible only if the trace and the determinant of the Hessian are negative and positive, respectively:
    \begin{align}
        \nonumber
        & \phi^{4,0,0} + \phi^{2,2,0} + \frac{2 (\phi^{1,2,0})^2}{1 - \phi^{0,2,0}}  < 0,\\
        \label{eq:maxima_cond}
        &\phi^{4,0,0} \left( \phi^{2,2,0} + \frac{2 (\phi^{1,2,0})^2}{1 - \phi^{0,2,0}} \right) > (\phi^{3,1,0})^2,
    \end{align}
    which together imply that $\phi^{4,0,0} < 0$ and $\phi^{2,2,0} < 0 $. The remaining derivatives $\phi^{3,1,0}, \phi^{1,3,0}, \phi^{0,4,0}$ are unconstrained.
    
    Furthermore, the shell-crossing point $(\vec{q}_c, t_c)$ automatically satisfies the condition for an $A_3^+$ singularity: $\bar{\nabla}_{\vec{q}} \alpha(\vec{q}, t_c) \cdot \hat{n}^{\alpha}(\vec{q}, t_c) = 0$, as well as that for a ridge point: $\bar{\nabla}_{\vec{q}} \alpha(\vec{q}, t_c) \cdot \hat{n}_H^{\beta}(\vec{q}, t_c) = 0$, by the virtue of being a local maximum of $\alpha(\vec{q}, t_c)$.  Refer to the fig. \ref{fig:phi_310}. The tangent to the locus of $A_3^+$ points at $(\vec{q}_c, t_c)$ is $-\phi^{3,1,0} \hat{n}^{\alpha}(\vec{q}_c,t_c) + \phi^{4,0,0} \hat{n}^{\beta}(\vec{q}_c,t_c)$, eq. \eqref{eq:A3_eq}, which is aligned along neither the reference axis $\hat{n}^{\beta}$ nor the ridge $\hat{n}_H^{\alpha}$ in Lagrangian space unless $\phi^{3,1,0} = 0$. However, it is aligned with the reference axis in Eulerian space. This is because $\hat{n}^{\alpha}(\vec{q}_c,t_c)$ gives the direction of shell-crossing of particles, or the folding of the CDM sheet, which is quasi one-dimensional for an infinitesimally small neighborhood and time interval. Therefore, the only portion of the CDM sheet in the immediate neighborhood that can continue folding lies along the transverse direction, which is $\hat{n}^{\beta}(\vec{q}_c,t_c)$. It is thus more straightforward to characterize the Eulerian shape of the shell-crossing structure in the frame of $(\hat{n}^{\alpha}(\vec{q}_c,t_c), \hat{n}^{\beta}(\vec{q}_c,t_c))$, and therein lies the reason behind our choice of the reference axes.
\end{itemize}

The final coefficient matrices $\phi^{i, j, k}$ after compiling all the constraints discussed, are:
\begin{align}
    \nonumber&\phi^{i, j, 0} =
    \begin{bmatrix}
        0 & 0 & (-\infty, 1) & (-\infty, \infty) & (-\infty, \infty) \\
        0 & 0 & (-\infty, \infty) & (-\infty, \infty) & \_ \\
        1 & 0 & (-\infty,0) & \_ & \_ \\
        0 & (-\infty, \infty) & \_ & \_ & \_ \\
        (-\infty, 0) & \_ & \_ & \_ & \_ \\
    \end{bmatrix},\\
    \label{eq:coeff_matrix} &\phi^{i, j, 1} =
    \begin{bmatrix}
        0 & 0 & (-\infty, \infty) \\
        0 & 0 &  \_ \\
        (0, \infty) & \_ &  \_ \\
    \end{bmatrix},
\end{align}
where the dashes '\_' correspond to the higher order derivatives we do not consider in our expansion.

To reiterate, these constraints are derived for shell-crossing points where pancakes emerge --- the local maxima of the eigenvalue field $\alpha(\vec{q},t)$. The pancakes can be further classified into halos and filaments based on the sign of the lower eigenvalue $\beta({\vec{q}_c, t_c}) = \phi^{0,2,0}$. If $\phi^{0,2,0} \in (0,1)$ and $\phi^{0,2,1} \in (0, \infty)$, shell-crossing occurs along the transverse axis $\hat{n}^{\beta}(\vec{q}_c, t_c)$, and the pancake collapses into a 2D halo. Conversely, if $\phi^{0,2,0}, \phi^{0,2,1} \in (- \infty, 0]$, the particles continue to drift apart along the transverse axis without undergoing a second shell-crossing, resulting in a 2D filament.

Extension to local minima and saddle points of the eigenvalue field $\alpha(\vec{q},t)$ is obtained simply by changing the sign constraints on the eigenvalues $\alpha_H > \beta_H$ of the Hessian $H_{ij}(\alpha(\vec{q}_c,t_c))$. For saddle points, $\alpha_H(\vec{q}_c, t_c) > 0 $ and $\beta_H(\vec{q}_c, t_c) < 0$, which is possible only when the determinant of the Hessian is negative. If $\phi^{0,2,0} > 0$, the saddle point undergoes shell-crossing along the transverse axis by merging into a halo. If $\phi^{0,2,0} \le 0$, the saddle point does not undergo further shell-crossing and becomes part of a filament. For local minima, $\alpha_H(\vec{q}_c, t_c), \: \beta_H(\vec{q}_c, t_c) > 0$, which is possible only when both the trace and the determinant of the Hessian are positive.

\subsection{Pancake structure }

After plugging in the constraints, eq. \eqref{eq:coeff_matrix}, into our expansion eq. \eqref{eq:taylor_exp}, we obtain the following reduced expression with a total of ten remaining coefficients:
\begin{align}
    \label{eq:eq_motion_taylor_exp}
    x( \vec{q}, t) &= -\phi^{2,0,1} q_x (t - t_c) - \frac{1}{2} \phi^{1,2,0} q_y^2 - \frac{1}{6} \phi^{4,0,0} q_x^3 \\
    \nonumber
    & - \frac{1}{2} \phi^{3,1,0} q_x^2 q_y - \frac{1}{2} \phi^{2,2,0} q_x q_y^2 - \frac{1}{6} \phi^{1,3,0} q_y^3 ,\\
    \nonumber
    y( \vec{q}, t) &= (1 - \phi^{0,2,0}) q_y - \phi^{0,2,1} q_y (t - t_c) - \frac{1}{2} \phi^{0,3,0} q_y^2\\
    \nonumber
    & - \phi^{1,2,0} q_x q_y - \frac{1}{6} \phi^{0,4,0} q_y^3 - \frac{1}{2} \phi^{1,3,0} q_x q_y^2\\
    \nonumber
    & - \frac{1}{2} \phi^{2,2,0} q_x^2 q_y - \frac{1}{6} \phi^{3,1,0} q_x^3.
\end{align}
In terms of catastrophe theory, these are the normal form derivatives corresponding to the subset of $A_3^+$ singularities that appear at the local maxima of the eigenvalue $\alpha(\vec{q}, t)$. From eq. \eqref{eq:eq_motion_taylor_exp}, we derive the loci of $A_3^+$ and $A_2$ singularities, which define the structure of the emerging pancake as shown in fig \ref{fig:shell_crossing}.

The deformation matrix $d_{ij} = \delta^D_{ij} - J_{ij}$ can be rewritten substituting our expansion form, eq. \eqref{eq:eq_motion_taylor_exp}, into eq. \eqref{eq:J}. Its elements, up to $O(q^2, t)$, are: 

\begin{equation}
    \label{eq:d_ij}
    \begin{bmatrix}
        1+\phi^{2,0,1} (t - t_c) + \frac{1}{2} \phi^{4,0,0} q_x^2  & \phi^{1,2,0} q_y + \frac{1}{2} \phi^{3,1,0} q_x^2 \\
        + \phi^{3,1,0} q_x q_y + \frac{1}{2} \phi^{2,2,0} q_y^2 & + \phi^{2,2,0} q_x q_y +\frac{1}{2} \phi^{1,3,0} q_y^2 \\
        &\\
        \phi^{1,2,0} q_y + \frac{1}{2} \phi^{3,1,0} q_x^2  &  \phi^{0,2,0} + \phi^{0,2,1} (t - t_c) \\
        + \phi^{2,2,0} q_x q_y + \frac{1}{2} \phi^{1,3,0} q_y^2 & + \phi^{0,3,0} q_y + \phi^{1,2,0} q_x + \phi^{1,3,0} q_x q_y\\
        & + \frac{1}{2} \phi^{0,4,0} q_y^2 +  \frac{1}{2} \phi^{2,2,0} q_x^2
    \end{bmatrix}.
\end{equation}
Its eigenvalues $\alpha > \beta$ and corresponding eigenvectors $\hat{n}^{\alpha}, \hat{n}^{\beta}$ are expanded up to $O(q^2, t)$ to be consistent:

\begin{align}
    \label{eq:alpha}
    \alpha(\vec{q},t) =& 1 + \phi^{2,0,1} (t - t_c) + \frac{1}{2} \phi^{4,0,0} q_x^2 + \phi^{3,1,0} q_x q_y\\
    \nonumber&  + \frac{1}{2} \Bigg( \phi^{2,2,0} +\frac{2(\phi^{1,2,0})^2}{(1 - \phi^{0,2,0})}  \Bigg) q_y^2,\\
    \nonumber\\
    \label{eq:n_alpha}
    \hat{n}^{\alpha}(\vec{q},t) =& 
    \begin{bmatrix}
        1 - \phi^{0,2,0} + (\phi^{2,0,1} - \phi^{0,2,1}) (t - t_c) -  \phi^{0,3,0} q_y \\
         - \phi^{1,2,0} q_x + (\phi^{3,1,0} - \phi^{1,3,0}) q_x q_y  -\frac{1}{2} (\phi^{2,2,0} - \phi^{4,0,0}) q_x^2 \\
        - \frac{1}{2} \Big( \phi^{0,4,0} - \phi^{2,2,0} - \frac{2(\phi^{1,2,0})^2 }{(1 - \phi^{0,2,0})}  \Big) q_y^2\\
        \\
        \\
        \phi^{1,2,0} q_y + \frac{1}{2} \phi^{3,1,0} q_x^2 +  \phi^{2,2,0} q_x q_y + \frac{1}{2} \phi^{1,3,0} q_y^2
    \end{bmatrix}.
\end{align}
The higher the value of $1 - \phi^{0,2,0}$, the faster the convergence of these expansions. Thus, our formalism works better for collapse scenarios that have a greater difference between the eigenvalues, and hence, are closer to being quasi one-dimensional. Substituting eq. \eqref{eq:alpha} in eq. \eqref{eq:A2_cond}, we obtain the expression for the locus of $A_2$ singularities at a given instant $t > t_c$ up to $O(q^2, t)$ \citep[refer to eq. (17) in][]{Arnold_1982_ASZ}:
\begin{align}
    \label{eq:A2_eq}
      \phi^{4,0,0} q_x^2 + 2\phi^{3,1,0} q_x q_y + \left( \phi^{2,2,0} +\frac{2(\phi^{1,2,0})^2}{(1 - \phi^{0,2,0})}  \right) q_y^2 = -2 \phi^{2,0,1} (t - t_c),
\end{align}
which is of the form $A q_x^2 + B q_x q_y + C q_y^2 + F = 0$. It represents an ellipse if $4AC - B^2 > 0$. Since $F > 0$ for $t > t_c$, the ellipse is real only if $A, C < 0$. These conditions are equivalent to eq. \eqref{eq:maxima_cond} for the shell-crossing point to be a local maximum of the eigenvalue field $\alpha(\vec{q}, t)$. For $A, C < 0$,
\begin{align}
    \label{eq:A2_theta}
    &2\theta_{A2} = \arctan{\left(\frac{B}{|C-A|} \right)} = \arctan{\left( \frac{2\phi^{3,1,0}}{|\phi^{4,0,0} - \phi^{2,2,0} - \frac{2(\phi^{1,2,0})^2}{(1 - \phi^{0,2,0})} |} \right)},\\
    \nonumber
    &a = \left( \frac{2F \left( -(A+C) + \sqrt{ (C-A)^2 + B^2 } \right)}{ 4AC - B^2 } \right)^{1/2} \propto (t - t_c)^{1/2}\\
    \label{eq:A2_axis_lengths}
    &b = \left( \frac{2F \left( -(A+C) - \sqrt{ (C-A)^2 + B^2 } \right)}{ 4AC - B^2} \right)^{1/2} \propto (t - t_c)^{1/2},
\end{align}
where $\theta_{A2}$ is the angle by which the axes of the $A_2$ ellipse $(\hat{n}_H^{\alpha}, \hat{n}_H^{\beta})$ are rotated with respect to the reference frame $(\hat{n}^{\alpha}, \hat{n}^{\beta})$ at the shell-crossing point $(\vec{q}_c, t_c)$. The semi-major and semi-minor axes, of lengths $a$ and $b$, are oriented along the directions of less and more negative curvature of $\alpha$ at $(\vec{q}_c, t_c)$, denoted by $\hat{n}_H^{\alpha}$ and $\hat{n}_H^{\beta}$, respectively, and satisfy $0 < b/a \leq 1$. Consequently, the $\alpha$-ridge aligns with the semi-major axis of the $A_2$ ellipse. For the case where $\phi^{3,1,0} = B = \theta_{A2} = 0$, that is, the axes of the $A_2$ ellipse being aligned with the reference frame,  the shell-crossing direction $\hat{n}^{\alpha}$ coincides with the minor axis (more negative curvature) if $A < C < 0$. Interestingly, if $C < A < 0$, it coincides with the major axis (less negative curvature).

From eq. \eqref{eq:A2_axis_lengths}, all points on the $A_2$ ellipse satisfy $|\vec{q}_{A2}-\vec{q}_c| \propto (t - t_c)^{1/2}$. Using the parameterization $h = t-t_c $ and $q_{A2} \propto h^{1/2}$ in eqs. \eqref{eq:eq_motion_taylor_exp} to obtain the Eulerian shape of the $A_2$ caustic,
\begin{align}
    \label{eq:A2_eq_Eul}x_{A2}(h) &= P \: h + Q \: h^{3/2}\\
    \nonumber y_{A2}(h) &= R \: h^{1/2} + S \: h + T \: h^{3/2}.
\end{align}
The $A_2$ caustic demarcates the multi-streaming region in Eulerian space, within which our expansion form of the motion around the shell-crossing point, eq. \eqref{eq:eq_motion_taylor_exp}, remains valid. Therefore, eq. \eqref{eq:eq_motion_taylor_exp} is consistent up to $O(h^{3/2})$. Terms of $O(q^2)$ in $d\mathbf{x}/dt$ would have introduced some, but not all the terms of $O(h^2)$ in $\mathbf{x}_{A2}$ leading to inconsistency. Hence, they were neglected.

For saddle points of the eigenvalue field $\alpha(\vec{q}, t)$, the determinant of the Hessian eq. \eqref{eq:H} is negative, which implies $4AC - B^2 < 0$. Eq. \eqref{eq:A2_eq} then represents a hyperbola. For points of minima, the determinant is positive $\implies 4AC - B^2 > 0$. However, the trace $A + C$ being positive implies $A, C > 0$ for which eq. \eqref{eq:A2_eq} does not have real solutions after shell-crossing $t > t_c$.

Returning to our shell-crossing point, we substitute eqs. \eqref{eq:alpha} and \eqref{eq:n_alpha} in eq. \eqref{eq:A3_cond} and keep only the terms up to $O(q, t)$ to obtain \citep[refer to eq. (5.12) in][]{Feldbrugge_2023_feb}:
\begin{align}
    \nonumber &
    \begin{bmatrix}
        \phi^{4,0,0} q_{x} + \phi^{3,1,0} q_{y} \\
        \phi^{3,1,0} q_{x} + \Big( \phi^{2,2,0} +\frac{2(\phi^{1,2,0})^2}{(1 - \phi^{0,2,0})}  \Bigg) q_{y}
    \end{bmatrix}
    \cdot \hat{n}^{\alpha} = 0 \\
    \label{eq:A3_eq} & \implies \phi^{4,0,0} q_{x} + \phi^{3,1,0} q_{y} = 0.
\end{align}
This is the locus of $A_3^+$ points in Lagrangian space, regardless of the time when they turn singular. It is a straight line at an angle $\theta_{A3}$ with respect to the reference axis $\hat{q}_y \equiv \hat{n}^{\beta}(\vec{q}_c, t_c)$
\begin{equation}
    \label{eq:theta_A3}
    \theta_{A3} = \arctan{\left( - \phi^{3,1,0} / \phi^{4,0,0} \right)},
\end{equation}
and is time-independent for our expansion form, eq. \eqref{eq:eq_motion_taylor_exp}, of the order $O(q^3, t)$. Higher order expansions would indeed introduce $O(q^{\ge2}, t^{\ge1})$ corrections. Refer to fig. \ref{fig:phi_310} to discern the alignments between $A_2$ ellipse, $A_3^+$ line and reference axes.

To obtain the Eulerian shape of the $A_3^+$ line, we substitute eq. \eqref{eq:A3_eq} in our expansion form eq. \eqref{eq:eq_motion_taylor_exp}. The expressions for $x_{A3}(\vec{q}, t)$ and $y_{A3}(\vec{q}, t)$ in parametric form are provided in the appendix, eq. \eqref{eq:A3_eq_Eul_full}. It is to be noted that eq. \eqref{eq:A3_eq} is of $O(q)$ and eq. \eqref{eq:eq_motion_taylor_exp} is of $O(q^3)$. So, upon substitution, eq. \eqref{eq:A3_eq_Eul_full} for the $A_3^+$ spine is strictly correct only up to $O(q)$. To have the complete expression of eq. \eqref{eq:A3_eq_Eul_full} correct up to $O(q^3)$, we need to start with expansion form, eq. \eqref{eq:eq_motion_taylor_exp}, of the order $O(q^5)$ that will result in eq. \eqref{eq:A3_eq} of the order $O(q^3)$, that is, quadratic and cubic corrections to the $A_3^+$ line in Lagrangian space. An $O(q^5)$ expansion in space necessitates $O(t^2)$ in time so that our expansion form is consistent up to $O(h^{5/2})$ (refer to eq. \eqref{eq:A2_eq_Eul}). However, $O(t^2)$ expansion requires PCPT for force field corrections, which is beyond the scope of this work. Nevertheless, only at shell-crossing $(t = t_c)$, the $O(q)$ term in expansion of $x(\vec{q},t)$, eq. \eqref{eq:eq_motion_taylor_exp}, vanishes and hence, the $A_3^+$ spine's trajectory in Eulerian space:
\begin{align}
    \label{eq:A3_eq_Eul}
    x_{A3}(y_{A3}, t_c) = &- \left[ \frac{\phi^{1,2,0}}{ 2 \left( 1 - \phi ^{0,2,0} \right)^2} \right] \: y_{A3}^2 \\
    \nonumber
    &- \Bigg[ \frac{\phi^{1,3,0} + \phi^{3,1,0} \left( 2\left(\frac{\phi^{3,1,0}}{\phi^{4,0,0}}\right)^2 - 3\frac{\phi^{2,2,0}}{\phi^{4,0,0}} \right) }{6\left( 1 - \phi ^{0,2,0} \right)^3} \\
    \nonumber
    & \hspace{0.6 cm}+ \frac{ \phi^{1,2,0} }{\left( 1 - \phi ^{0,2,0} \right)^4} \left(\frac{1}{2} \phi ^{0,3,0}-\frac{\phi ^{1,2,0} \phi ^{3,1,0}}{\phi ^{4,0,0}}\right) \Bigg] \: y_{A3}^3,
\end{align}
is indeed correct up to $O(y^2)$. The only missing term is at the order $O(y^3)$ and involves $O(q^5)$ derivative of the potential $\phi$. On physical grounds, we can argue that if the field $\phi$ is Gaussian distributed, its $O(q^5)$ derivatives and their effects on the pancake shape are typically suppressed compared to those of the lower order derivatives for a relevant range of power spectra (refer to the appendix fig. \ref{fig:sigma_n} and the discussion therein). Eq. \eqref{eq:A3_eq_Eul} is therefore an accurate enough approximation at shell-crossing, which is where we focus the derivation of expressions for properties related to the pancake shape.

We can transform the intersection of eqs. \eqref{eq:A2_eq} and \eqref{eq:A3_eq} through eq. \eqref{eq:A3_eq_Eul_full} to obtain the Eulerian location of the tips of the growing pancake at instants shortly after shell-crossing $t > t_c$. From  eq. \eqref{eq:A3_eq_Eul_full}, the leading order term $O(q_y)$ in $x_{A3}(q, t)$ is time dependent suggesting that the spine gradually rotates. At shell-crossing $t = t_c$, however, it vanishes, which means the spine is aligned with our reference axis $\hat{q}_y \equiv \hat{n}^{\beta}(\vec{q}_c, t_c)$. The order $O(q_y^2)$ term depends on $\phi^{1,2,0}$ and contributes to the C-shape of the spine. Lastly, the order $O(q_y^3)$ term depends on $\phi^{1,3,0}$ and $\phi^{3,1,0}$, and contributes to the S-shape of the spine.

From eq. \eqref{eq:A3_eq_Eul}, we compute a few properties related to the shape of the pancake at shell-crossing:
\begin{enumerate}
    \item The curvature of the $A_3^+$ spine:
    \begin{equation}
        \label{eq:A_3_curvature}
        \frac{d^2 \: x_{A3}}{d \: y_{A3}^2} \Bigg|_{q_y = 0} (t_c) = \frac{\phi^{1,2,0}}{\left( 1 - \phi^{0,2,0} \right)^2}.
    \end{equation}
   Previously \citet{Doroshkevich_1978} (in section 5.3) approximated the mean curvature of the lateral surfaces (equivalent to the $A_2$ boundary in 2D) of the pancake in 3D Eulerian space, assuming it to be infinitely thin and symmetric. It is different from the $A_3^+$ spine's curvature eq. \eqref{eq:A_3_curvature}, which is related to the leading order correction to the presumed symmetrical shape of the pancake. 

    \item From eq. \eqref{eq:A3_eq_Eul}, we note that close to the shell-crossing point, the $A_3^+$ spine is C-shaped given the $O(y^2)$ term. At larger scales, it turns S-shaped as the $O(y^3)$ term starts to dominate. The transition scale provides an estimate of the extent up to which the pancake can be considered C-shaped, which we estimate using the point $y$ at which the spine starts to bend from C to S in Eulerian space at shell-crossing, that is $d x_{A3}/dy_{A3} \: (t_c) = 0$, which implies
    \begin{align}
        \label{eq:A_3_C_S_trans}
        y_{C\rightarrow S}(t_c) &= -2\left( 1 - \phi ^{0,2,0} \right) \phi^{1,2,0} \times \\ 
        \nonumber
        &\Bigg[ \phi^{1,3,0} + \phi^{3,1,0} \left( 2\left(\frac{\phi^{3,1,0}}{\phi^{4,0,0}}\right)^2 - 3\frac{\phi^{2,2,0}}{\phi^{4,0,0}} \right)\\
        \nonumber
        & \hspace{0.3 cm} + \frac{6 \phi^{1,2,0}}{\left( 1 - \phi ^{0,2,0} \right)} \left(\frac{1}{2} \phi ^{0,3,0}-\frac{\phi ^{1,2,0} \phi ^{3,1,0}}{\phi ^{4,0,0}}\right) \Bigg]^{-1}.
    \end{align}
    For halos, $\phi^{0,2,0}, \phi^{0,2,1} > 0$, which means the transverse axis contracts. Thus, the curvature increases and the $y_{C\rightarrow S}$ decreases with time. On the other hand, for filaments, $\phi^{0,2,0}, \phi^{0,2,1} \le 0$, which means the transverse axis stretches, leading to decreasing curvature and increasing $y_{C\rightarrow S}$. 
\end{enumerate}

\subsection{Parameters and their effects}
The shape of the shell-crossing structure in Lagrangian space is captured by the loci of $A_2$ and $A_3^+$ points, which are governed by eqs. \eqref{eq:A2_eq} and \eqref{eq:A3_eq}, respectively. Their transformation to Eulerian space is governed by eq. \eqref{eq:eq_motion_taylor_exp}. These three sets of equations constitute our analytical model and contain ten parameters in total. Constraints on the signs of some of these parameters are listed in eq. \eqref{eq:coeff_matrix}. We further isolate and investigate the effect of each of these parameters on the shape of the pancake. The values of parameters used in the figures \ref{fig:phi_201_020}, \ref{fig:phi_120_phi_130}, \ref{fig:phi_220_phi_400}, and \ref{fig:phi_030_phi_040} are only for the purpose of presentation.

\begin{itemize}
    \item $\phi^{2,0,1}$: velocity gradient along shell-crossing axis. It determines how quickly the particles cross trajectories and turn singular, and hence, the rate at which the $A_2$ caustic grows in Lagrangian, eq. \eqref{eq:A2_eq}, and Eulerian spaces.

    \item $\phi^{0,2,0}, \phi^{0,2,1}$: displacement and velocity gradient along the transverse axis. They determine if the pancake stretches or shrinks longitudinally in Eulerian space, eq. \eqref{eq:eq_motion_taylor_exp}.
    \begin{figure}
        \centering
        \includegraphics[width=\linewidth]{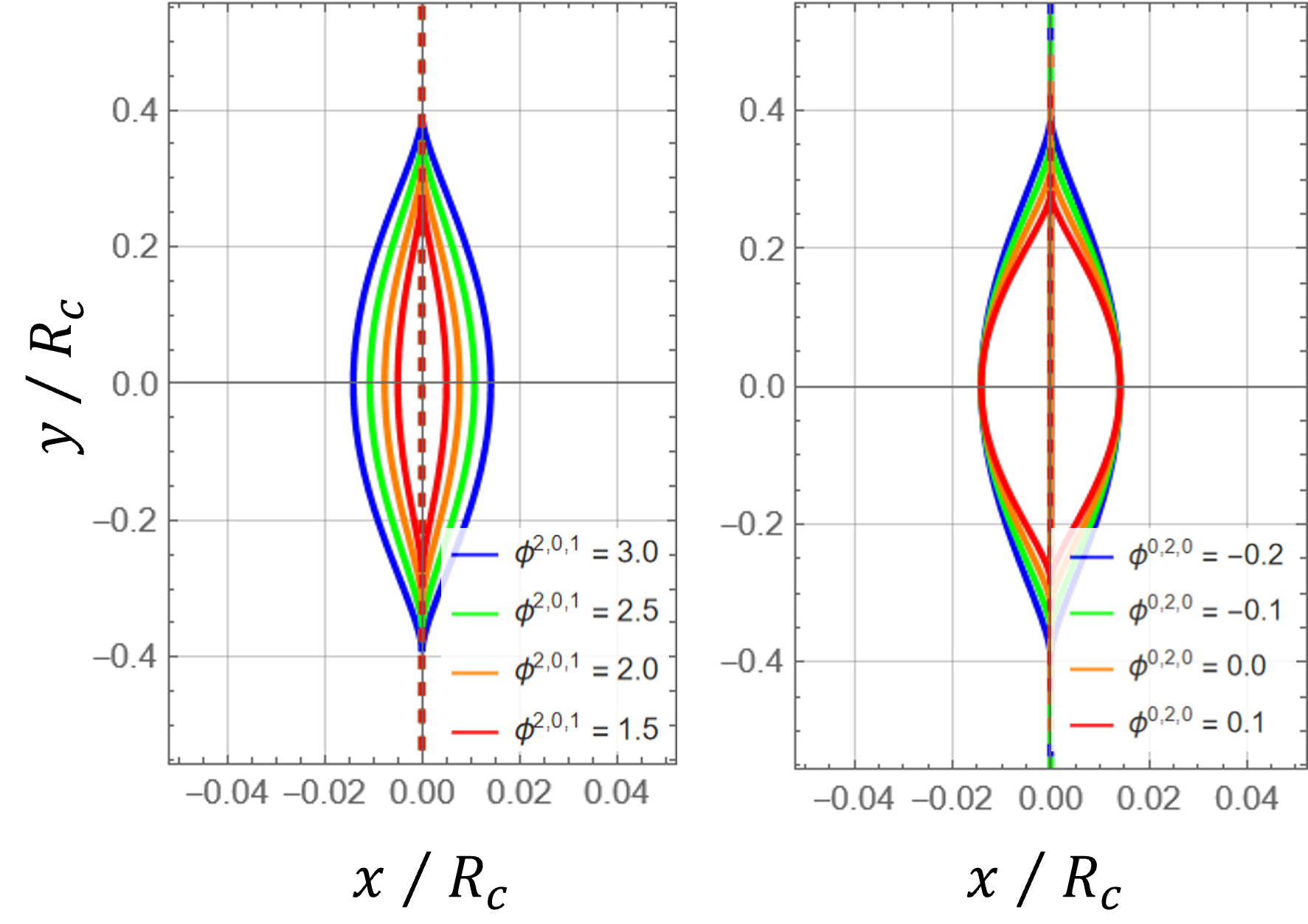}
        \caption{Effect of varying $\phi^{2,0,1}$ (left) and $\phi^{0,2,0}$ (right) on the shape of $A_2$ caustic in Eulerian space, while keeping the other parameters and the time same.}
        \label{fig:phi_201_020}
    \end{figure}

    \item $\phi^{4,0,0}, \phi^{2,2,0}$: they are related to the two eigenvalues --- the principal curvatures --- of $H_{ij}(\alpha(\vec{q}_c,t_c))$, eq. \eqref{eq:H}. They determine the axis ratio of the $A_2$ caustic, eq. \eqref{eq:A2_axis_lengths}, which in turn affects how flat the pancake is in Eulerian space.
    \begin{figure}
        \centering
        \includegraphics[width=\linewidth]{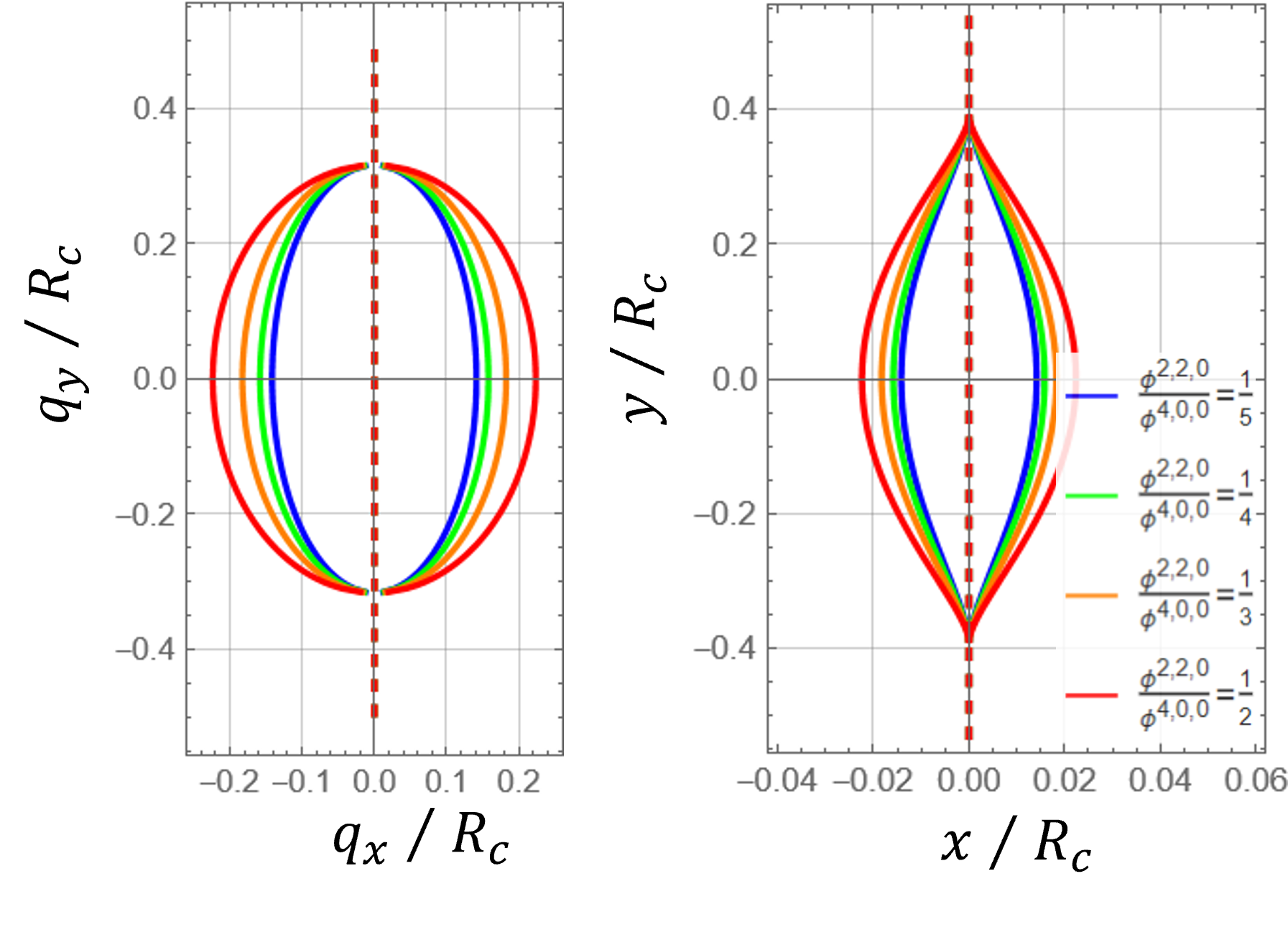}
        \caption{Effect of varying the ratio $\phi^{2,2,0} / \phi^{4,0,0}$ on the shape of $A_2$ caustic in Lagrangian (left) and Eulerian (right) spaces, while keeping the other parameters and the time same.}
        \label{fig:phi_220_phi_400}
    \end{figure}

    \item $\phi^{1,2,0}, \phi^{1,3,0}$: they affect the shape of the $A_3$ spine eq. \eqref{eq:A3_eq_Eul} in Eulerian space. The former is responsible for the 'C' shape, and the latter is responsible for the 'S' shape.
    \begin{figure}
        \centering
        \includegraphics[width=\linewidth]{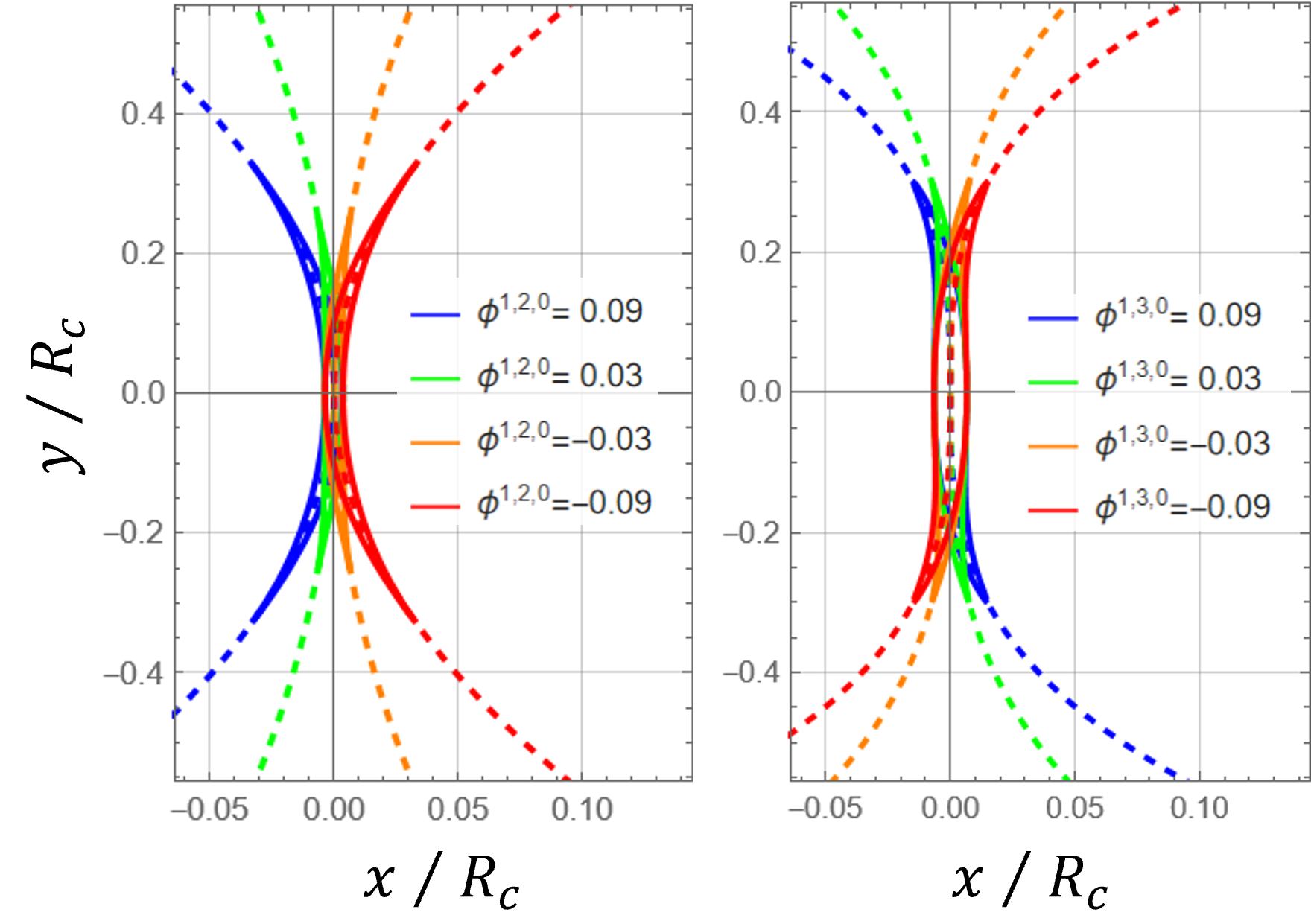}
        \caption{Effect of varying $\phi^{1,2,0}$ (left) and $\phi^{1,3,0}$ (right) on the shape of $A_2$ caustic and $A_3$ spine in Eulerian space, while keeping the other parameters and the time same.}
        \label{fig:phi_120_phi_130}
    \end{figure}

    \item $\phi^{0,3,0}, \phi^{0,4,0}$: the former adds asymmetric and the latter adds symmetric higher order corrections to the longitudinal stretch of the pancake in Eulerian space, eq. \eqref{eq:eq_motion_taylor_exp}.
    \begin{figure}
        \centering
        \includegraphics[width=\linewidth]{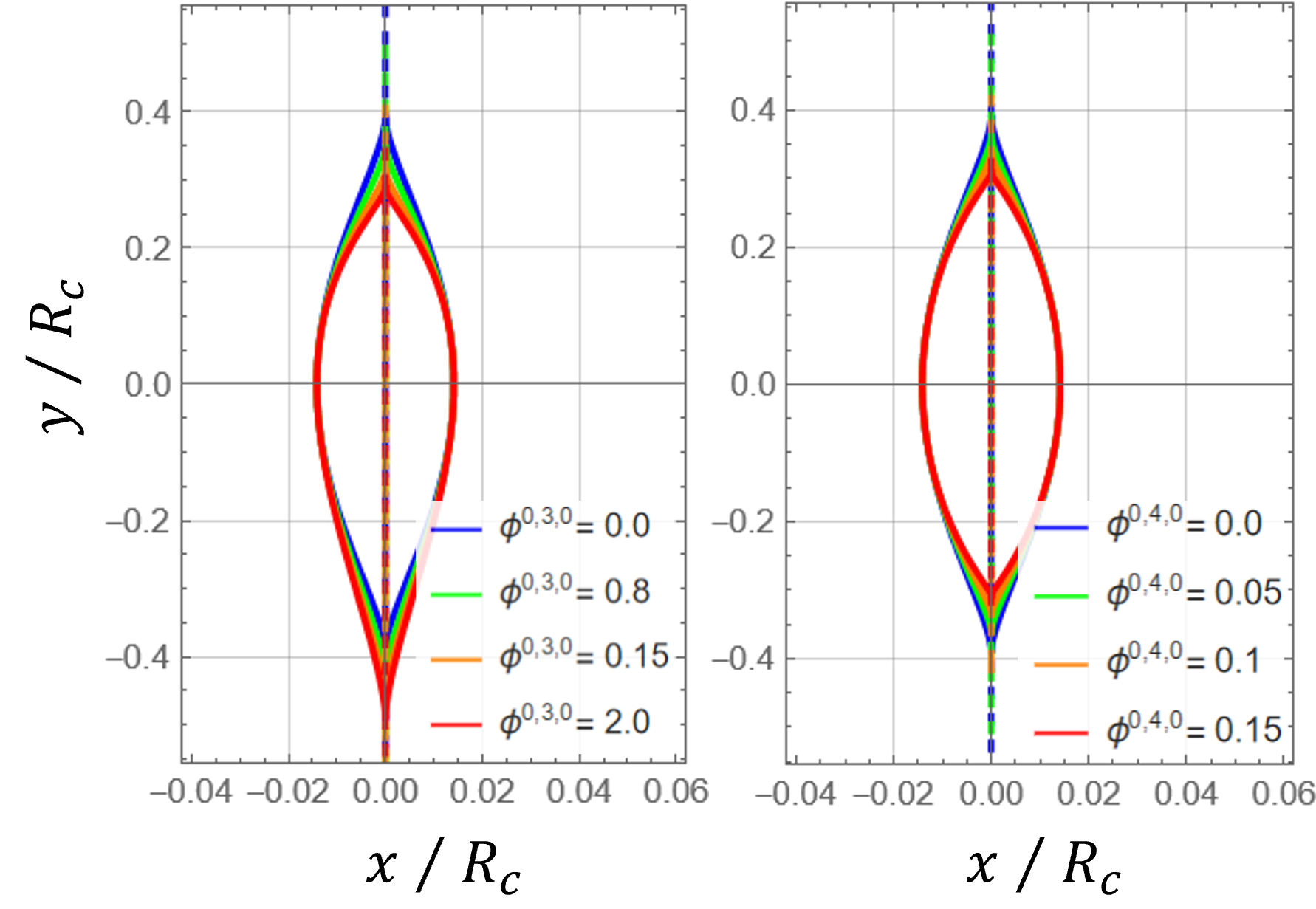}
        \caption{Effect of varying $\phi^{0,3,0}$ (left) and $\phi^{0,4,0}$ (right) on the shape of $A_2$ caustic in Eulerian space, while keeping the other parameters and the time same.}
        \label{fig:phi_030_phi_040}
    \end{figure}

    \item $\phi^{3,1,0}$: leads to rotation of the $A_3^+$ spine at leading order and contributes to its S-shape at order $O(y^3)$ in Eulerian space, eq. \eqref{eq:A3_eq_Eul}.
    \begin{figure}
        \centering
        \includegraphics[width=\linewidth]{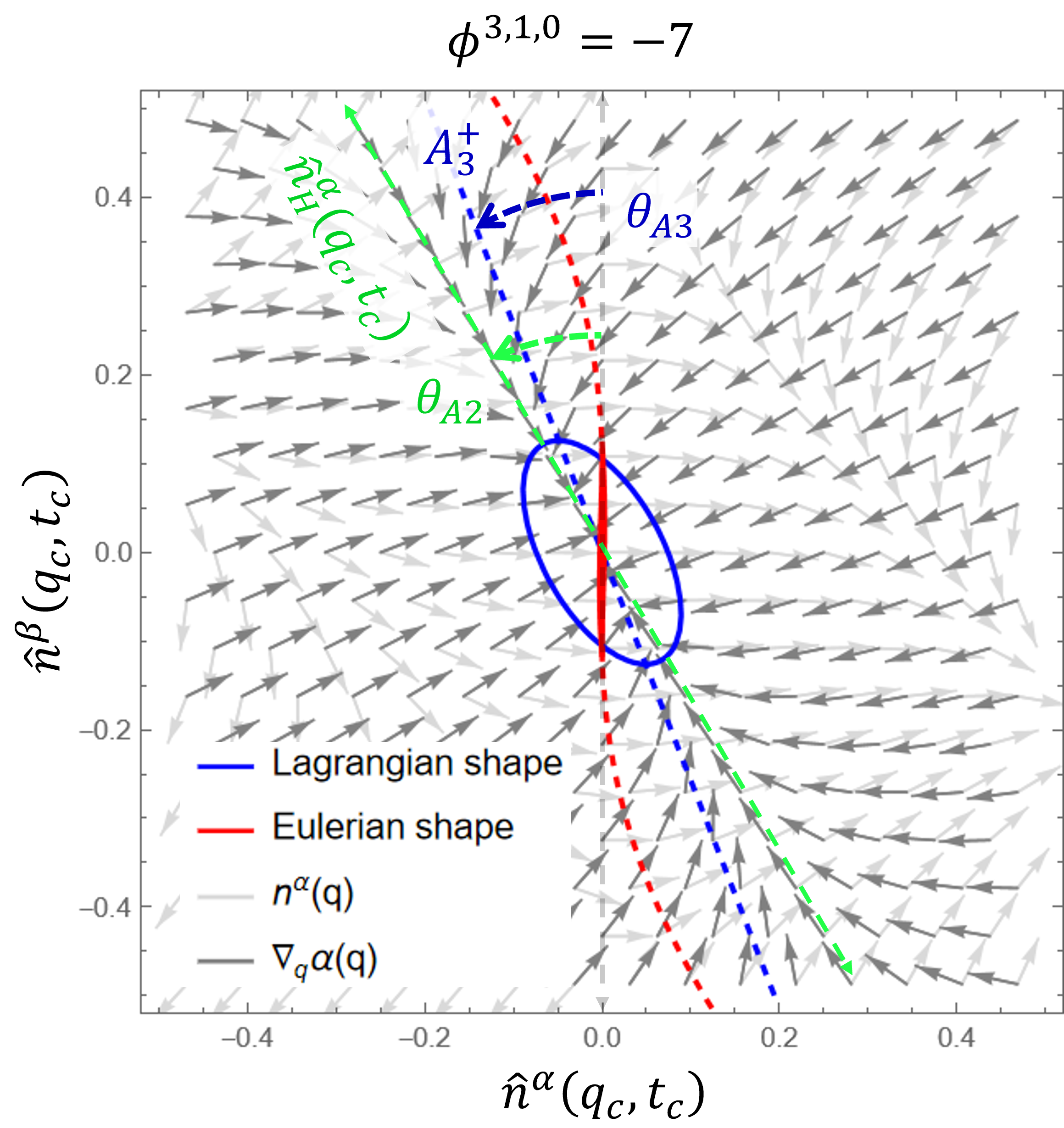}
        \caption{Effect of $\phi^{3,1,0}$ on the shape of the $A_2$ caustic (solid) and the $A_3^+$ spine (dashed) in Lagrangian (blue) and Eulerian (red) spaces. The angles $\theta_{A2}$, eq. \eqref{eq:A2_theta}, and $\theta_{A3}$, eq. \eqref{eq:theta_A3}, have been indicated in green and blue, respectively. The vector fields of eigenvector $\hat{n}^{\alpha}$ and gradient $\nabla_{\vec{q}} \alpha$ are shown in light and dark gray arrows, respectively.}
        \label{fig:phi_310}
    \end{figure}
\end{itemize}

\section{Gaussian statistics}
\label{sec:gaussian_stats}

Using arguments from catastrophe theory, we modeled how the derivatives of the displacement potential $\phi$ at the shell-crossing point $(\vec{q}_c, t_c)$ relate to the shape of the emerging pancake. There are ten such derivatives, or, as we prefer to call them, the parameters in our model that appear in our expansion form, eq. \eqref{eq:eq_motion_taylor_exp}. In this section, starting from a Gaussian random field for the displacement potential $\phi(\vec{q})$ and its derivatives, we compute the conditional and marginal probabilities of our model parameters subject to the constraints of a shell-crossing point, and obtain the distribution and expectation values of the features defining the shape of the pancakes.

We first construct our random variables. As discussed earlier, the shape of the pancake is captured by the $A_2$ caustic and the $A_3$ spine. Properties like the curvature, eqs. \eqref{eq:A_3_curvature}, and the transition scale from C to S-shape depend on $O(q^2), O(q^3)$ and $O(q^4)$ derivatives of the displacement potential $\phi$ at shell-crossing. For the purpose of computing statistics, we assume Zel'dovich flow (ZA) before shell-crossing. \citet{Coles_1993} and \citet{Melott_1994} showed that the ZA is a good approximation down to the non-linearity scale, by which stage most pancakes have already formed yet remain relatively thin. So, ZA can be safely used to capture the time-evolution of the potential $\phi$ up to shell-crossing: $\phi(\vec{q},t) = D_+(t) \phi(\vec{q})$. With this approximation, the time derivatives $\phi^{i,j,1}$'s are proportional to $ \phi^{i,j,0}$'s, which makes their constraints redundant in the computation of conditional probabilities at the shell-crossing point. For our random variables, it is thus sufficient to consider the set of $O(q^2), O(q^3)$ and $O(q^4)$ spatial derivatives for computing the properties we are interested in. The Zel'dovich approximation (ZA) also allows for a rescaling of these derivatives, thereby eliminating the explicit time dependence.
:
\begin{equation}
    \label{eq:phi_scaling_Dtc}
    \phi^{i,j}(\vec{q}) = D_+^{-1}(t) \phi^{i,j,0}(\vec{q}, t).
\end{equation}

We define two random variables: $\tilde{Y}$, which comprises the derivatives at a point in the generic frame, and $Y$, which comprises the derivatives in the reference frame (same as used in section \ref{sec:catastrophe_theory}) rotated to align with the eigenvectors $\hat{n}^{\alpha}, \hat{n}^{\beta}$ at that point, eq. \eqref{eq:n_alpha_LPT}.

\begin{align}
    \nonumber
    &\tilde{Y} = \left[ \tilde{\phi}^{2,0} ,  \tilde{\phi}^{1,1} ,  \tilde{\phi}^{0,2} ,  \tilde{\phi}^{3,0} ,  \tilde{\phi}^{2,1} ,  \tilde{\phi}^{1,2} ,  \tilde{\phi}^{0,3} ,  \tilde{\phi}^{4,0} ,  \tilde{\phi}^{3,1} ,  \tilde{\phi}^{2,2} ,  \tilde{\phi}^{1,3} ,  \tilde{\phi}^{0,4} \right],\\
    \label{eq:Y_tilda_and_Y}&Y = \left[ \phi^{2,0} ,  \theta ,  \phi^{0,2} , \phi^{3,0},\phi^{2,1} ,  \phi^{1,2} ,  \phi^{0,3} ,  \phi^{4,0} ,  \phi^{3,1} ,  \phi^{2,2} ,  \phi^{1,3} ,  \phi^{0,4} \right],
\end{align}
where $\theta$ is the rotation angle between the two frames, and it diagonalizes the deformation matrix:
\begin{equation}
    \begin{bmatrix}
        \tilde{\phi}^{2,0} & \tilde{\phi}^{1,1} \\
        \tilde{\phi}^{1,1} & \tilde{\phi}^{0,2}
    \end{bmatrix}
    =
    \begin{bmatrix}
        \cos{\theta} & \sin{\theta} \\
        -\sin{\theta} & \cos{\theta}
    \end{bmatrix}
    \begin{bmatrix}
        \phi^{2,0} & 0 \\
        0 & \phi^{0,2}
    \end{bmatrix}
    \begin{bmatrix}
        \cos{\theta} & -\sin{\theta} \\
        \sin{\theta} & \cos{\theta}
    \end{bmatrix}.
\end{equation}
The transformations of the $O(q^3)$ and $O(q^4)$ derivatives under rotation, as well as the Jacobian $d\tilde{Y}_i/dY_j$, are given in the appendix (eqs.~\eqref{eq:Y_tilda_Y} and \eqref{eq:J_Y_tilda_Y}; see also eqs.~E.3--E.5 in \citealt{Feldbrugge_2023_feb}). The determinant of the Jacobian is
\begin{equation}
    \label{eq:Det_J_Y_tilda_Y}
    \left\lVert \frac{d \tilde{Y}}{dY} \right\rVert 
    = \big| \phi^{2,0} - \phi^{0,2} \big| .
\end{equation}
We know that $\tilde{\phi}(\vec{q})$ is Gaussian distributed with the power spectrum $P_{\tilde{\phi}}(k)$, which is related to the matter power spectrum $P_{\delta}(k)$ through eq. \eqref{eq:P_phi_delta_k}. Therefore, $\tilde{Y}$ comprising derivatives of $\phi(\vec{q})$ is Gaussian distributed as well. Its probability density is
\begin{equation}
    \label{eq:P_Ytilda}
    P(\tilde{Y}) = \frac{1}{\sqrt{ (2\pi)^N |M|}} \exp{ \left( -\frac{1}{2} \tilde{Y}^T M^{-1} \tilde{Y} \right) },
\end{equation}
where $N$ is the number of elements in $\tilde{Y}$ and $M$ is the covariance matrix between $\tilde{Y}$ and $Y$, given by 
\citep[see also eqs.~57--65 in][]{Feldbrugge_2025}:
\begin{align}
\nonumber 
M_{2,2} &= \frac{\sigma_2^2}{8}
\begin{bmatrix}
    3 & 0 & 1 \\
    0 & 1 & 0 \\
    1 & 0 & 3
\end{bmatrix},
\\
\nonumber 
M_{2,4} &= -\frac{\sigma_3^2}{16}
\begin{bmatrix}
    5 & 0 & 1 & 0 & 1 \\
    0 & 1 & 0 & 1 & 0 \\
    1 & 0 & 1 & 0 & 5
\end{bmatrix},
\\
\nonumber
M_{3,3} &= \frac{\sigma_3^2}{16}
\begin{bmatrix}
    5 & 0 & 1 & 0 \\
    0 & 1 & 0 & 1 \\
    1 & 0 & 1 & 0 \\
    0 & 1 & 0 & 5
\end{bmatrix},\\
\label{eq:covariance_M} 
M_{4,4} &= \frac{\sigma_4^2}{128}
\begin{bmatrix}
    35 & 0 & 5 & 0 & 3 \\
    0  & 5 & 0 & 3 & 0 \\
    5  & 0 & 3 & 0 & 5 \\
    0  & 3 & 0 & 5 & 0 \\
    3  & 0 & 5 & 0 & 35
\end{bmatrix},
\end{align}
where $M_{i,j}$ denotes the covariance matrix between $O(q^i)$ and 
$O(q^j)$ derivatives of the potential $\phi$. The $O(q^3)$ derivatives have no correlation with $O(q^2)$ or $O(q^4)$ derivatives. $\sigma_i$ refers to the $i^{th}$ moment of the power spectrum $P_{\tilde{\phi}}(k)$:
\begin{align}
    \label{eq:sigma}
    \sigma_i^2 &= \frac{1}{2\pi} \int_0^{\infty} k^{2i+1}P_{\tilde{\phi}}(k) dk \\
    \nonumber
    \propto  &\frac{1}{R_c^{n+2(i-1)}}
    \begin{cases}
        \Gamma(n+2(i-1)) \text{ ; Exponential filter } P_{\tilde{\phi}} \propto k^{n-4} e^{-kR_c}\\
        \Gamma(n/2+i-1) \text{ ; Gaussian filter } P_{\tilde{\phi}} \propto k^{n-4} e^{-k^2 R_c^2}
    \end{cases}
\end{align}
where $n$ and $R_c$ are the spectral index and smoothing scale, respectively. These $\sigma_i$'s encapsulate the dependence on the power spectrum parameters $n, R_c$ as well as the smoothing filter used, thereby capturing all the statistical information about the initial potential field $\phi(\vec{q})$ that we require for our computations. Their dependence on the spectral index $n$ is shown in fig. \ref{fig:sigma_n}.

To obtain $P(Y)$, we use eqs. \eqref{eq:P_Ytilda} and \eqref{eq:Det_J_Y_tilda_Y}:
\begin{align}
    \nonumber &P(Y) \: dY = P(\tilde{Y}) \: d\tilde{Y}\\
    \nonumber \implies &P(Y) = \left\Vert \frac{d\tilde{Y}}{dY} \right \Vert \: P(\tilde{Y})\\
    \label{eq:PY} \implies &P(Y) \propto \left| \phi ^{2,0} - \phi ^{0,2} \right| \: \exp{ \left( -\frac{1}{2} \tilde{Y}^T(Y) \: M^{-1} \: \tilde{Y}(Y) \right) }
\end{align}
The full expression after substituting $\tilde{Y}$ in terms of $Y$, eq. \eqref{eq:Y_tilda_Y}, and the covariance matrix $M$, eq. \eqref{eq:covariance_M} is provided in the appendix, eq. \eqref{eq:P_Y_full}. This is the probability density of $Y$ at a generic spatial point $\vec{q}$. We are interested in the conditional probability of $Y$ given that it is a shell-crossing point $\vec{q}_c$ --- a peak in the eigenvalue field $\alpha(\vec{q})$ which seeds a pancake. For this, we take an approach similar to the computation of the number density of peaks in \citet{Bardeen_1986}. Refer to the constraints, eq. \eqref{eq:coeff_matrix}.
\begin{itemize}
    \item At shell-crossing, the higher eigenvalue $\alpha(\mathbf{q_c}, t_c) = \phi^{2,0,0}$ is equal to $1$, which implies $\phi^{2,0} = D_+^{-1}(t_c) > 0$. For pancakes, referring to the total set of halos and filaments, it is sufficient to have $\alpha(\mathbf{q_c},t_c) > \beta(\mathbf{q_c}, t_c) \implies  \phi^{2,0} > \phi^{0,2}$. For the subset of halos, $\phi^{0,2}$ must be further restricted to $\phi^{0,2} > 0$ to have shell-crossing along the second axis as well. On the other hand, for the subset of filaments, $\phi^{0,2} \leq 0$ so that the second axis does not collapse. All these constraints are introduced as Heaviside functions.
    \item The shell-crossing point is a local maximum in the eigenvalue field $\alpha(\vec{q}, t)$. This implies constraints on its gradient, $\phi^{3,0}, \phi^{2,1} = 0$, which appear as Dirac-delta functions, and on the eigenvalues of its Hessian, eq. \eqref{eq:maxima_cond}, which appear as Heaviside functions in the conditional probability.
\end{itemize}
The conditional probability of $Y$ can be expressed as:
\begin{align}
    \label{eq:CP_Y}
    P \: (Y \: |& \: \text{Pancake}) \propto \: P(Y) \: \delta_D(\phi^{3,0}) \:\delta_D(\phi^{2,1})\\
    \nonumber
    &  H[\phi^{2,0}] \: H[\phi^{2,0} - \phi^{2,0}] \: H\left[ -\phi^{4,0} - \phi^{2,2} - \frac{2 (\phi^{1,2})^2}{\phi^{2,0} - \phi^{0,2}} \right] \\
    \nonumber
    & H\left[ \phi^{4,0} \left( \phi^{2,2} + \frac{2 (\phi^{1,2})^2}{\phi^{2,0} - \phi^{0,2}} \right) - (\phi^{3,1})^2 \right],
\end{align}
where $H$ refers to the Heaviside function. For halo and filament populations, we need to further multiply $H[\phi^{0,2}]$ and $H[-\phi^{0,2}]$, respectively.

\subsection{Marginal probabilities}

The conditional probability eq. \eqref{eq:CP_Y} is completely parameterized by the moments of power spectrum $\sigma_2, \sigma_3$, and $\sigma_4$ through eq. \eqref{eq:P_Y_full}. The choice of the power spectrum parameters $n, R_c$, and the smoothing filter affects the probability distribution only through the values of $\sigma$'s. If $\phi^{i,j}$ is uncorrelated with other derivatives, the expectation value $\langle | \phi^{i,j} | \rangle $ is proportional to $\sigma_{i+j}$. Even for the correlated ones, $\phi^{i,j} \: / \: \sigma_{i+j}$ is only weakly dependent on the spectral index $n$ (refer to the appendix fig. \ref{fig:MP_phi_n} and the discussion therein). It is thus useful to present the distributions in terms of quantities that are scaled by their corresponding $\sigma_{i+j}$'s, so that they are dimensionless and effectively independent of the choice of power spectrum parameters and the smoothing filter.

\begin{figure*}
    \centering
    \includegraphics[width=0.4\linewidth]{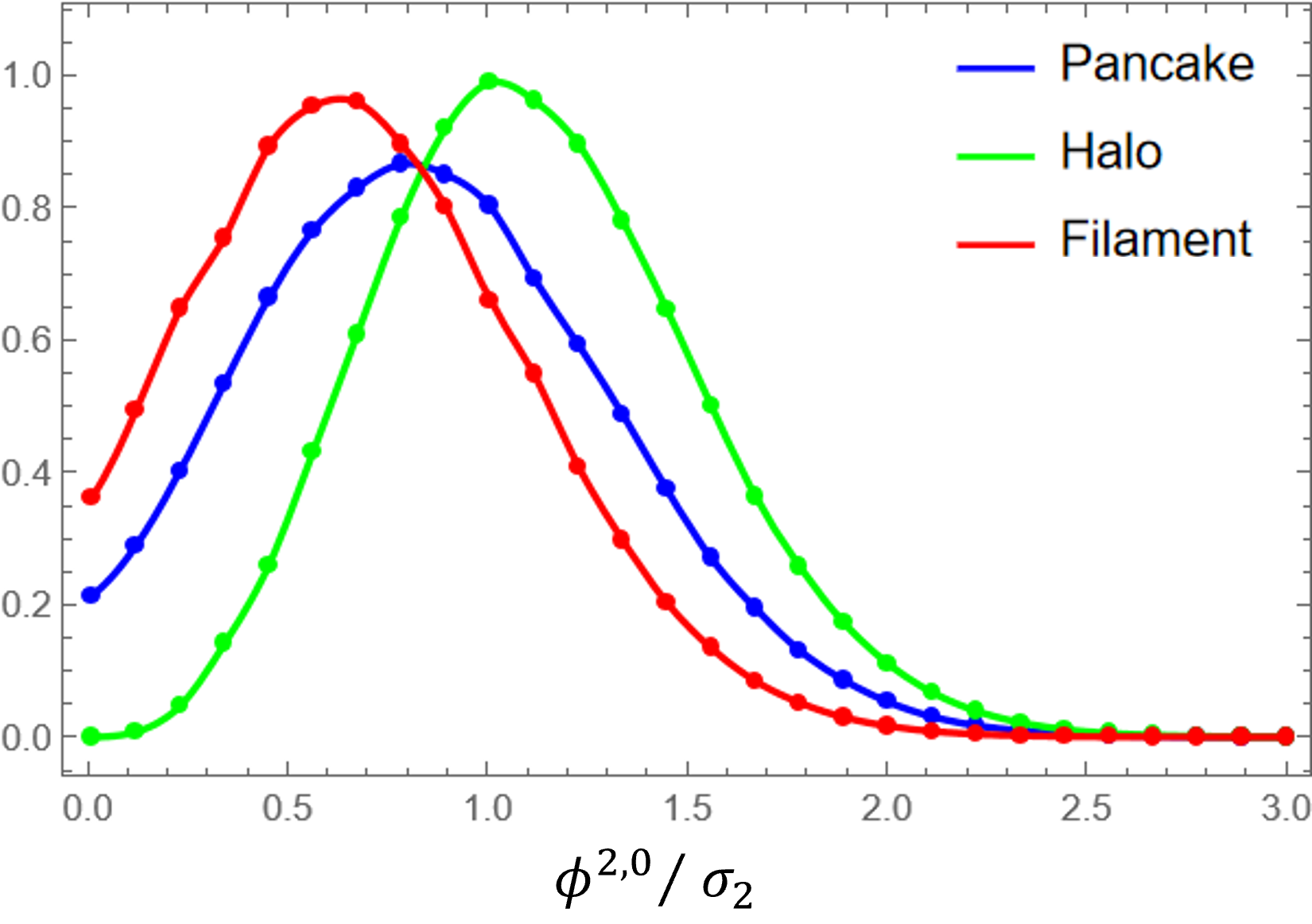}
    \includegraphics[width=0.4\linewidth]{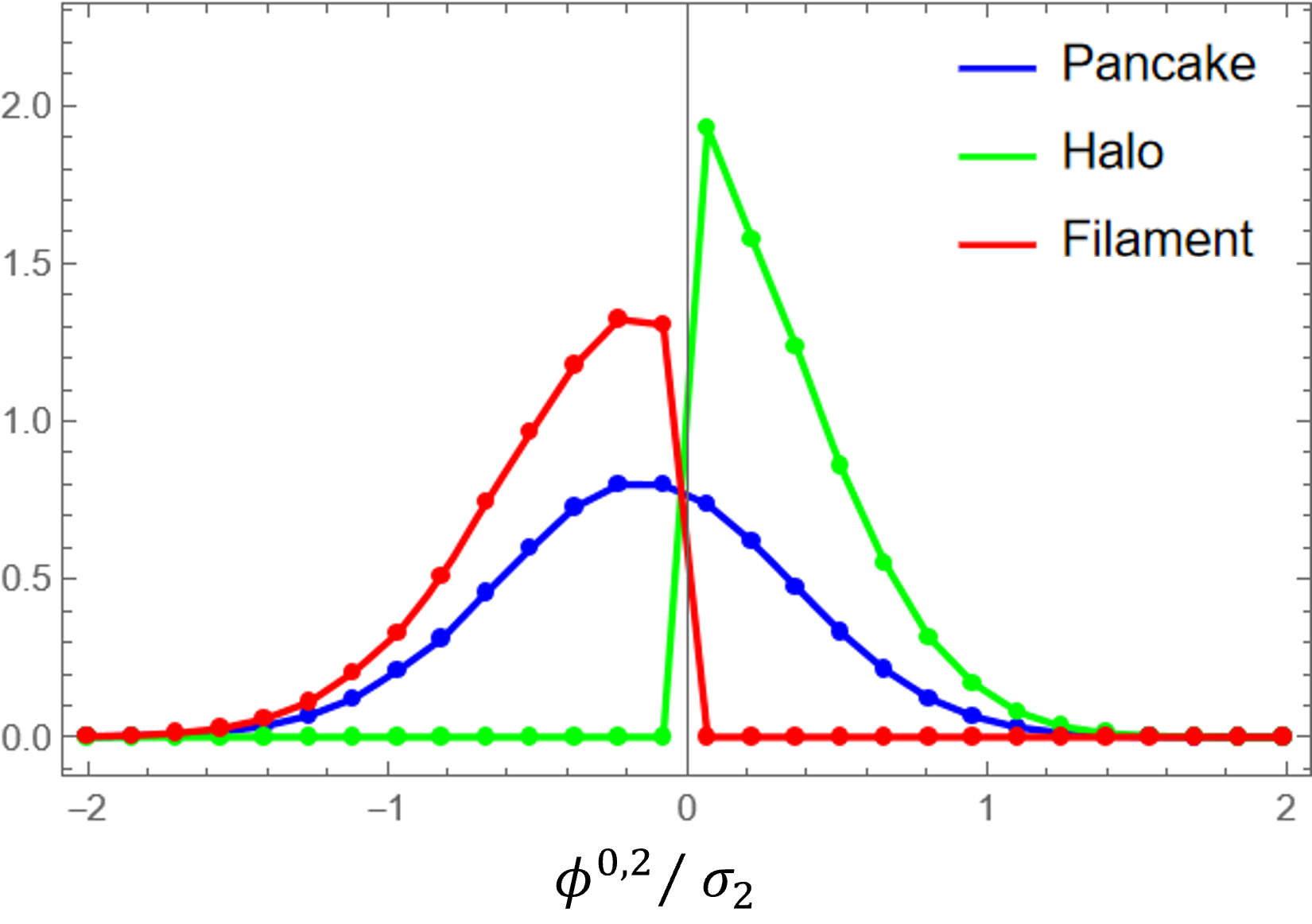}
    \caption{Marginal probability densities of $\phi^{2,0}$ (left) and $\phi^{0,2}$ (right) scaled by $\sigma_2$.}
    \label{fig:MP_q2}
\end{figure*}

\begin{figure*}
    \centering
    \includegraphics[width=0.4\linewidth]{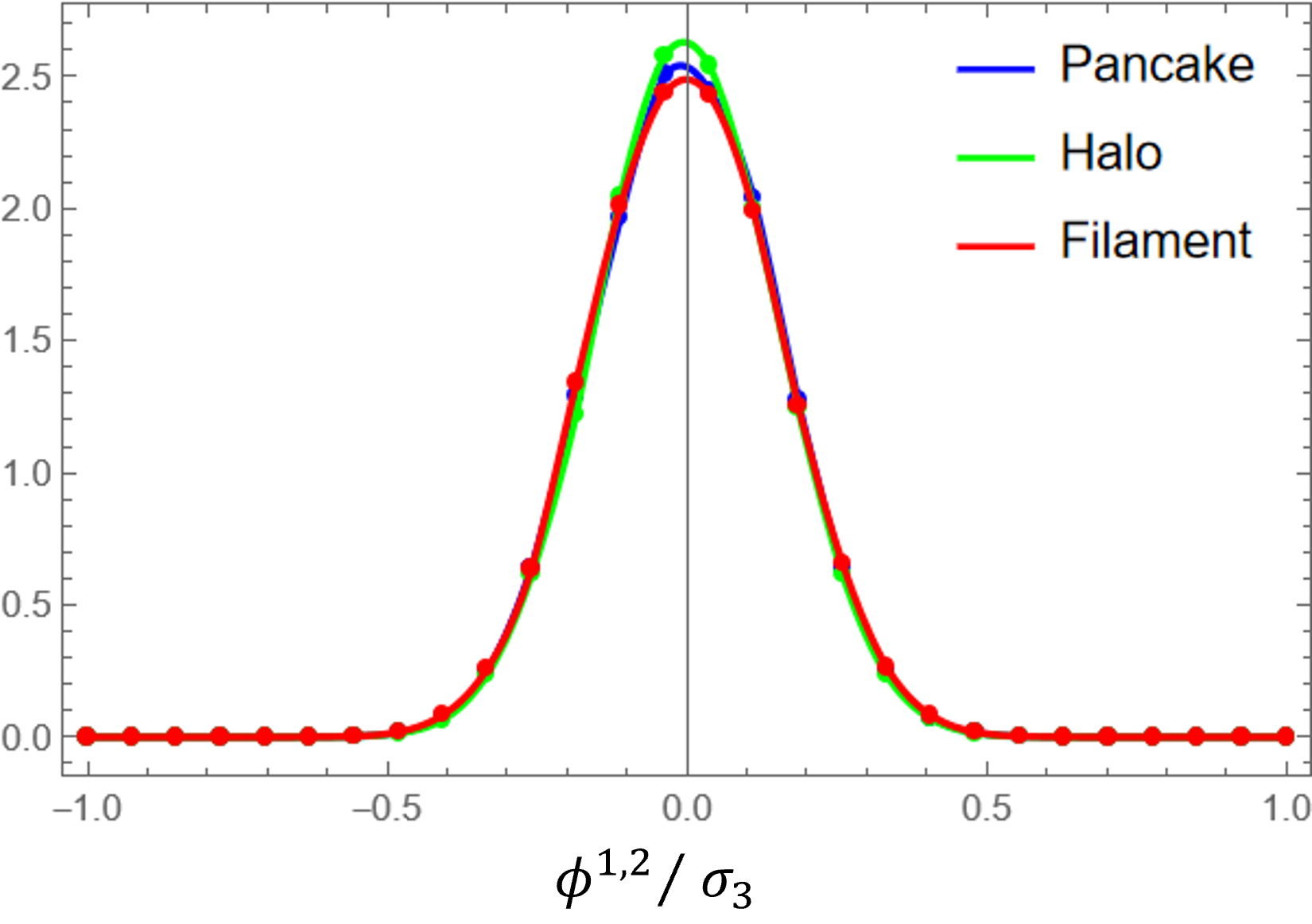}
    \includegraphics[width=0.4\linewidth]{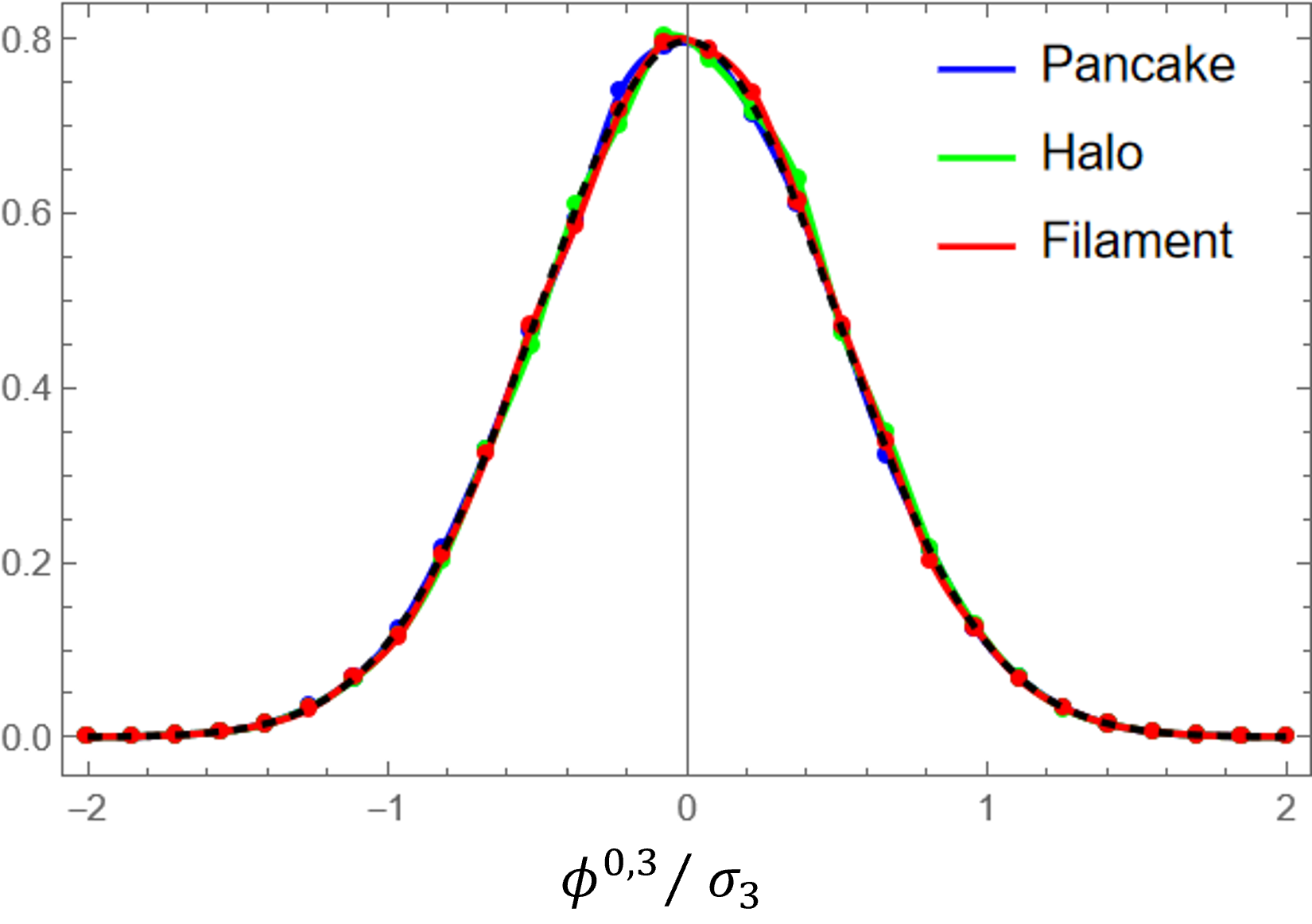}
    \caption{Marginal probability densities of $\phi^{1,2}$ (left) and $\phi^{0,3}$ (right) scaled by $\sigma_3$. Eq. \eqref{eq:MP_phi03} is shown in dashed black curve for comparison.}
    \label{fig:MP_q3}
\end{figure*}

\begin{figure*}
    \centering
    \includegraphics[width=0.4\linewidth]{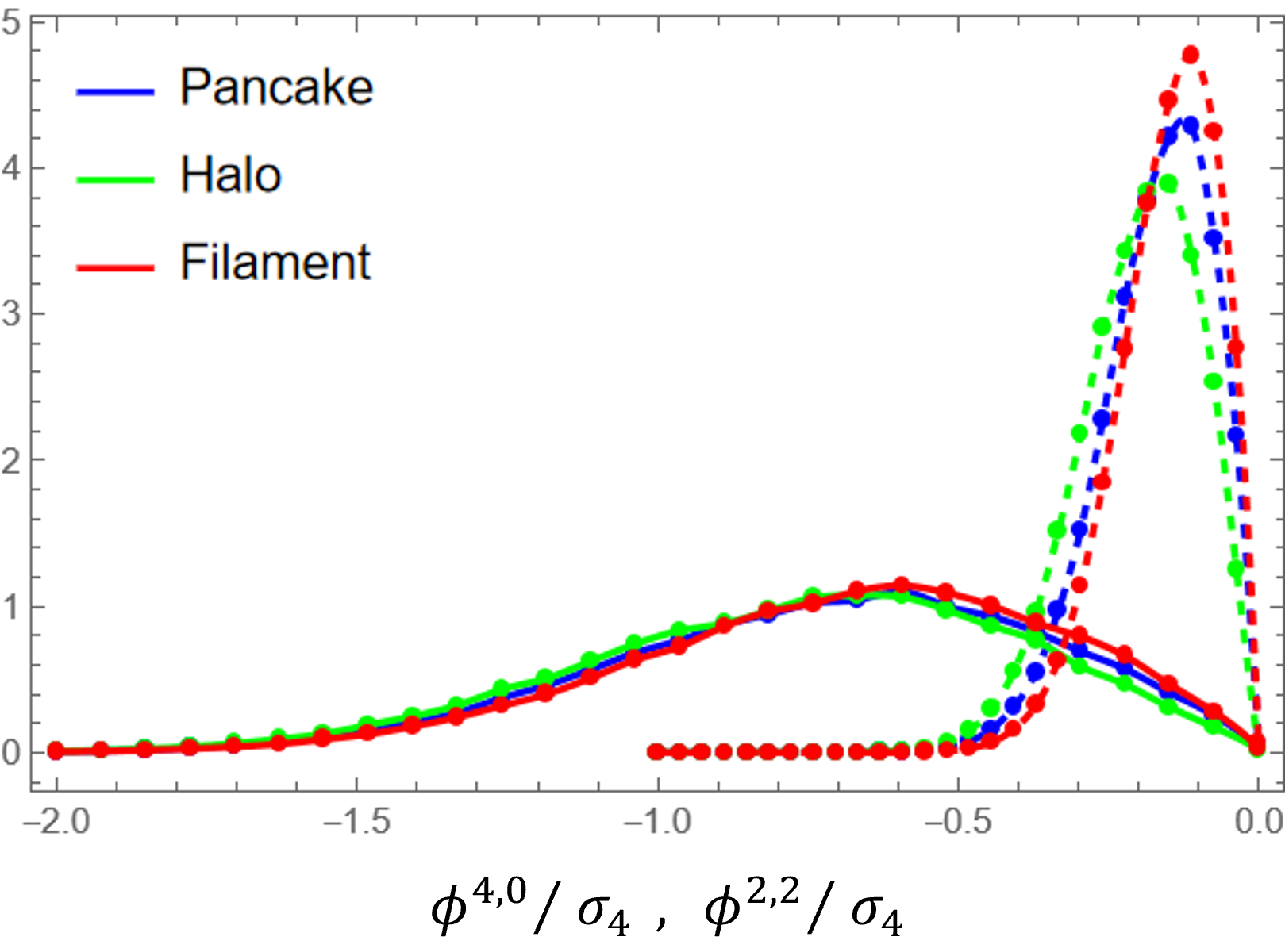}
    \includegraphics[width=0.4\linewidth]{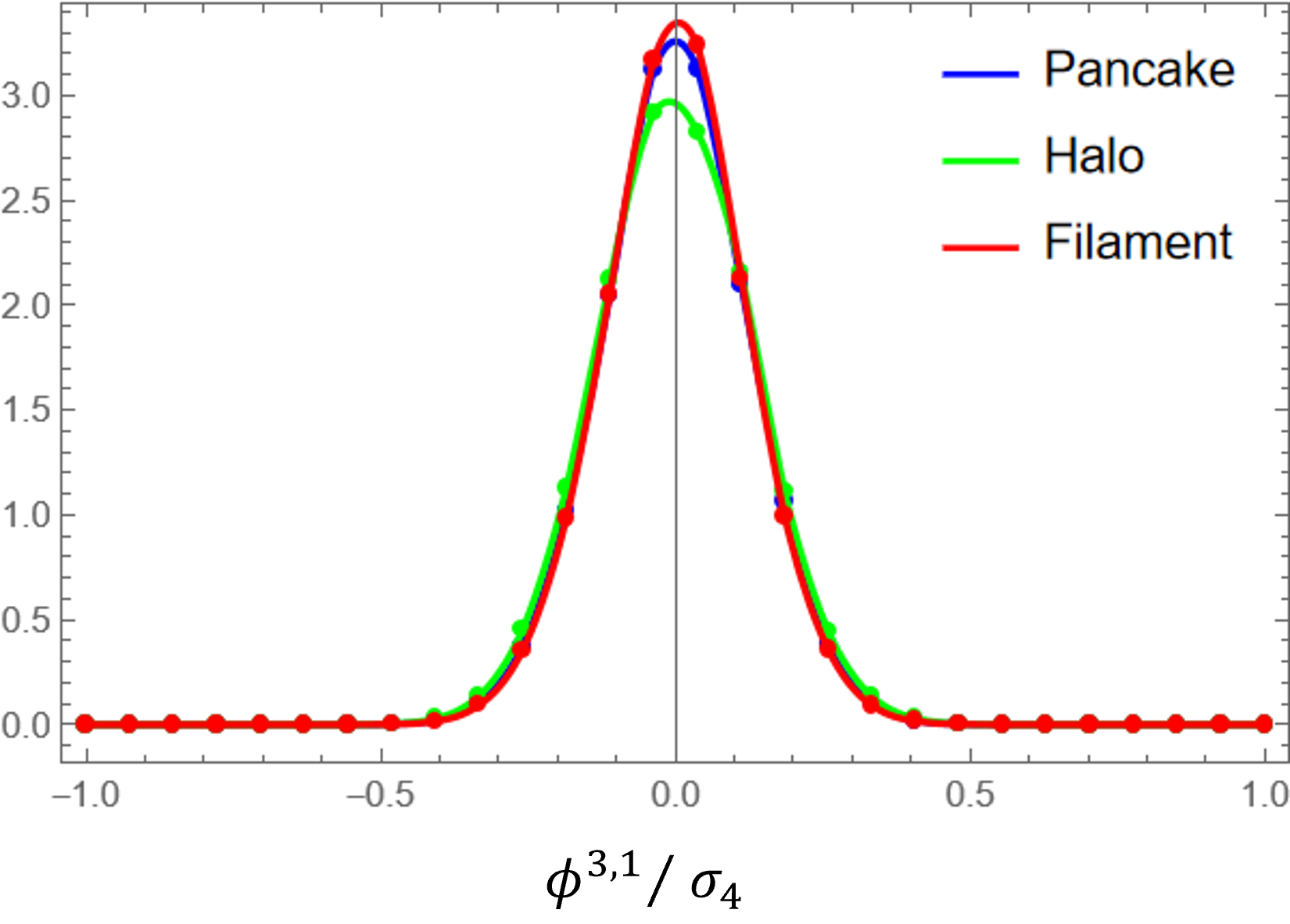}
    \includegraphics[width=0.4\linewidth]{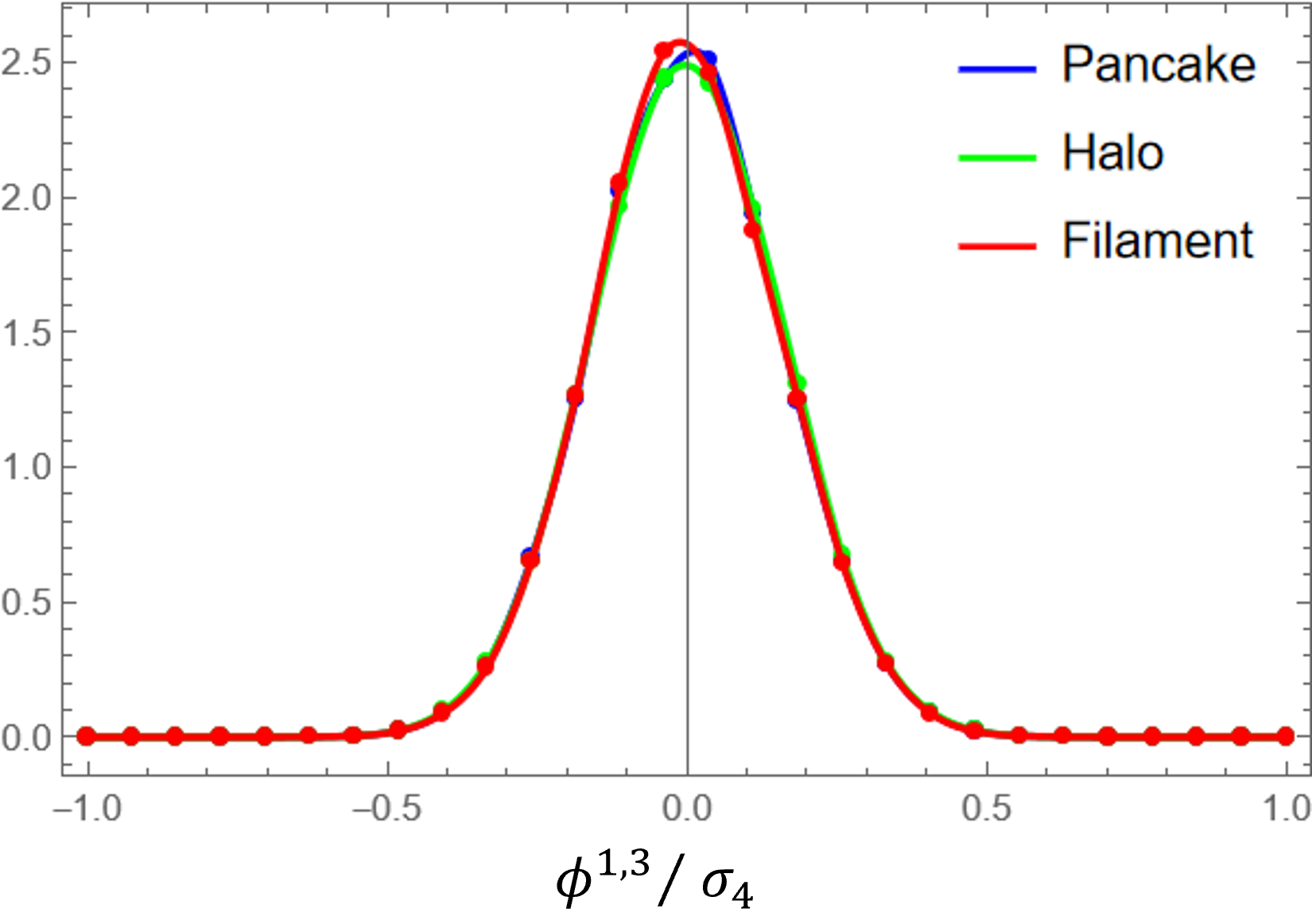}
    \includegraphics[width=0.4\linewidth]{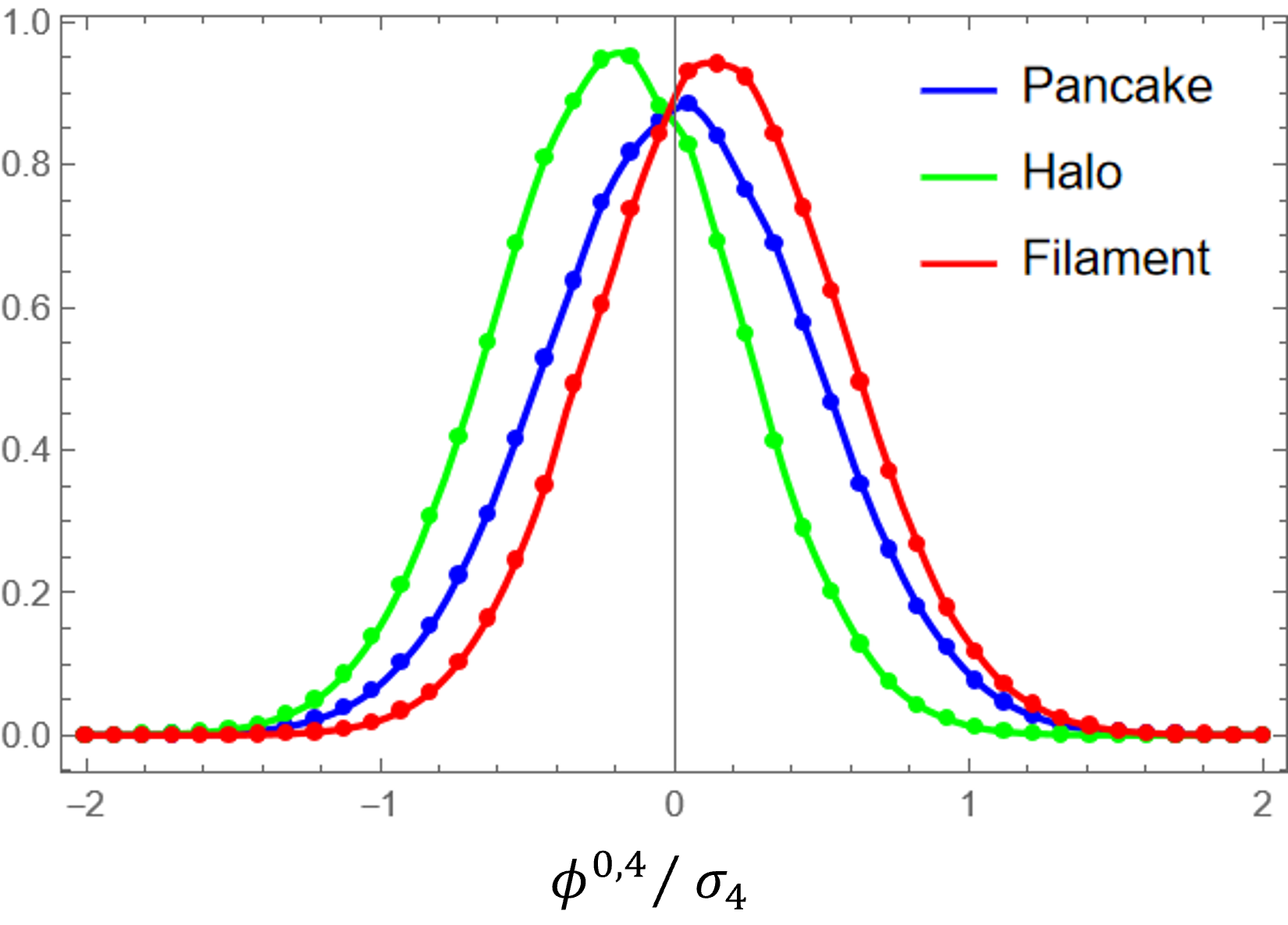}
    \caption{Marginal probability densities of $\phi^{4,0}$ (solid) and $\phi^{2,2}$ (dashed) in the top-left panel, $\phi^{3,1}$ in the top-right panel, $\phi^{1,3}$ in the bottom-left panel, and $\phi^{0,4}$ in the bottom-right panel, all scaled by $\sigma_4$.
}
    \label{fig:MP_q4}
\end{figure*}

From eq. \eqref{eq:CP_Y}, we note that the marginal distribution of $\theta$ is simply constant since $P(Y)$ eq. \eqref{eq:P_Y_full} is independent of $\theta$, and those of $\phi^{2,1}$ and $\phi^{3,0}$ are Dirac-delta functions centered at zero. So, we integrate them over to obtain:
\begin{align}
   \nonumber    & P (\phi^{2,0},\phi^{0,2},\phi^{1,2}, \phi^{0,3}, \phi^{4,0}, \phi^{3,1}, \phi^{2,2},\phi^{1,3}, \phi^{0,4} \: | \: \text{Pancake})  \propto \\
    \nonumber 
    & \text{Exp} \Bigg[ -10\left( \frac{\phi^{1,2}}{\sigma_3} \right)^2 - 2\left( \frac{\phi^{0,3}}{\sigma_3} \right)^2 - \frac{1}{2} \Big( \sum_{\mu, \nu} f_{\mu\nu}(\sigma_2, \sigma_3, \sigma_4) \phi_{\mu} \phi_{\nu} \Big)\Bigg]\\
     \nonumber
    & |\phi^{2,0}  - \phi^{0,2}| \: H[\phi^{2,0}] \: H[\phi^{2,0} - \phi^{2,0}]\: H\left[ -\phi^{4,0} - \phi^{2,2} - \frac{2 (\phi^{1,2})^2}{\phi^{2,0} - \phi^{0,2}} \right] \\
    \label{eq:CP_Y_9}
    &  H\left[ \phi^{4,0} \left( \phi^{2,2} + \frac{2 (\phi^{1,2})^2}{\phi^{2,0} - \phi^{0,2}} \right) - (\phi^{3,1})^2 \right],
\end{align}
where $\phi_{\mu}, \phi_{\nu} \in \{\phi^{2,0}, \phi^{0,2}, \phi^{4,0}, \phi^{3,1}, \phi^{2,2}, \phi^{1,3}, \phi^{0,4}\}$ and $f_{\mu\nu}(\sigma_2, \sigma_3, \sigma_4)$ is the coefficient of the product pair $\phi_{\mu}, \phi_{\nu}$, which can be read off eq. \eqref{eq:P_Y_full}. Again, $H[\phi^{0,2}]$ and $H[-\phi^{0,2}]$ have to be multiplied for halo and filament populations, respectively. We can analytically calculate the marginal distribution of $\phi^{0,3}$, since it is uncorrelated:
\begin{align}
    \label{eq:MP_phi03}
    P(\phi^{0,3} \: | \: \text{Pancake/Halo/Filament} ) = &\sqrt{ \frac{4 }{2 \pi \sigma_3^2}} \exp \left(-2 \left( \frac{\phi^{0,3}}{\sigma_3} \right)^2\right),
\end{align}
which is the same across pancake, halo, and filament populations. The expectation value of its magnitude is:
\begin{align}
    \label{eq:exp_phi03}
    \langle | \phi^{0,3} | \rangle& = \sigma_3 \sqrt{\frac{1}{2 \pi }} \propto \frac{1}{R_c^{n/2+2}}.
\end{align}
The other derivatives are correlated, and analytical expressions for their distributions are not straightforward. Instead, we use the Monte Carlo method to integrate eq. \eqref{eq:CP_Y_9} and present their marginal distributions in figures \ref{fig:MP_q2}, \ref{fig:MP_q3}, and \ref{fig:MP_q4}. The data points from numerical integration are cubically interpolated. Distributions across pancake, halo, and filament populations are shown in blue, green, and red curves, respectively. The expectation values, also evaluated numerically, are noted in table \ref{table:exp_MP_phi}. For those with symmetrical distributions --- $\phi^{1,2}, \phi^{0,3}, \phi^{3,1}, \phi^{1,3}$ --- we record the expectation value of their magnitude.

\begin{table}
    \caption{Expectation values of the derivatives of the displacement potential, scaled with respect to their corresponding $\sigma$'s, computed using Monte-Carlo integrals for pancake, halo, and filament populations individually.}
    \label{table:exp_MP_phi}
    \centering
    \begin{tabular}{c c c c}
    \hline\hline
    $\langle \phi^{i,j} \rangle /\sigma_{i+j}$& Pancake   & Halo      & Filament  \\ \hline
    $\langle \phi^{2,0}\rangle / \sigma_2$     & 0.879 & 1.125    & 0.723 \\
    $\langle \phi^{0,2} \rangle / \sigma_2$    & -0.150   & 0.340 & -0.458  \\ 
    $\langle | \phi^{1,2}| \rangle / \sigma_3$ & 0.125  & 0.122  & 0.127  \\ 
    $\langle | \phi^{0,3}| \rangle / \sigma_4$ & 0.396  & 0.397  & 0.391  \\ 
    $\langle \phi^{4,0} \rangle / \sigma_4$ & -0.726  & -0.766  & -0.699  \\ 
    $\langle | \phi^{3,1}| \rangle / \sigma_4$ & 0.101  & 0.108  & 0.099  \\ 
    $\langle \phi^{2,2} \rangle / \sigma_4$ & -0.174  & -0.201  & -0.156  \\ 
    $\langle | \phi^{1,3}| \rangle / \sigma_4$ & 0.127  & 0.128  & 0.126  \\ 
    $\langle \phi^{0,4} \rangle / \sigma_4$ & 0.021  & -0.200  & 0.156  \\ \hline
    \end{tabular}
\end{table}

Fig. \ref{fig:MP_q2} shows the distributions of $O(q^2)$ derivatives: $\phi^{2,0} / \sigma_2$ and $\phi^{0,2} / \sigma_2$, which correspond to the eigenvalues $\alpha(\vec{q}_c)$ and $\beta(\vec{q}_c)$. From eq. \eqref{eq:rho_alpha}, the density, $\rho(\mathbf{x}_c, t_c) \: / \: \bar{\rho}(t_c) \approx \alpha(\vec{q}_c, t_c) + \beta(\vec{q}_c, t_c)$ up to leading order, depends on both the eigenvalues. On the other hand, the shell-crossing time depends only on $\alpha(\vec{q}_c)$ as $D_+(t_c) = 1/\alpha(\vec{q}_c)$, eq. \eqref{eq:P_Dtc}. In a Gaussian random field, the eigenvalues are correlated: points with higher $\alpha$ typically have higher $\beta$. Thus, the shell-crossing points, which are the peaks of $\alpha$, may not correspond to the density peaks exactly, but are typically located in the overdense neighborhood. Shell-crossing points where halos appear, $\beta > 0$, have higher $\alpha$ and hence, are generally denser and undergo shell-crossing earlier than those where filaments appear, $\beta < 0$. This is what we note from the distributions and expectation values. However unlikely, it is not entirely impossible that a point, which is less dense than another, undergoes shell-crossing earlier. Also, $\langle \phi^{0,2} \rangle$ being negative for the pancake population suggests that the fraction of filaments is higher than that of halos.

Fig. \ref{fig:MP_q3} shows the distributions of $O(q^3)$ derivatives: $\phi^{1,2} / \sigma_3$ and $\phi^{0,3} / \sigma_3$. They are symmetrical and the same across pancake, halo, and filament populations. The numerically-integrated distribution and the expectation value of $\phi^{0,3}/ \sigma_3$ demonstrate good agreement with the analytical curve, eq. \eqref{eq:MP_phi03}, shown in black, and the expectation value, eq. \eqref{eq:exp_phi03}, respectively.

Fig. \ref{fig:MP_q4} shows the distributions of $O(q^4)$ derivatives: $\phi^{4,0}, \phi^{3,1}, \phi^{2,2}, \phi^{1,3}$, and $\phi^{0,4}$ scaled with respect to $\sigma_4$. The distributions for $\phi^{3,1}$ and $\phi^{1,3}$ are symmetrical and show little variation across pancake, halo, and filament populations. In eq. \eqref{eq:P_Y_full}, $\phi^{3,1}$ and $\phi^{1,3}$ are correlated only with each other and have the same variance; however, the distribution for $\phi^{3,1}$, from eq. \eqref{eq:CP_Y_9}, turns out narrower, and its expectation value correspondingly lower than that of $\phi^{1,3}$ because of the constraints on the eigenvalues of $H_{ij}(\alpha)$ at the shell-crossing (local maximum) point, eq. \eqref{eq:maxima_cond}, which appear in the Heaviside function and rule out higher values of $\phi^{3,1}$ that would instead lead to a saddle point.

The distribution of $\phi^{4,0}$ is wider than that of $\phi^{2,2}$, which suggests that generally $A < C < 0$ in eq. \eqref{eq:A2_axis_lengths}, that is, the minor axis of the $A_2$ ellipse is more or less aligned with the shell-crossing direction. Crudely approximated, refer to eq. \eqref{eq:P_ellip}, the semi-major and semi-minor axis lengths, $a$ and $b$, are proportional to $1/\sqrt{-\phi^{2,2}}$ and $1/\sqrt{-\phi^{4,0}}$, respectively. We may thus expect the $A_2$ caustics to be predominantly flat. Comparing between halo and filament populations, the variation in $\phi^{4,0}$ is less than that of $\phi^{2,2}$. Also, the absolute means are higher for halos. This suggests that, in general, the semi-minor axis length differs little between the two populations, but the major axis length tends to be smaller for the halos than the filaments of the same age.

The marginal distributions and expectation values do have a weak dependence on the spectral index $n$ (refer fig. \ref{fig:MP_phi_n}); the distributions in figures \ref{fig:MP_q2}-\ref{fig:MP_q4}, and the expectation values noted in table \ref{table:exp_MP_phi} were computed assuming $n = 1$ and $R_c = 10$ grid units as representative values, along with exponential smoothing. Neglecting the weak dependence on $n$, $\langle |\phi^{i,j}| \rangle \propto \sigma_{i+j} \propto R_c^{-n/2-(i+j-1)}$. It can thus be inferred that higher $n$ and $R_c$ correspond to smoother density and potential fields having narrower distributions of $\phi^{i,j}$.

\subsection{A typical pancake}

\begin{figure}
    \centering
    \includegraphics[width=\linewidth]{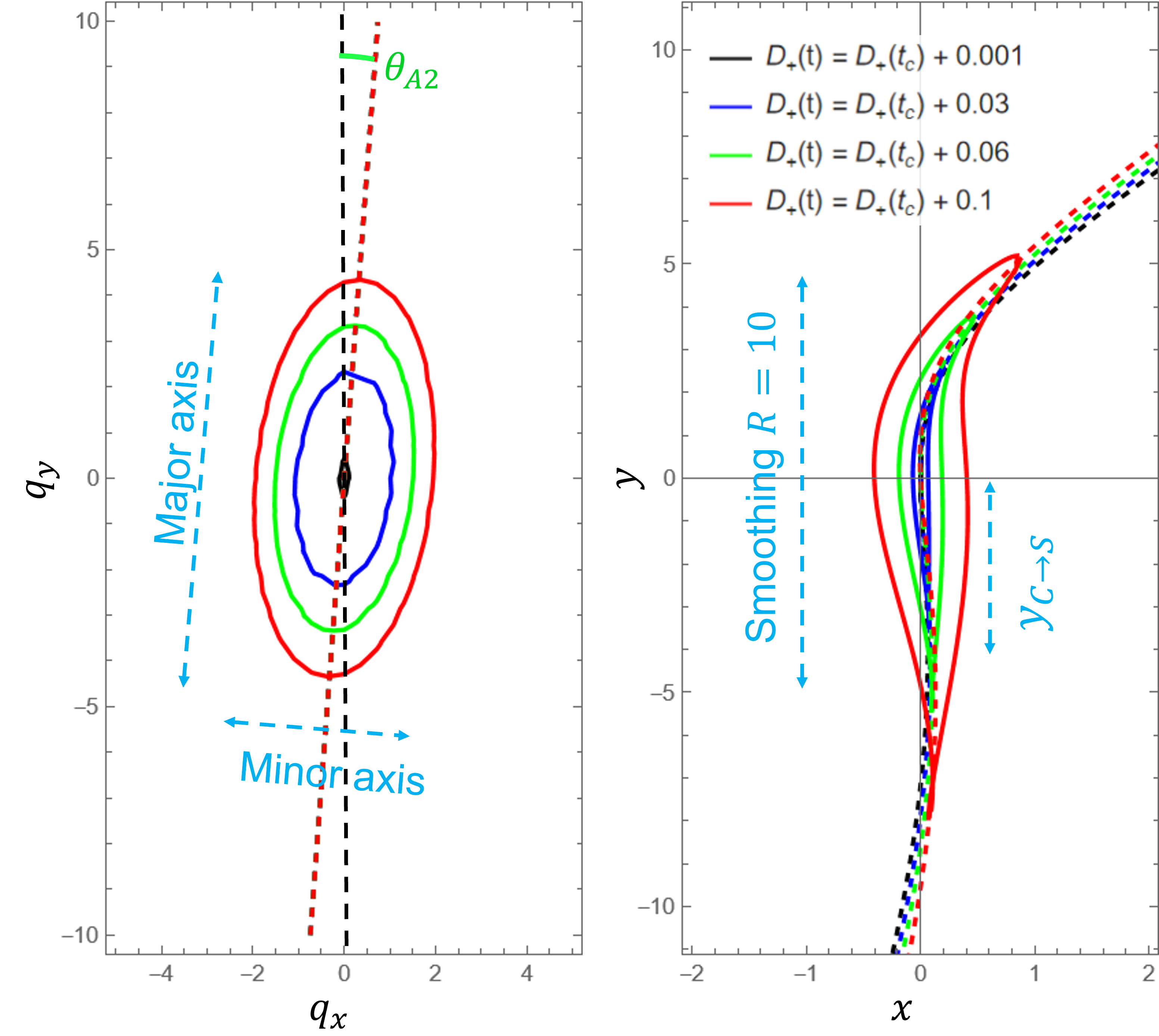}
    \caption{The $A_2$ caustic (solid lines) and the $A_3$ spine (dashed lines) of the average pancake at $D_+(t)-D_+(t_c) \in \{ 0.001, 0.05, 0.1, 0.15 \}$ in Lagrangian (left) and Eulerian (right) spaces. The $x$-axis in Eulerian space has been zoomed 200\% to make the features distinguishable.}
    \label{fig:avg_pancake}
\end{figure}

We make use of our statistical computations of the expectation values of $\phi^{i,j}$ (table \ref{table:exp_MP_phi}) in our analytical model of the pancake structure (section \ref{sec:catastrophe_theory}) to demonstrate a typical pancake in a cosmology of our choice: $n = 1, R_c = 10$ grid units with exponential smoothing. As a note, the qualitative features of the geometrical shape of the pancake, with length scale roughly set by the peak of the matter power spectrum ($\approx R_c/n$ for $n > 0$), remain nearly the same regardless of our choice of cosmology and smoothing filter (refer to eqs. \eqref{eq:P_curv} and \eqref{eq:P_CS_trans}). Fig. \ref{fig:avg_pancake} shows the $A_2$ caustic and $A_3$ spine of this pancake at different times after shell-crossing in both Lagrangian and Eulerian spaces. They are computed using eqs. \eqref{eq:A2_eq}, \eqref{eq:A3_eq}, and \eqref{eq:eq_motion_taylor_exp}.

Since $\langle \phi^{0,2} \rangle < 0$, this pancake evolves into a filament. Looking at the $A_3^+$ spine in Eulerian space, refer to eq. \eqref{eq:A3_eq_Eul}, it is C-shaped close to the center. Its curvature turns out to be $0.036$ (grid unit)$^{-1}$ at $(\mathbf{q_c}, t_c)$. As expected from eq. \eqref{eq:A_3_curvature}, for filaments, the curvature decreases with time, which can be observed as the $A_3^+$ spine stretching outwards from the shell-crossing point.

At larger scales, the pancake transitions to S-shape. The distance from the center at which the $A_3^+$ spine bends from C to S turns out to be $y_{C\rightarrow S} \approx 4.0$ grid units at shell-crossing, refer to eq. \eqref{eq:A_3_C_S_trans}. It increases with time as expected of filaments. Our choice of power spectrum peaks for structures of the size $\approx R_c/n = 10$ grid units, which is comparable to the extent of the C-shape of a typical pancake $\approx 2 y_{C\rightarrow S}$. We can thus expect that the pancakes appearing in 2D cosmologies seeded by Gaussian random fields are dominantly C-shaped. In Lagrangian space, the length ratio of semi-minor to semi-major axis of the $A_2$ caustic $b/a$ is approximately $0.48$, which changes little with time. Consequently, the Eulerian shape of the pancake turns out quite flat. 

In the following subsections, we present in greater detail the distributions of the following properties:
\begin{enumerate}
    \item Shell-crossing time --- $D_+(t_c)$
    \item Curvature of $A_3^+$ spine --- $d^2 x_{A3} / dy_{A3}^2\Big|_{\vec{q}_c, t_c}$
    \item Length scale at which $A_3^+$ spine transitions from C to S-shape --- $2 \: y_{C \rightarrow S}(t_c)$
    \item Axis length ratio of the $A_2$ caustic in Lagrangian space --- $b / a$
\end{enumerate}

Additionally, the differences in distributions between halo and filament populations, and the dependence of expectation values on the power spectrum parameters characterizing the 2D Gaussian random field $\phi(\vec{q})$ are discussed.

\begin{table}
    \caption{Expectation values of shell-crossing time, curvature and C-S transition of $A_3^+$ spine, and the axes length ratio of the $A_2$ caustic in units provided in brackets, computed using Monte-Carlo integrals for pancake, halo, and filament populations individually.}
    \label{table:exp_props}
    \centering
    \begin{tabular}{l c c c}
    \hline\hline
    Property (unit) & Pancake   & Halo      & Filament  \\ \hline
    Shell-crossing time $(\sigma_2^{-1})$     & 1.620 & 1.079    & 1.983 \\
    Curvature $ \left( \sigma_3 / \sigma_2 \right)$    & 0.182   & 0.368 & 0.067  \\\\
    C-S transition \hspace{0.3cm} (Exact)& 3.734  & 2.005  &  4.873\\
    scale $ \left( \sigma_3 / \sigma_4 \right)$ \hspace{0.1cm} (Approx.)& 4.107  & 2.205  &  5.360 \\\\
    Axes length \hspace{0.5cm} (Exact) & 0.469  & 0.493  & 0.445  \\
    ratio $ (\sigma_4) $ \hspace{0.65cm} (Approx.)& 0.521  & 0.547  & 0.505  \\ \hline
    \end{tabular}
\end{table}

\subsection{Distribution of shell-crossing times}

\begin{figure}
    \centering
    \includegraphics[width=\linewidth]{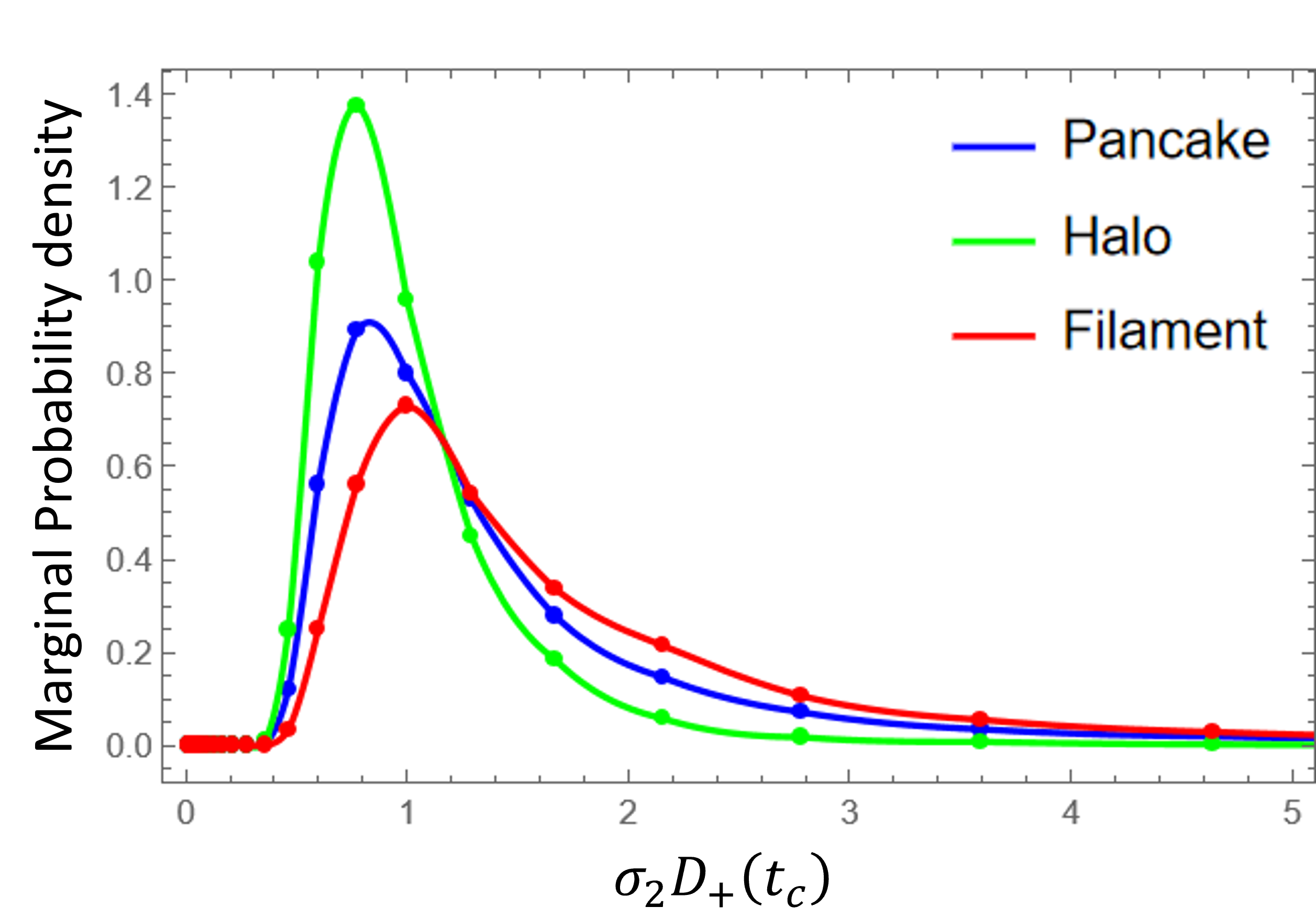}
    \caption{Distribution of shell-crossing times $D_+(t_c$ for pancakes (blue), halos (green), and filaments (red).}
    \label{fig:MP_Dtc}
\end{figure}

The shell-crossing time can be obtained from eq. \eqref{eq:A2_cond} assuming Zeldovich flow:
\begin{align}
    \nonumber &\alpha(\vec{q}_c, t_c) = \phi^{2,0,0} = D_+(t_c) \phi^{2,0} = 1\\
    \label{eq:P_Dtc} \implies &D_+(t_c) = \frac{1}{\phi^{2,0}}\propto \sigma_2^{-1} \propto R_c^{n/2 + 1}.
\end{align}
This suggests that a higher spectral index $n$ and a larger smoothing scale $R_c$ will push the initial shell-crossings to later times. We obtain the marginal distribution $P(\phi^{2,0}\: | \: \text{Pancake})$ using Monte Carlo integration, see fig. \ref{fig:MP_q2}, and transform it according to eq. \eqref{eq:P_Dtc} to obtain $P(D_+(t_c)\: | \: \text{Pancake})$. Fig. \ref{fig:MP_Dtc} shows the distribution of shell-crossing times for pancake, halo, and filament populations. Their expectation values are provided in table \ref{table:exp_props}.

Halos undergo first shell-crossing earlier than filaments on average. For our choice of cosmology $n = 1$ and $R_c = 10$ grid units with exponential smoothing, the average shell-crossing redshifts turn out to be $z \approx 1.2, 2.3, 0.8$ for pancakes, halos, and filaments, respectively.

\subsection{Distribution of curvature of the $A_3^+$ spine}

\begin{figure}
    \centering
    \includegraphics[width=\linewidth]{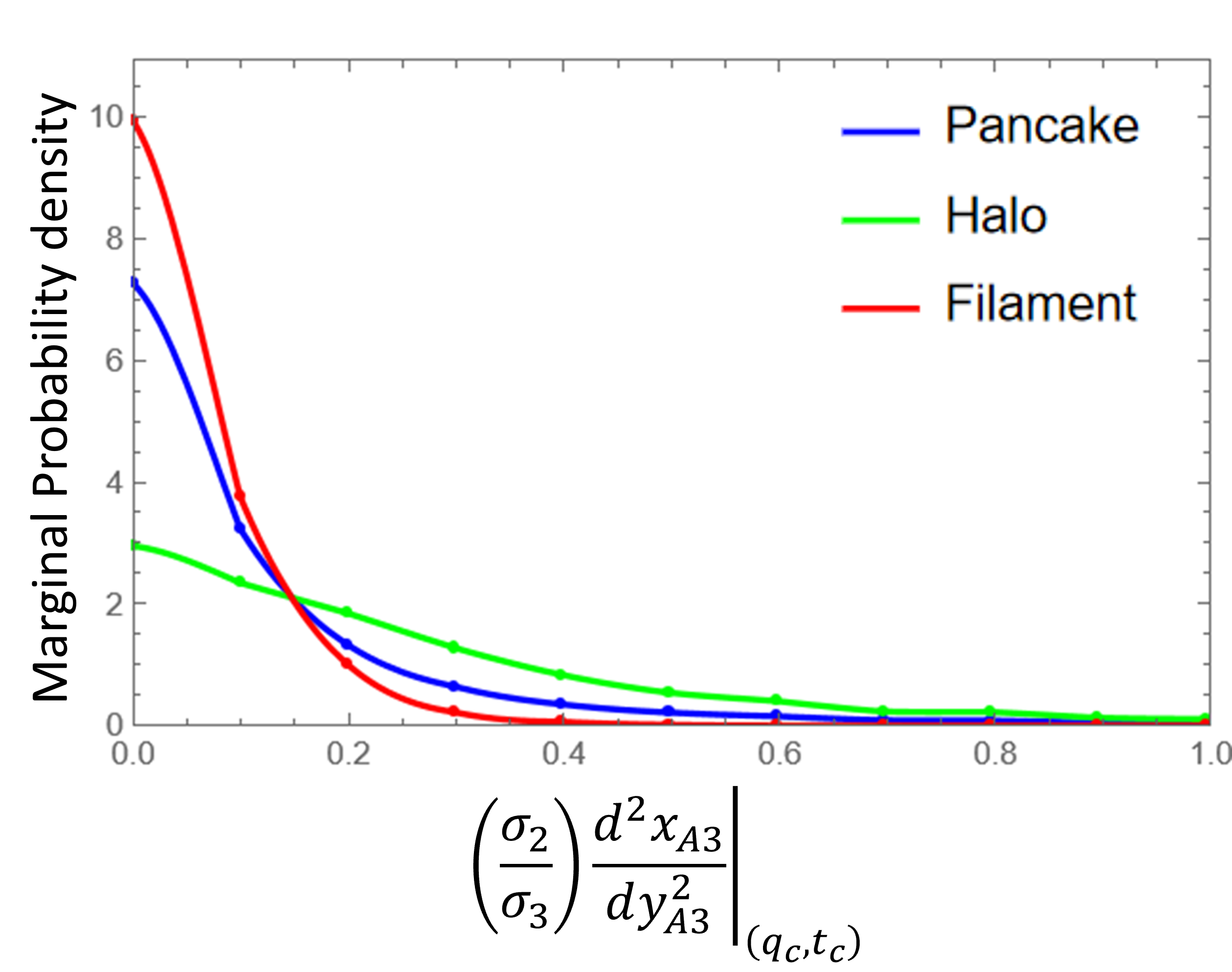}
    \caption{Distribution of curvature of the $A_3^+$ spine for pancakes (blue), halos (green), and filaments (red).}
    \label{fig:MP_curv}
\end{figure}

The curvature of the $A_3^+$ spine at shell-crossing $(\vec{q}_c, t_c)$, eq. \eqref{eq:A_3_curvature}, can be re-written as:

\begin{align}
    \nonumber \Bigg| \frac{d^2 \: x_{A3}}{d \: y_{A3}^2} \Bigg|_{\vec{q}_c, t_c} &= \frac{ | \phi^{1,2,0} | }{\left( 1 - \phi^{0,2,0} \right)^2} = \frac{ | D_+(t_c)\phi^{1,2} |}{\left( 1 - D_+(t_c)\phi^{0,2} \right)^2} = \frac{ | \phi^{2,0}\phi^{1,2} | }{\left( \phi^{2,0} - \phi^{0,2} \right)^2}\\
    \label{eq:P_curv} &  \propto \frac{\sigma_3}{\sigma_2} \propto \frac{1}{R_c}\sqrt{\frac{\Gamma(n + 4)}{\Gamma(n + 2)}} \approx \frac{n}{R_c} \text{ (if $n \gg 0$)}.
\end{align}
This suggests that a lower spectral index $n$ or a larger smoothing scale $R_c$ will result in pancakes with lower curvature. We obtain the joint distribution $P(\phi^{2,0},\phi^{0,2},\phi^{1,2} \: | \: \text{Pancake})$ using Monte Carlo integration of eq. \eqref{eq:CP_Y_9} and transform it according to eq. \eqref{eq:P_curv}. Fig. \ref{fig:MP_curv} shows the distribution of curvature for pancake, halo, and filament populations. Their expectation values are provided in table \ref{table:exp_props}

Even though the distribution of $P(\phi^{1,2})$ is nearly the same across pancake, halo, and filament populations, fig. \ref{fig:MP_q3}, halos have higher curvature than filaments. This is because they have $\beta(\vec{q}_c, t_c) = \phi^{0,2,0} > 0$, which indicates that the $A_3^+$ spine contracts along the $y$-axis, contrary to what happens in the case of filaments, refer to eq. \eqref{eq:A_3_curvature}.

For our choice of cosmology $n = 1$ and $R_c = 10$ grid units with exponential smoothing, the average curvatures turn out to be $\approx 0.067, 0.127, 0.023$ (grid unit)$^{-1}$ for pancakes, halos, and filaments, respectively. The maximum possible curvature of a pancake of size corresponding to the power spectrum peak is $\approx 4n / R_c = 0.4$ (grid unit)$^{-1}$. Compared to this, the spines of pancakes at shell-crossing, in particular the filaments, are mostly straight, but with noticeable bends.

\subsection{Distribution of the length scale at which the $A_3^+$ spine transitions from C to S-shape}

\begin{figure}
    \centering
    \includegraphics[width=\linewidth]{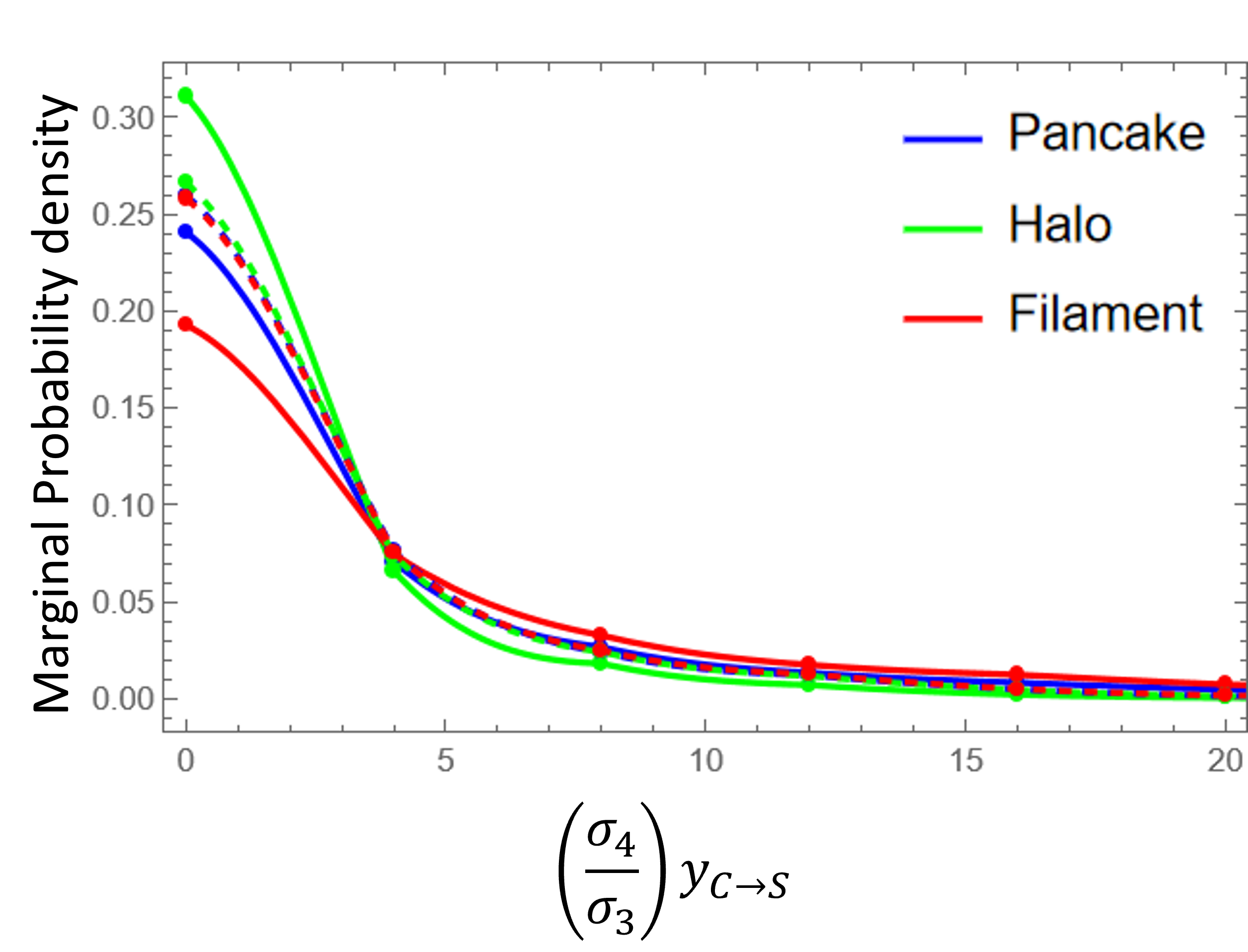}
    \caption{Distribution of $y_{C\rightarrow S}(t_c)$ (in solid lines) for pancakes (blue), halos (green), and filaments (red). For comparison, the distributions of $q_{y,C\rightarrow S}(t_c)$ are shown in dashed lines.}
    \label{fig:MP_CS_trans}
\end{figure}

The length scale at which the $A_3^+$ spine transitions from C to S-shape is about $2 \: y_{C\rightarrow S}$ given by eq. \eqref{eq:A_3_C_S_trans}. However, computing the distribution $P(y_{C\rightarrow S})$ from it is tedious. We make a few crude approximations:
\begin{align}
     &\left| \phi^{3,1,0} \left( 2\left(\frac{ \phi^{3,1,0}}{\phi^{4,0,0} }\right)^2 - 3\frac{ \phi^{2,2,0} }{ \phi^{4,0,0} } \right) \right|\ll |\phi^{1,3,0}|
     , \\
     &\left| \frac{6 \phi^{1,2,0}}{\left( 1 - \phi ^{0,2,0} \right)} \left(\frac{1}{2} \phi ^{0,3,0}-\frac{\phi ^{1,2,0} \phi ^{3,1,0}}{\phi ^{4,0,0}}\right) \right| \ll |\phi^{1,3,0}|,
\end{align}
which reduce eq. \eqref{eq:A_3_C_S_trans} down to:
\begin{align}
    \nonumber 
    |q_{y,C\rightarrow S}(t_c)| &= 2 \left| \frac{\phi^{1,2,0}}{\phi^{1,3,0}} \right| = 2 \left| \frac{\phi^{1,2}}{\phi^{1,3}} \right|,\\
    \nonumber
    |y_{C\rightarrow S}(t_c)| &= 2 \left( 1 - \frac{\phi^{0,2}}{\phi^{2,0}} \right) \: \left| \frac{\phi^{1,2}}{\phi^{1,3}} \right|\\
    \label{eq:P_CS_trans} &\propto \frac{\sigma_3}{\sigma_4} \propto R_c\sqrt{\frac{\Gamma(n + 4)}{\Gamma(n + 6)}} \approx \frac{R_c}{n}  \text{ (if $n \gg 0$)}.
\end{align}
This suggests that a lower spectral index $n$ or a larger smoothing scale $R_c$ will result in pancakes with a greater extent of C-shape. We obtain the joint distribution $P(\phi^{2,0},\phi^{0,2},\phi^{1,2},\phi^{1,3} \: | \: \text{Pancake})$ using Monte Carlo integration of eq. \eqref{eq:CP_Y_9} and transform it according to eq. \eqref{eq:P_CS_trans} to obtain $P( y_{C\rightarrow S}(t_c) \: | \: \text{Pancake})$. Fig. \ref{fig:MP_CS_trans} shows the distribution of $y_{C\rightarrow S}(t_c)$ for pancake, halo, and filament populations. The expectation values using the exact and approximate formulae, eqs. \eqref{eq:A_3_C_S_trans} and \eqref{eq:P_CS_trans}, are provided in the table \ref{table:exp_props}. Our approximation somewhat overpredicts the extent of C-shape. We further cross-check our approximate $P(y_{C\rightarrow S}(t_c))$ against the distribution measured from simulations (refer to fig. \ref{fig:hist_MP_props}).

The extent of C-shape of the $A_3^+$ spine is inversely related to that of curvature. This is because the distribution of $q_{y,C\rightarrow S}(t_c)$ is the same across pancakes, halos, and filaments, and the difference in distributions of $y_{C\rightarrow S}(t_c)$ comes from the sign of the lower eigenvalue $\beta(\vec{q}_c)=\phi^{0,2}$ (refer to eq. \eqref{eq:P_CS_trans}). If a halo and a filament start off with the same $q_{y,C\rightarrow S}(t_c)$, the contraction of the $A_3^+$ spine for the halo due to its positive $\beta(\vec{q}_c)$ results in lower $y_{C\rightarrow S}(t_c)$, whereas the opposite holds for the filament. Filaments, therefore, have $A_3^+$ spines that appear C-shaped up to larger scales than halos.

For our choice of cosmology $n = 1$ and $R_c = 10$ grid units with exponential smoothing, the average values of $y_{C\rightarrow S}(t_c)$ based on the exact eq. \eqref{eq:A_3_C_S_trans} turn out to be $\approx 6.82, 3.66, 8.90$ grid units for pancakes, halos, and filaments, respectively. The extent of the C-shape $\approx 2 y_{C\rightarrow S}$ is comparable to the scale of structures where the power spectrum peaks $R_c/n \approx 10$ grid units. Thus, the pancakes, in particular the filaments, are mostly C-shaped.

\subsection{Distribution of axis length ratio of the $A_2$ caustic}

\begin{figure}
    \centering
    \includegraphics[width=\linewidth]{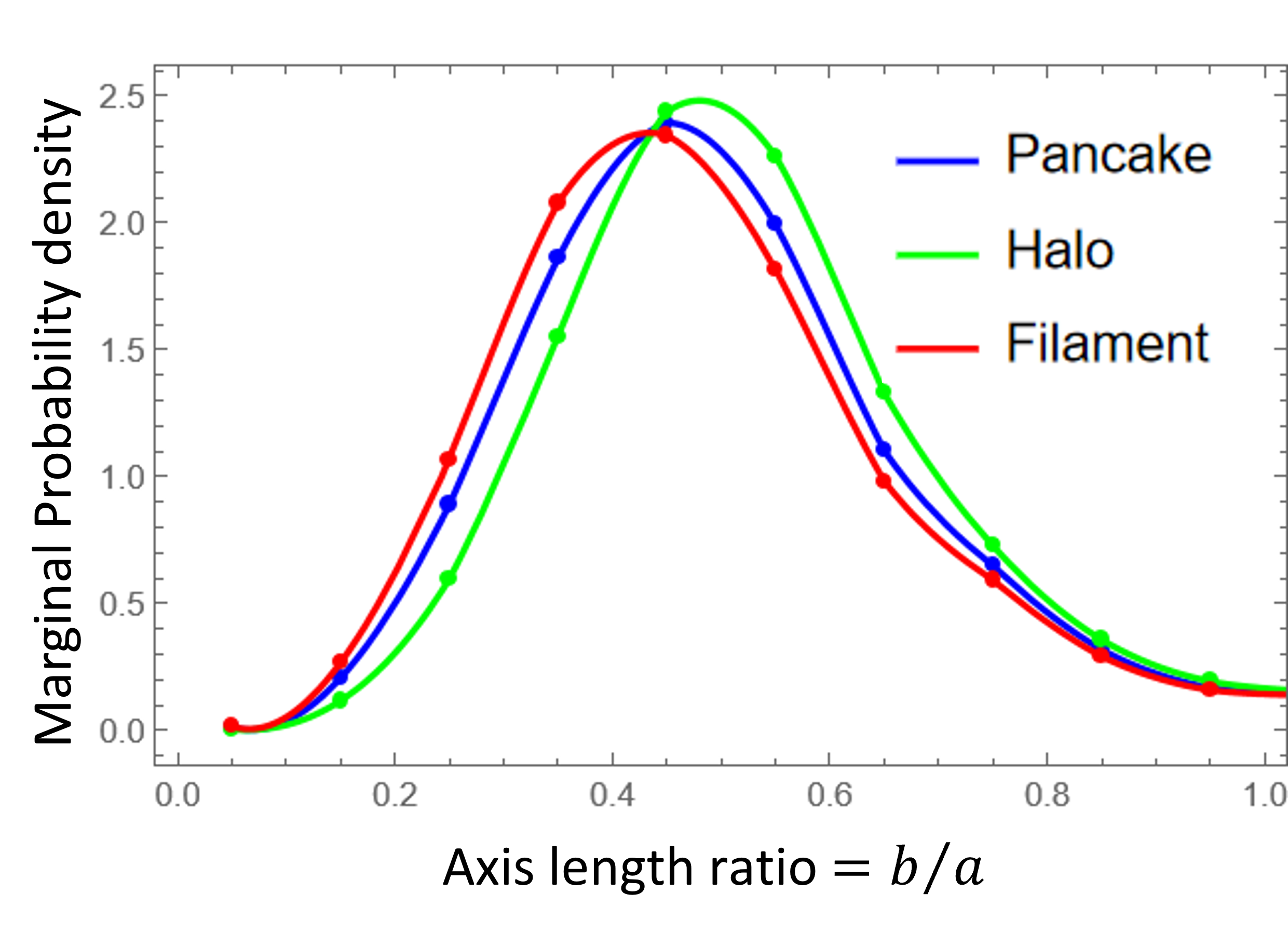}
    \caption{Distribution of axis length ratio of the $A_3^+$ spine for pancakes (blue), halos (green), and filaments (red).}
    \label{fig:MP_ellip}
\end{figure}

The axis length ratio $b / a$ of the $A_2$ ellipse in Lagrangian space can be computed from eq. \eqref{eq:A2_axis_lengths}. Here too, we make a few assumptions:
\begin{align}
    & 2 |\phi^{3,1,0}| \ll \left| \phi^{2,2,0} + \frac{2(\phi^{1,2,0})^2}{(1 - \phi^{0,2,0})} - \phi^{4,0,0} \right|,
    \\
    &\frac{2(\phi^{1,2,0})^2}{(1 - \phi^{0,2,0})} \ll |\phi^{2,2,0}|.
\end{align}
The axis length ratio reduces to
\begin{align}
    \label{eq:P_ellip} & \frac{b}{a} = 
    \begin{cases}
        \sqrt{ \phi^{2,2} / \phi^{4,0} } \text{ ; } \phi^{4,0} < \phi^{2,2} < 0\\
        \sqrt{ \phi^{4,0} / \phi^{2,2} } \text{ ; } \phi^{2,2} < \phi^{4,0} < 0
    \end{cases},
\end{align}
which is time-independent and scale-free. We obtain the joint distribution $P(\phi^{4,0},\phi^{2,2} \: | \: \text{Pancake})$ using Monte Carlo integration of eq. \eqref{eq:CP_Y_9} and transform it according to eq. \eqref{eq:P_ellip} to obtain $P( \: b/a \: \: | \: \text{Pancake})$. Fig. \ref{fig:MP_ellip} shows the distribution of axis length ratio for pancake, halo, and filament populations. Their expectation values using exact and approximate formulae, eqs. \eqref{eq:A2_axis_lengths} and \eqref{eq:P_ellip}, are provided in table \ref{table:exp_props}. Our approximation somewhat overpredicts the axis ratio of $A_2$ caustics in Lagrangian space. We further cross-check our approximate $P(b/a)$ against the distribution measured from simulations (refer to fig \ref{fig:hist_MP_props}).

Since $\phi$ is Gaussian distributed, the $O(q^2)$ and $O(q^4)$ are so correlated that the ordering of eigenvalues (by rotation) $\alpha =\phi^{2,0} \ge \beta = \phi^{0,2} $ at the local maxima of eigenvalue $\alpha$ generally leads to the ordering of curvatures $\phi^{4,0} \le \phi^{2,2} < 0$, which means that the shell-crossing direction has the higher magnitude of curvature and is more or less aligned with the minor axis of the $A_2$ ellipse in Lagrangian space. Therefore, for these cases in general, the axis ratio $0 < b/a \le 1$ can be seen as a proxy for axis symmetry in the shell-crossing dynamics. The lower the axis ratio, the greater the difference in eigenvalues $\alpha, \beta$, the more quasi one-dimensional (or anisotropic) the shell-crossing, the more eccentric the $A_2$ caustic, and the flatter the pancake in Eulerian space. Axis-symmetric collapse corresponding to $b/a = 1$ is nearly impossible because the probability of the two eigenvalues $\alpha, \beta$ being equal is zero (refer to eq. \eqref{eq:CP_Y_9}). From fig \ref{fig:MP_ellip}, we observe that halos tend to have slightly more axis-symmetric $A_2$ caustics than filaments. The average axis ratios, based on the exact eq. \eqref{eq:A2_axis_lengths} and nearly independent of cosmology and time evolution, are $\approx 0.469, 0.493, 0.445$ for pancakes, halos, and filaments, respectively, which are still fairly anisotropic.

The analytical and statistical models we have thus presented allow us to describe the shape of the pancake given the derivatives of the displacement potential $\phi$ at the shell-crossing point $(\vec{q}_c, t_c)$, together with the probability distribution of those derivatives and several properties of pancakes that arise in Gaussian random realizations, given the power spectrum parameters $n, R_c$. To verify these results, we carry out simulations of Zeldovich flow initialized with Gaussian random fields, make measurements of the shape of the shell-crossing structure and the distribution of derivatives of $\phi$ at shell-crossing points, and compare them to our models. The details of this analysis are presented in the appendix \ref{sec:numerical_verification}.

\section{Conclusions}
\label{sec:conclusions}

Under the broader goal of understanding the cosmic web of the universe, we have aimed to study through this work the structure of pancakes that emerge out of the first shell-crossings as well as their distribution in 2D cosmologies seeded by Gaussian random fields. Based on arguments from catastrophe theory, we have developed an analytical model to capture the motion and emerging structure around the point of singularity formed at shell-crossing in 2D. We have also studied the distribution of a number of observable properties of the pancakes related to their formation and shape by modeling the distribution of parameters characterizing them using Gaussian statistics. For the verification of our models, we have performed 1-LPT simulations and compared the predicted shapes of pancakes and the distribution of parameters against those measured from the simulations. The main results and inferences from our work are as follows:

\begin{enumerate}
    \item The 2D cosmic web can indeed be traced using the locus of $A_3^+$ singularities and the shape of this locus can be pieced together by studying the critical points that lie on it \citep{Hidding_2014, Feldbrugge_2018}. In particular, we focus our work on the points of local maxima, as these are the points where pancakes emerge out of first shell-crossings, and later evolve into halos and filaments - the key structures defining the overall cosmic web. The models for the analytical shape and statistical distribution developed for pancakes can be further extended to the saddle points on the $A_3^+$ spine through simple changes to the constraints on model parameters (refer to subsection \ref{sec:catastrophe_theory}.2). 

    \item We have presented a parameterized expansion form, correct up to $O(q^3, t^1)$, for the motion around a shell-crossing point in eq. \eqref{eq:eq_motion_taylor_exp} and studied the structure of the emerging pancake by tracing the $A_3^+$ spine and $A_2$ caustic in Lagrangian and Eulerian spaces (refer to eqs. \eqref{eq:A2_eq}, \eqref{eq:A3_eq}, and \eqref{eq:eq_motion_taylor_exp}). The effect of each parameter on the shape of the pancake has been mapped out in subsection \ref{sec:catastrophe_theory}.4. The most notable are the leading $O(y^2)$ and next-to-leading $O(y^3)$ order terms in eq. \eqref{eq:A3_eq_Eul}, which are responsible for the C- and S-shape of the pancake, respectively.

    \item We have also derived an analytical expression eq. \eqref{eq:CP_Y_9} for the joint probability distribution of the parameters for pancake, halo, and filament populations appearing in 2D cosmologies seeded by Gaussian random fields with power spectrum characterized by the index $n$ and smoothing scale $R_c$.
    
    \item Combining Gaussian statistics with our analytical model for pancake structure and formation, we have obtained the distributions of several observable features of the pancakes --- shell-crossing times (fig. \ref{fig:MP_Dtc}), curvature (fig. \ref{fig:MP_curv}), shape transition scale (fig. \ref{fig:MP_CS_trans}), and axis ratio (fig. \ref{fig:MP_ellip}). We have also noted in table \ref{table:exp_props} how their expectation values vary across halo and filament populations and scale with power spectrum parameters $n$ and $R_c$. On average, the points on the $A_3^+$ spine that evolve into halos undergo first shell-crossing earlier than those that evolve into filaments. The $A_3^+$ spine typically has a higher curvature at points of halos than that of filaments. Moreover, for halos, this curvature increases with time, whereas for filaments, it decreases. In general, the pancakes are dominantly C-shaped, filaments more so than halos. The average axes ratio of the $A_2$ caustic of the pancakes is significantly anisotropic, again, more so for filaments than halos.
\end{enumerate}

Improving this work will involve including vorticity and using expansion forms of order higher than $O(q^3, t^1)$. $O(q^{>3})$ terms will introduce higher order corrections to the shape of $A_3^+$ spine in Lagrangian space, which is only a straight line up to $O(q^{3})$. They will also introduce asymmetry in the width of the $A_2$ caustic along the shell-crossing axis on either side of the central point $q_c$. Extension to 3D is conceptually similar, but analytically cumbersome, as the number of coefficients, constraints, and parameters will increase. More importantly, it would allow us to test our predictions against observations of the large-scale structure. Deviations, if any, would be indicators of non-Gaussianity at corresponding scales.

We have thus demonstrated through this work that analytical models using catastrophe theory are relevant to the study of the cosmic web. Combined with Gaussian statistics, it allows us to make powerful predictions about the observable features of the large-scale structure of our universe.

\begin{acknowledgements}
This work is supported in part by the French Doctoral school ED127, Astronomy and Astrophysics Ile de France (AP), the Programme National
Cosmology et Galaxies (PNCG) of CNRS/INSU with INP and IN2P3, co-funded by CEA and CNES, the « action th\'ematique » Cosmology-Galaxies (ATCG) of the CNRS/INSU PN Astro (SC \& AP), Grant ISHIZUE 2025 of Kyoto University (AT), the Japan Society for the Promotion of Science (JSPS) Overseas Research Fellowships, the JSPS KAKENHI Grant No. 23K19050 and No. 24K17043 (SS); JP23K20844 and JP23K25868 (AT), and by the “PHC Sakura” program (project number: 51203TL, grant number: JPJSBP120243208), implemented by the French Ministry for Europe and Foreign Affairs, the French Ministry of Higher Education and Research, and the Japan Society for Promotion of Science (JSPS).

\end{acknowledgements}

\bibliographystyle{aa}
\bibliography{lib.bib}

\onecolumn

\appendix

\section{Analytical model using catastrophe theory}
\label{sec:appendix_A}

The expressions for the eigenvalues $\alpha > \beta$ and eigenvectors $\vec{n}^{\alpha}, \vec{n}^{\beta}$ of $d_{ij}$ are
\begin{align}
    \nonumber
    \alpha(\vec{q}, t) &= \frac{1}{2} \Big( \phi ^{2,0}(\vec{q},t) + \phi ^{0,2}(\vec{q},t)
    + \sqrt{ \left( \phi ^{2,0}(\vec{q},t) - \phi ^{0,2}(\vec{q},t) \right)^2 + 4 \phi ^{1,1}(\vec{q},t)^2 } \Big),\\
    \label{eq:alpha_LPT}
    \beta(\vec{q}, t) &=  \frac{1}{2} \Big( \phi ^{2,0}(\vec{q},t) + \phi ^{0,2}(\vec{q},t)
    - \sqrt{ \left( \phi ^{2,0}(\vec{q},t) - \phi ^{0,2}(\vec{q},t) \right)^2 + 4 \phi ^{1,1}(\vec{q},t)^2 } \Big),\\
    \nonumber
    \vec{n}^{\alpha}(\vec{q},t) &=
    \begin{bmatrix}
        \phi ^{2,0}(\vec{q}, t)-\phi ^{0,2}(\vec{q}, t)
        +\sqrt{ \left( \phi ^{2,0}(\vec{q},t) - \phi ^{0,2}(\vec{q},t) \right)^2 + 4 \phi ^{1,1}(\vec{q},t)^2 }
        \\
        2 \phi ^{1,1}(\vec{q}, t)
    \end{bmatrix},\\
    \label{eq:n_alpha_LPT}
    \vec{n}^{\alpha}(\vec{q},t) &=
    \begin{bmatrix}
        2 \phi ^{1,1}(\vec{q}, t)
        \\
        -\phi ^{2,0}(\vec{q}, t) + \phi ^{0,2}(\vec{q}, t)
        - \sqrt{ \left( \phi ^{2,0}(\vec{q},t) - \phi ^{0,2}(\vec{q},t) \right)^2 + 4 \phi ^{1,1}(\vec{q},t)^2 }
    \end{bmatrix}.
\end{align}
The expressions for the components of $\nabla_{\vec{q}}\alpha(\vec{q}, t)$ are
\begin{align}
    \nonumber
    & \frac{\partial \alpha}{\partial q_x}(\vec{q},t) = \frac{1}{2} \Bigg(\phi ^{1,2}(\vec{q},t) + \phi ^{3,0}(\vec{q},t) +\frac{4 \phi ^{1,1}(\vec{q},t) \phi ^{2,1}(\vec{q},t)+\left(\phi ^{2,0}(\vec{q},t)-\phi ^{0,2}(\vec{q},t)\right) \left(\phi ^{3,0}(\vec{q},t)-\phi ^{1,2}(\vec{q},t)\right)}{\sqrt{4 \phi ^{1,1}(\vec{q},t)^2+\left(\phi ^{2,0}(\vec{q},t)-\phi ^{0,2}(\vec{q},t)\right)^2}}\Bigg),\\
    \label{eq:grad_alpha_LPT}
    & \frac{\partial \alpha}{\partial q_y}(\vec{q},t) = \frac{1}{2} \Bigg(\phi ^{2,1}(\vec{q},t) + \phi ^{0,3}(\vec{q},t) +\frac{4 \phi ^{1,1}(\vec{q},t) \phi ^{1,2}(\vec{q},t)+\left(\phi ^{2,0}(\vec{q},t)-\phi ^{0,2}(\vec{q},t)\right) \left(\phi ^{2,1}(\vec{q},t)-\phi ^{0,3}(\vec{q},t)\right)}{\sqrt{4 \phi ^{1,1}(\vec{q},t)^2+\left(\phi ^{2,0}(\vec{q},t)-\phi ^{0,2}(\vec{q},t)\right)^2}}\Bigg).
\end{align}
The full expression for the Hessian of the eigenvalue field $\alpha(\vec{q},t)$ is
\begin{align}
    \nonumber
    H_{11} = &\frac{1}{2} \Bigg(-\frac{\left(4 \phi ^{1,1}(\vec{q},t) \phi ^{2,1}(\vec{q},t)+\left(\phi ^{2,0}(\vec{q},t)-\phi ^{0,2}(\vec{q},t)\right) \left(\phi ^{1,2}(\vec{q},t)-\phi ^{3,0}(\vec{q},t)\right)\right)^2}{\left(4 \phi ^{1,1}(\vec{q},t)^2+\left(\phi ^{2,0}(\vec{q},t)-\phi ^{0,2}(\vec{q},t)\right)^2\right)^{3/2}} + \phi ^{2,2}(\vec{q},t) + \phi ^{4,0}(\vec{q},t) \\
    \nonumber
    & + \frac{4 \phi ^{2,1}(\vec{q},t)^2+\left(\phi ^{1,2}(\vec{q},t)-\phi ^{3,0}(\vec{q},t)\right)^2+4 \phi ^{1,1}(\vec{q},t) \phi ^{3,1}(\vec{q},t)+\left(\phi ^{0,2}(\vec{q},t)-\phi ^{2,0}(\vec{q},t)\right) \left(\phi ^{2,2}(\vec{q},t)-\phi ^{4,0}(\vec{q},t)\right)}{\sqrt{4 \phi ^{1,1}(\vec{q},t)^2+\left(\phi ^{2,0}(\vec{q},t)-\phi ^{0,2}(\vec{q},t)\right)^2}}\Bigg), \\
    \nonumber
    H_{12} = &\frac{1}{2} \Bigg[ \phi ^{1,3}(\vec{q},t) + \phi ^{3,1}(\vec{q},t) + \frac{1}{\left(4 \phi ^{1,1}(\vec{q},t)^2+\left(\phi ^{2,0}(\vec{q},t)-\phi ^{0,2}(\vec{q},t)\right)^2\right)^{3/2}} \Bigg( \left( 4 \phi ^{1,1}(\vec{q},t)^2+\left(\phi ^{2,0}(\vec{q},t)-\phi ^{0,2}(\vec{q},t)\right)^2 \right)\\
    \nonumber
    & \Big(4 \phi ^{1,1}(\vec{q},t) \phi ^{2,2}(\vec{q},t)+\phi ^{0,3}(\vec{q},t) \left(\phi ^{1,2}(\vec{q},t)-\phi ^{3,0}(\vec{q},t)\right)+\phi ^{2,1}(\vec{q},t) \left(3 \phi ^{1,2}(\vec{q},t)+\phi ^{3,0}(\vec{q},t)\right)\\
    \nonumber
    &+\left(\phi ^{0,2}(\vec{q},t)-\phi ^{2,0}(\vec{q},t)\right) \left(\phi ^{1,3}(\vec{q},t)-\phi ^{3,1}(\vec{q},t)\right)\Big) \\
    \nonumber
    &-\left(4 \phi ^{1,1}(\vec{q},t) \phi ^{1,2}(\vec{q},t)+\left(\phi ^{0,2}(\vec{q},t)-\phi ^{2,0}(\vec{q},t)\right) \left(\phi ^{0,3}(\vec{q},t)-\phi ^{2,1}(\vec{q},t)\right)\right) \\
    \nonumber
    &\left(4 \phi ^{1,1}(\vec{q},t) \phi ^{2,1}(\vec{q},t)+\left(\phi ^{0,2}(\vec{q},t)-\phi ^{2,0}(\vec{q},t)\right) \left(\phi ^{1,2}(\vec{q},t)-\phi ^{3,0}(\vec{q},t)\right)\right) \Bigg) \Bigg],\\
    \nonumber
    H_{22} = &\frac{1}{2} \Bigg(-\frac{\left(4 \phi ^{1,1}(\vec{q},t) \phi ^{1,2}(\vec{q},t)+\left(\phi ^{0,2}(\vec{q},t)-\phi ^{2,0}(\vec{q},t)\right) \left(\phi ^{0,3}(\vec{q},t)-\phi ^{2,1}(\vec{q},t)\right)\right)^2}{\left(4 \phi ^{1,1}(\vec{q},t)^2+\left(\phi ^{2,0}(\vec{q},t)-\phi ^{0,2}(\vec{q},t)\right)^2\right)^{3/2}}-\phi ^{0,4}(\vec{q},t) -\phi ^{2,2}(\vec{q},t) \\
    \label{eq:H_alpha_LPT}
    &+\frac{\left(\phi ^{0,3}(\vec{q},t)-\phi ^{2,1}(\vec{q},t)\right)^2+4 \left(\phi ^{1,2}(\vec{q},t)^2+\phi ^{1,1}(\vec{q},t) \phi ^{1,3}(\vec{q},t)\right)+\left(\phi ^{0,2}(\vec{q},t)-\phi ^{2,0}(\vec{q},t)\right) \left(\phi ^{0,4}(\vec{q},t)-\phi ^{2,2}(\vec{q},t)\right)}{\sqrt{4 \phi ^{1,1}(\vec{q},t)^2+\left(\phi ^{2,0}(\vec{q},t)-\phi ^{0,2}(\vec{q},t)\right)^2}}\Bigg).
\end{align}
The expressions for the eigenvalues $\alpha_H > \beta_H$ and eigenvectors $\vec{n}_H^{\alpha}, \vec{n}_H^{\beta}$ of $H_{ij}(\alpha)$ at $(\vec{q}_c , t_c)$ are
\begin{align}
    \nonumber
    \alpha_H(\vec{q}_c,t_c) &= \frac{1}{2} \Bigg[ \left( \phi^{4,0,0} + \phi^{2,2,0} + \frac{2 (\phi^{1,2,0})^2}{1 - \phi^{0,2,0}} \right) + \sqrt{ \left( \phi^{4,0,0} - \phi^{2,2,0} - \frac{2 (\phi^{1,2,0})^2}{1 - \phi^{0,2,0}} \right)^2 + 4 (\phi^{3,1,0})^2} \Bigg],\\
    \label{eq:alpha_H_LPT}
    \beta_H(\vec{q}_c,t_c) &= \frac{1}{2} \Bigg[ \left( \phi^{4,0,0} + \phi^{2,2,0} + \frac{2 (\phi^{1,2,0})^2}{1 - \phi^{0,2,0}} \right) - \sqrt{ \left( \phi^{4,0,0} - \phi^{2,2,0} - \frac{2 (\phi^{1,2,0})^2}{1 - \phi^{0,2,0}} \right)^2 + 4 (\phi^{3,1,0})^2} \Bigg],\\
    \nonumber
    \vec{n}^\alpha_H(\vec{q}_c,t_c) &= 
    \begin{bmatrix}
        2 \phi^{3,1,0}\\
        -\phi^{4,0,0} + \phi^{2,2,0} + 2 (\phi^{1,2,0})^2 / (1 - \phi^{0,2,0}) + \sqrt{ \left( -\phi^{4,0,0} + \phi^{2,2,0} + 2 (\phi^{1,2,0})^2 / (1 - \phi^{0,2,0}) \right)^2 + 4 (\phi^{3,1,0})^2 }
    \end{bmatrix},\\
    \label{eq:n_alpha_H_LPT}
    \vec{n}^\beta_H(\vec{q}_c,t_c) &=
    \begin{bmatrix}
        \phi^{4,0,0} - \phi^{2,2,0} - 2 (\phi^{1,2,0})^2 / (1 - \phi^{0,2,0}) - \sqrt{ \left( -\phi^{4,0,0} + \phi^{2,2,0} + 2 (\phi^{1,2,0})^2 / (1 - \phi^{0,2,0}) \right)^2 + 4 (\phi^{3,1,0})^2 }\\
        2 \phi^{3,1,0}
    \end{bmatrix}.
\end{align}
The complete equations for the shape of the $A_3^+$ spine about a shell-crossing point in Eulerian space are
\begin{align}
    \label{eq:A3_eq_Eul_full}
    x_{A3}(\vec{q}, t) &= \left( \frac{ \phi ^{2,0,1} \phi ^{3,1,0}}{\phi^{4,0,0}} \right) (t-t_c) q_y - \frac{1}{2} \phi^{1,2,0} q_y^2 - \frac{1}{6} \left( \phi^{1,3,0} + \phi^{3,1,0} \left( 2\left(\frac{\phi^{3,1,0}}{\phi^{4,0,0}}\right)^2 - 3\frac{\phi^{2,2,0}}{\phi^{4,0,0}} \right) \right)q_y^3,\\
    \nonumber
    y_{A3}(\vec{q}, t) &= \left( 1 - \phi ^{0,2,0} - \phi ^{0,2,1} (t-t_c) \right) q_y - \left(\frac{1}{2} \phi ^{0,3,0}-\frac{\phi ^{1,2,0} \phi ^{3,1,0}}{\phi ^{4,0,0}}\right)q_y^2 -  \frac{1}{6} \left(\phi^{0,4,0} - \phi ^{3,1,0} \left( \left(\frac{ \phi^{3,1,0} }{\phi^{4,0,0}}\right)^3 - 3 \frac{ (\phi^{2,2,0} \phi ^{3,1,0}) }{(\phi^{4,0,0})^2} + 3 \left( \frac{\phi ^{1,3,0} }{\phi ^{4,0,0}} \right) \right) \right)q_y^3.
\end{align}

\section{Gaussian statistics}
\label{sec:appendix_B}

The transformation of $O(q^3)$ and $O(q^4)$ derivatives of the potential $\phi$ under rotation by $\theta$ is given by:
\begin{align}
    \label{eq:Y_tilda_Y} &
    \begin{bmatrix}
         \tilde{\phi}^{3,0,0} \\  \tilde{\phi}^{2,1,0} \\  \tilde{\phi}^{1,2,0} \\  \tilde{\phi}^{0,3,0}
    \end{bmatrix}
    =
    \begin{bmatrix}
        \sin ^3(\theta ) \phi ^{0,3,0}+\cos ^3(\theta ) \phi ^{3,0,0}+3 \sin (\theta ) \cos (\theta ) \left(\sin (\theta ) \phi ^{1,2,0}+\cos (\theta ) \phi ^{2,1,0}\right) \\
        -\frac{1}{4} (\sin (\theta )-3 \sin (3 \theta )) \phi ^{1,2,0}+\sin ^2(\theta ) \cos (\theta ) \phi ^{0,3,0}+\frac{1}{2} \cos (\theta ) \left((3 \cos (2 \theta )-1) \phi ^{2,1,0}-\sin (2 \theta ) \phi ^{3,0,0}\right)\\
        \sin (\theta ) \cos ^2(\theta ) \phi ^{0,3,0}+\frac{1}{4} \left((\sin (\theta )-3 \sin (3 \theta )) \phi ^{2,1,0}+(\cos (\theta )+3 \cos (3 \theta )) \phi ^{1,2,0}+4 \sin ^2(\theta ) \cos (\theta ) \phi ^{3,0,0}\right)\\
        -\sin ^3(\theta ) \phi ^{3,0,0}+\cos ^3(\theta ) \phi ^{0,3,0}+3 \sin (\theta ) \cos (\theta ) \left(\sin (\theta ) \phi ^{2,1,0}-\cos (\theta ) \phi ^{1,2,0}\right)
    \end{bmatrix}
    \\
    \nonumber
    &
    \begin{bmatrix}
        \tilde{\phi}^{4,0,0} \\  \tilde{\phi}^{3,1,0} \\  \tilde{\phi}^{2,2,0} \\  \tilde{\phi}^{1,3,0} \\  \tilde{\phi}^{0,4,0}
    \end{bmatrix}
    =
    \begin{bmatrix}
        \sin ^4(\theta ) \phi ^{0,4,0}+\cos (\theta ) \left(4 \sin ^3(\theta ) \phi ^{1,3,0}+\cos (\theta ) \left(6 \sin ^2(\theta ) \phi ^{2,2,0}+\cos ^2(\theta ) \phi ^{4,0,0}+4 \sin (\theta ) \cos (\theta ) \phi ^{3,1,0}\right)\right) \\
 \frac{3}{4} \sin (4 \theta ) \phi ^{2,2,0}+\cos ^2(\theta ) \left((2 \cos (2 \theta )-1) \phi ^{3,1,0}-\sin (\theta ) \cos (\theta ) \phi ^{4,0,0}\right)+\sin ^3(\theta ) \cos (\theta ) \phi ^{0,4,0}+\sin ^2(\theta ) (2 \cos (2 \theta )+1) \phi ^{1,3,0} \\
 \sin ^2(\theta ) \cos ^2(\theta ) \phi ^{4,0,0}+\frac{1}{4} \left(\sin ^2(2 \theta ) \phi ^{0,4,0}+2 \sin (4 \theta ) \left(\phi ^{1,3,0}-\phi ^{3,1,0}\right)+(3 \cos (4 \theta )+1) \phi ^{2,2,0}\right) \\
 -\frac{3}{4} \sin (4 \theta ) \phi ^{2,2,0}+\sin (\theta ) \sin (3 \theta ) \phi ^{3,1,0}+\frac{1}{2} (\cos (2 \theta )+\cos (4 \theta )) \phi ^{1,3,0}+\sin (\theta ) \cos ^3(\theta ) \phi ^{0,4,0}-\sin ^3(\theta ) \cos (\theta ) \phi ^{4,0,0} \\
 \cos ^4(\theta ) \phi ^{0,4,0}+\sin (\theta ) \left(\sin (\theta ) \left(\sin ^2(\theta ) \phi ^{4,0,0}+6 \cos ^2(\theta ) \phi ^{2,2,0}-4 \sin (\theta ) \cos (\theta ) \phi ^{3,1,0}\right)-4 \cos ^3(\theta ) \phi ^{1,3,0}\right)
    \end{bmatrix}.
\end{align}
The Jacobian matrices for $O(q^2)$, $O(q^3)$, and $O(q^4)$ derivatives of $\phi$, individually, are:
\begin{align}
    \label{eq:J_Y_tilda_Y}
    & J_{2,2} = 
    \begin{bmatrix}
        \cos ^2(\theta ) & \sin (\theta ) (-\cos (\theta )) & \sin ^2(\theta ) \\
        2 \sin (\theta ) \cos (\theta ) (\phi ^{0,2,0} - \phi ^{2,0,0})   & ( \cos ^2(\theta ) -\sin ^2(\theta ) ) \left(\phi ^{0,2,0}-\phi ^{2,0,0}\right) & 2 \sin (\theta ) \cos (\theta ) (\phi ^{2,0,0} - \phi ^{0,2,0}) \\
        \sin ^2(\theta ) & \sin (\theta ) \cos (\theta ) & \cos ^2(\theta )
    \end{bmatrix},
    \\
    \nonumber
    & J_{3,3} = 
    \begin{bmatrix}
        \cos ^3(\theta ) & -\frac{1}{2} \sin (2 \theta ) \cos (\theta ) & \sin ^2(\theta ) \cos (\theta ) & -\sin ^3(\theta )\\
        3 \sin (\theta ) \cos ^2(\theta ) & \frac{1}{2} \cos (\theta ) (3 \cos (2 \theta )-1) & \frac{1}{4} (\sin (\theta )-3 \sin (3 \theta )) & 3 \sin ^2(\theta ) \cos (\theta )\\
        3 \sin ^2(\theta ) \cos (\theta ) & \frac{1}{4} (3 \sin (3 \theta )-\sin (\theta )) & \frac{1}{4} (\cos (\theta )+3 \cos (3 \theta )) & -3 \sin (\theta ) \cos ^2(\theta ) \\
        \sin ^3(\theta ) & \sin ^2(\theta ) \cos (\theta ) & \sin (\theta ) \cos ^2(\theta ) & \cos ^3(\theta )
    \end{bmatrix},
    \\
    \nonumber
    & J_{4,4} =
    \begin{bmatrix}
         \cos ^4(\theta ) & \sin (\theta ) \left(-\cos ^3(\theta )\right) & \sin ^2(\theta ) \cos ^2(\theta ) & \sin ^3(\theta ) (-\cos (\theta )) & \sin ^4(\theta ) \\
         4 \sin (\theta ) \cos ^3(\theta ) & \cos ^2(\theta ) (2 \cos (2 \theta )-1) & -\frac{1}{2} \sin (4 \theta ) & \sin (\theta ) \sin (3 \theta ) & -4 \sin ^3(\theta ) \cos (\theta ) \\
         6 \sin ^2(\theta ) \cos ^2(\theta ) & \frac{3}{4} \sin (4 \theta ) & \frac{1}{4} (3 \cos (4 \theta )+1) & \frac{1}{4} (-3) \sin (4 \theta ) & 6 \sin ^2(\theta ) \cos ^2(\theta ) \\
         4 \sin ^3(\theta ) \cos (\theta ) & \sin ^2(\theta ) (2 \cos (2 \theta )+1) & \frac{1}{2} \sin (4 \theta ) & \frac{1}{2} (\cos (2 \theta )+\cos (4 \theta )) & -4 \sin (\theta ) \cos ^3(\theta ) \\
         \sin ^4(\theta ) & \sin ^3(\theta ) \cos (\theta ) & \frac{1}{4} \sin ^2(2 \theta ) & \sin (\theta ) \cos ^3(\theta ) & \cos ^4(\theta ) \\
    \end{bmatrix}.
\end{align}
Their determinants are $|\phi ^{2,0,0} - \phi ^{0,2,0}|$, $1$, and $1$, respectively.

The full expression for the probability density of $Y$:
\begin{align}
    \label{eq:P_Y_full}
    P(Y) = &\frac{|\phi ^{2,0,0} - \phi ^{0,2,0}| }{\pi ^6 \sqrt{8 \sigma _3^8 \left(\sigma _2^2 \sigma _4^4 - \sigma _3^4 \sigma _4^2 \right) \left( \sigma _4^2 \sigma _3^8- 2 \sigma _2^2 \sigma _4^4 \sigma _3^4 + \sigma _2^4 \sigma _4^6\right)}} \\
    \nonumber
    &\exp \Bigg( -\frac{1}{2\sigma _3^2 \sigma _4^2 \left(\sigma _3^4-\sigma _2^2 \sigma _4^2\right)}
    \Big( 
    -4 \sigma _2^2 \sigma _4^4 \left( (\phi ^{0,3,0})^2-2 \phi ^{2,1,0} \phi ^{0,3,0}+(\phi ^{3,0,0})^2+5 \left( (\phi ^{1,2,0})^2+(\phi ^{2,1,0})^2\right)-2 \phi ^{1,2,0} \phi ^{3,0,0}\right) \\
    \nonumber
    & +2 \sigma _3^6 \left(16 \left(\phi ^{1,3,0}-\phi ^{3,1,0}\right)^2+\left(\phi ^{0,4,0}-6 \phi ^{2,2,0}+\phi ^{4,0,0}\right)^2\right)\\
    \nonumber
    & +2 \sigma _3^4 \sigma _4^2 \Big(2 (\phi ^{0,3,0})^2-4 \phi ^{2,1,0} \phi ^{0,3,0}+2 \left((\phi ^{3,0,0})^2-2 \phi ^{1,2,0} \phi ^{3,0,0}+5 \left((\phi ^{1,2,0})^2+(\phi ^{2,1,0})^2\right)\right)\\
    \nonumber
    &\hspace{1.5cm}+\phi ^{0,2,0} \left(-3 \phi ^{0,4,0}-2 \phi ^{2,2,0}+\phi ^{4,0,0}\right)+\phi ^{2,0,0} \left(\phi ^{0,4,0}-2 \phi ^{2,2,0}-3 \phi ^{4,0,0}\right)\Big) \\
    \nonumber
    & -\sigma _3^2 \sigma _4^2 \big(\sigma _2^2 \left(5 (\phi ^{0,4,0})^2+2 \left(\phi ^{4,0,0}-10 \phi ^{2,2,0}\right) \phi ^{0,4,0}+40 (\phi ^{1,3,0})^2+76 (\phi ^{2,2,0})^2 \right.\\
    \nonumber
    &\hspace{1.5cm} \left. +40 (\phi ^{3,1,0})^2+5 (\phi ^{4,0,0})^2-48 \phi ^{1,3,0} \phi ^{3,1,0}-20 \phi ^{2,2,0} \phi ^{4,0,0}\right)\\
    \nonumber
    &\hspace{1cm}+\sigma _4^2 \left(3 (\phi ^{0,2,0})^2-2 \phi ^{2,0,0} \phi ^{0,2,0}+3 (\phi ^{2,0,0})^2\right)\big)
    \Big)
    \Bigg).
\end{align}

\begin{figure}
    \includegraphics[width=0.45\textwidth]{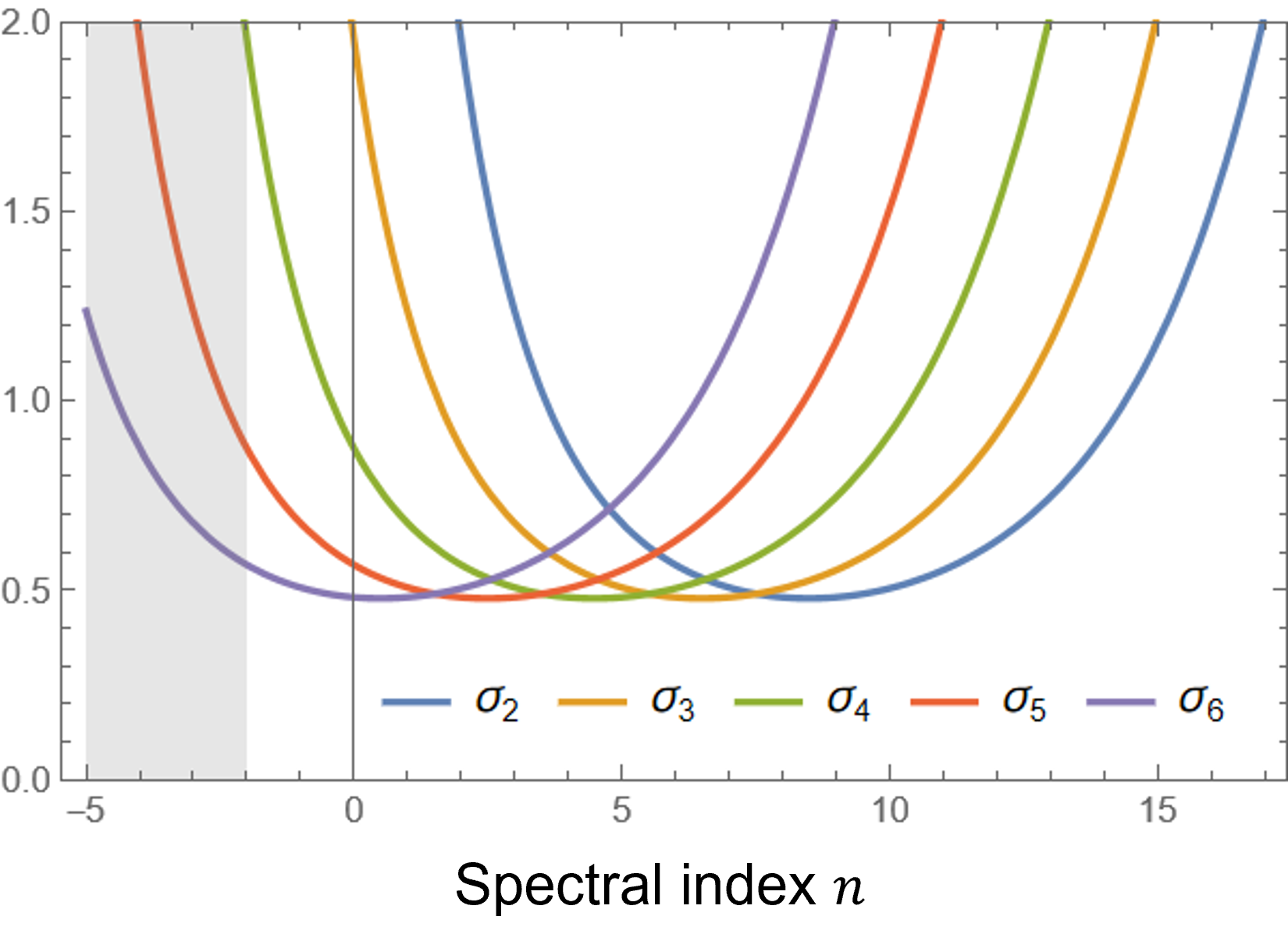}
    \includegraphics[width=0.45\textwidth]{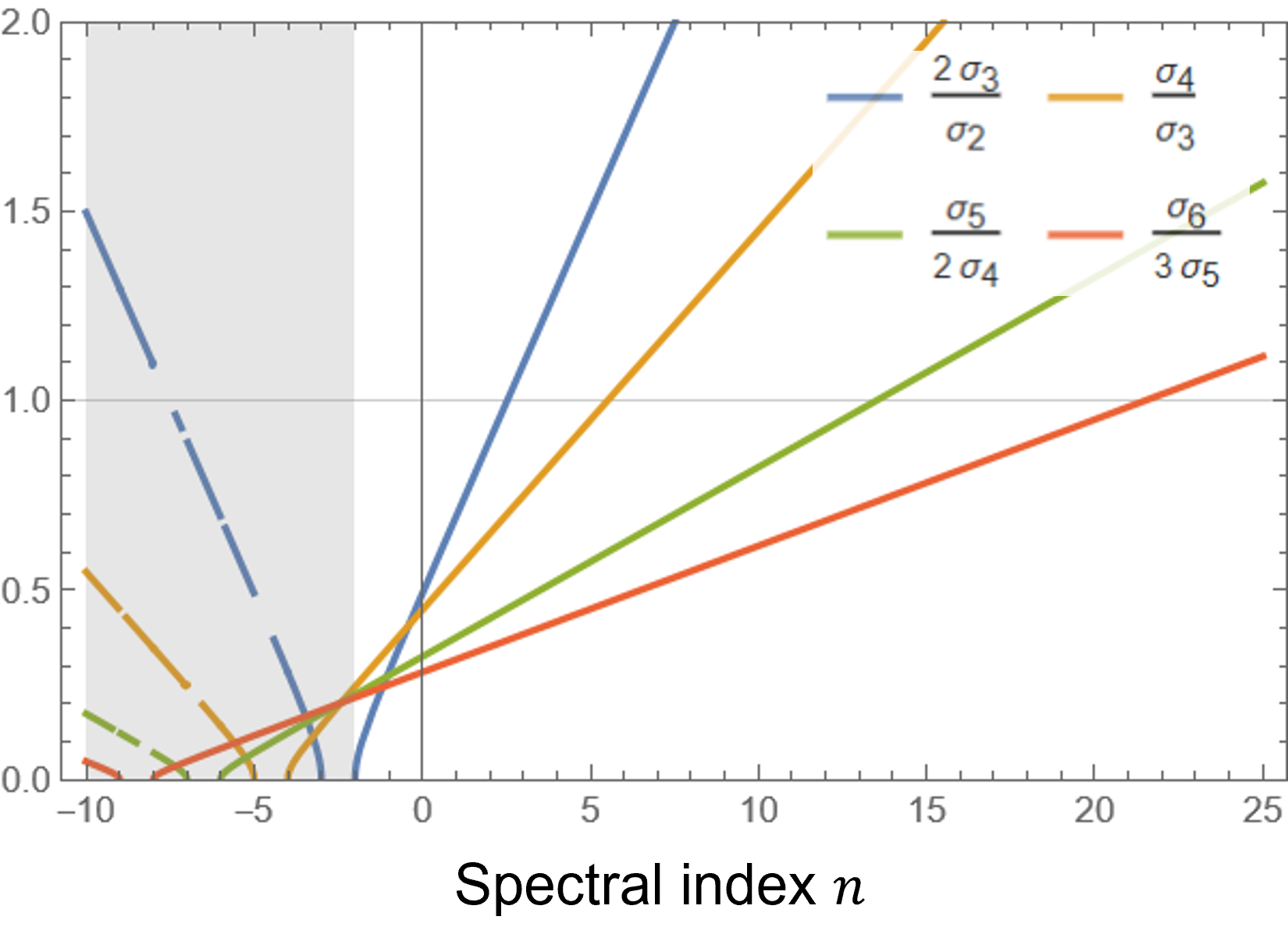}
    \centering
    \caption{Variation of $\sigma_i$ (left) and $\sigma_i / \sigma_{i-1}$ (right) with the spectral index $n$ (fixed smoothing scale $R$ and exponential filter in eq. \eqref{eq:sigma}).}
    \label{fig:sigma_n}
\end{figure}

\begin{figure}
    \includegraphics[width=0.45\textwidth]{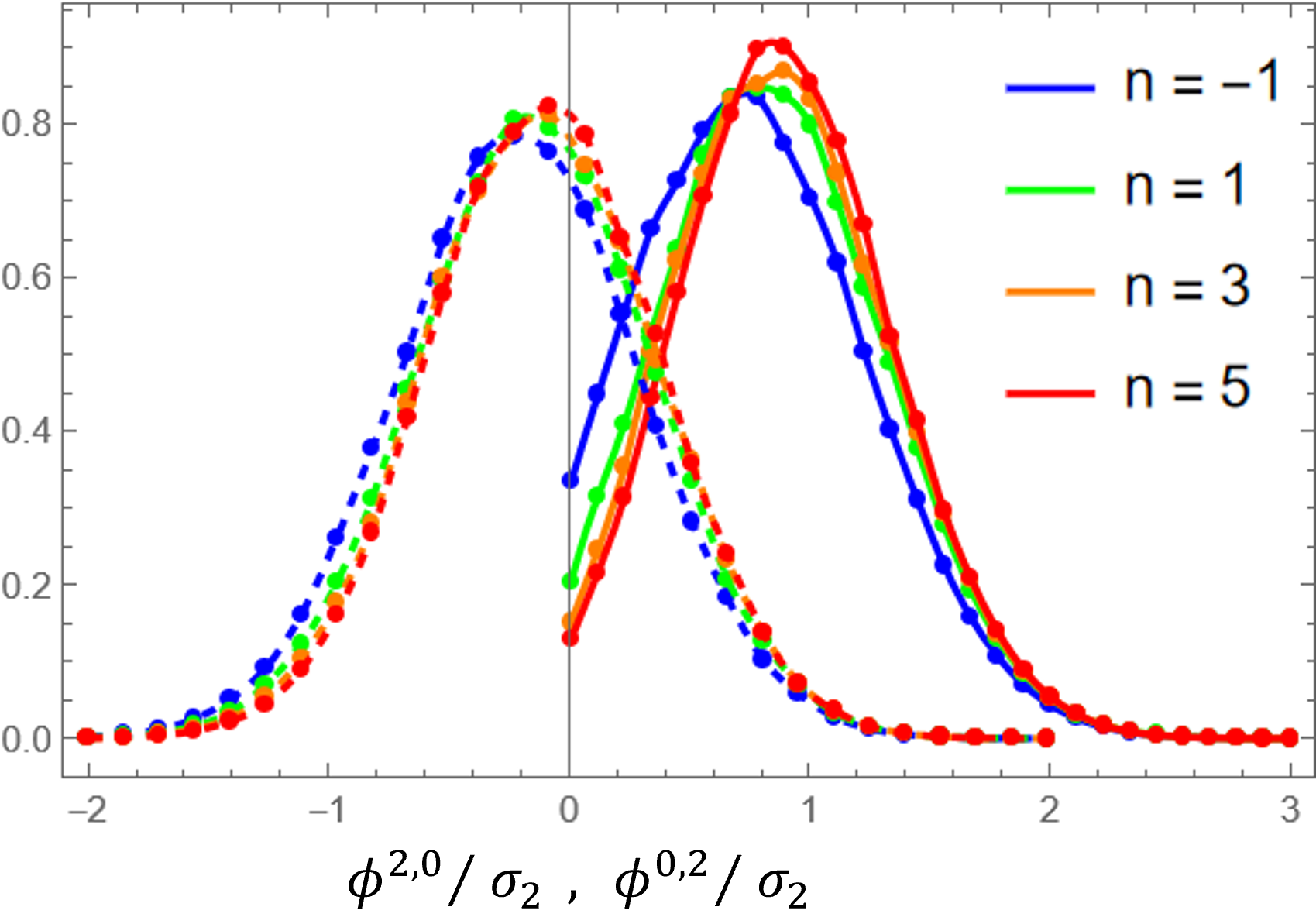}
    \includegraphics[width=0.45\textwidth]{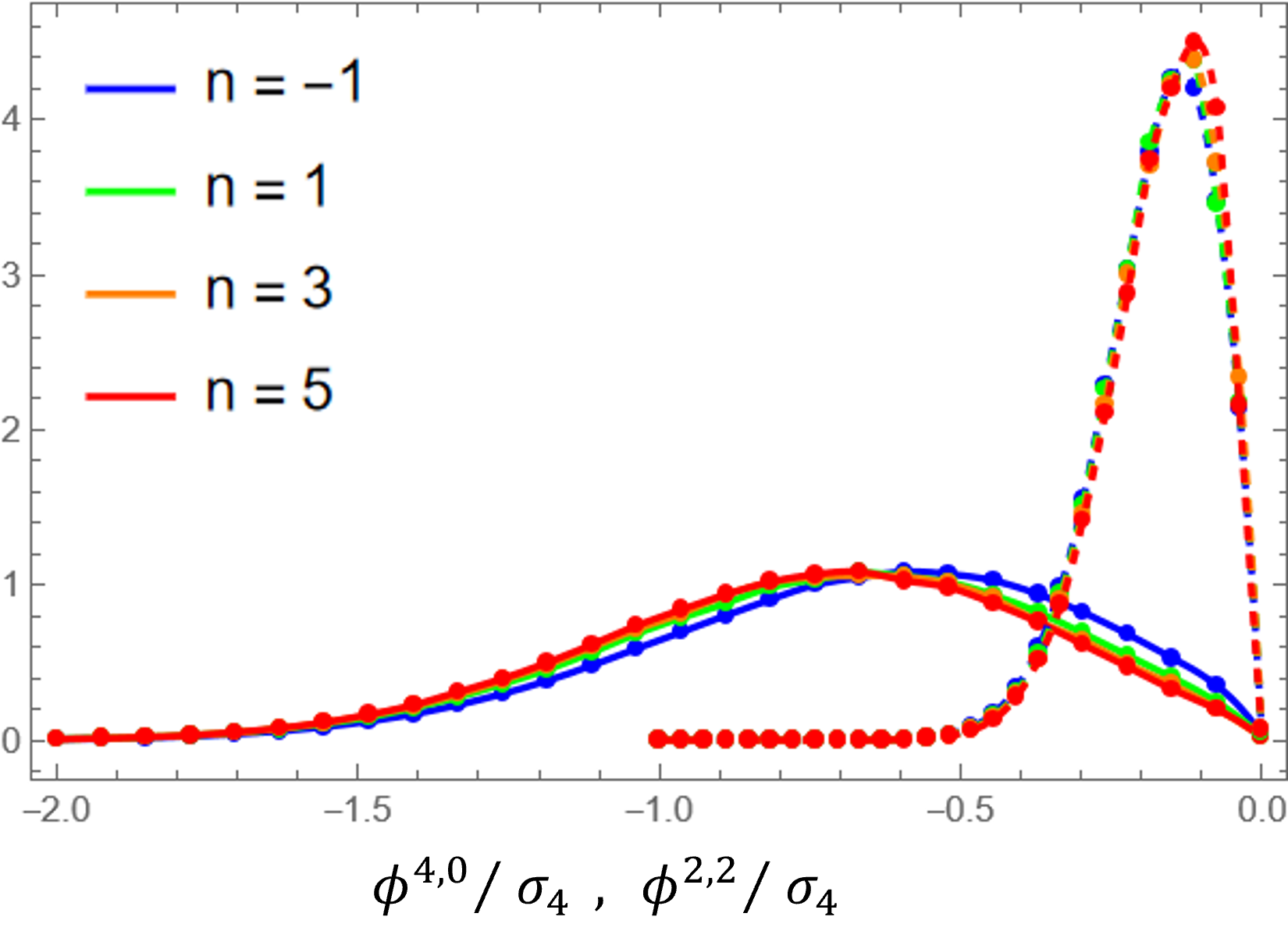}
    \centering
    \caption{The marginal distributions of $\phi^{2,0} / \sigma_2$ (solid) and $\phi^{0,2} / \sigma_2$ (dashed) in the left panel, and $\phi^{4,0} / \sigma_4$ (solid) and $\phi^{2,2} / \sigma_4$ (dashed) in the right panel for different values of the spectral index $n$.}
    \label{fig:MP_phi_n}
\end{figure}

Our statistical results --- distributions and expectation values --- depend on the power spectrum parameters $n, R_c$ solely through the $\sigma$'s (refer to eq. \eqref{eq:CP_Y_9}). Eq. \eqref{eq:sigma} shows that $\sigma_i$ has a simple power-law fall-off with the smoothing scale $R_c$. In the left panel of fig. \ref{fig:sigma_n}, we show the dependence of $\sigma$'s on the spectral index $n$, assuming exponential smoothing. The case $n = -2$ corresponds to the scale-invariant CDM power spectrum in 2D, while larger values of $n$ are 
characteristic of warm dark matter models. Consequently, the physically relevant range of $n$ is typically $(-2, 5]$, within which we adopt $n=1$ as our representative choice. 

The highest order term in eq. \eqref{eq:A3_eq} for the $A_3^+$ spine is proportional to $O(q^4)$ derivative of $\phi$, which is typically of the magnitude $\sigma_4$ since $\langle|\phi^{i,j}|\rangle \propto \sigma_{i+j}$. Higher order expansions would introduce terms proportional to $O(q^5)$ and $O(q^6)$ derivatives of $\phi$, which are of magnitudes $\approx \sigma_5 / 2$ and $\sigma_6 / 6$, respectively. These would successively turn significant when $\sigma_5 / 2 \sigma_4 \approx 1$ and $\sigma_6 / 3 \sigma_5 \approx 1$. In the right panel of fig. \ref{fig:sigma_n}, we show how the ratio of successive moments $\sigma_i / \sigma_{i-1}$ vary with the spectral index $n$ assuming exponential smoothing, from which we note that the $O(q^5)$ and $O(q^6)$ derivatives of $\phi$ turn significant for $n \gtrapprox 13$ and $n \gtrapprox 21$, respectively. Therefore, for $n \in (-2, 5]$, neglecting $O(q^5)$ derivatives in eq. \eqref{eq:A3_eq_Eul} (and in the derivations for properties like curvature and C-S transition scale) is a good enough approximation for the pancake shape.

By scaling our results with the corresponding $\sigma$'s, we render them independent of $R_c$. For uncorrelated derivatives, the distributions of $\phi^{i,j} / \sigma_{i+j}$ are also independent of $n$, whereas those of the correlated derivatives do show some $n$-dependence, as shown in fig. \ref{fig:MP_phi_n}. The variations in the distributions of $\phi^{2,0} / \sigma_{2}, \phi^{0,2} / \sigma_{2}$, and $ \phi^{4,0} / \sigma_{4}$ are indeed noticeable with the expectation values varying within $10\%$ of the FWHM, which is certainly not strong enough to alter our main inferences, at least for $n \in (-2, 5]$. The other correlated derivatives --- $\phi^{1,2} / \sigma_3, \phi^{1,3}/ \sigma_4, \phi^{3,1}/ \sigma_4, \phi^{0,4}/ \sigma_4$ --- show even less variation. We can thus expect that the distributions of derived properties, such as curvature and C-S transition scale (refer figures \ref{fig:MP_Dtc}-\ref{fig:MP_ellip}), when scaled by the appropriate $\sigma$'s, remain largely independent of $n$ within the physically relevant range $n \in (-2, 5]$.

\section{Numerical verification}
\label{sec:numerical_verification}

\begin{figure}
    \centering
    \includegraphics[width=\linewidth]{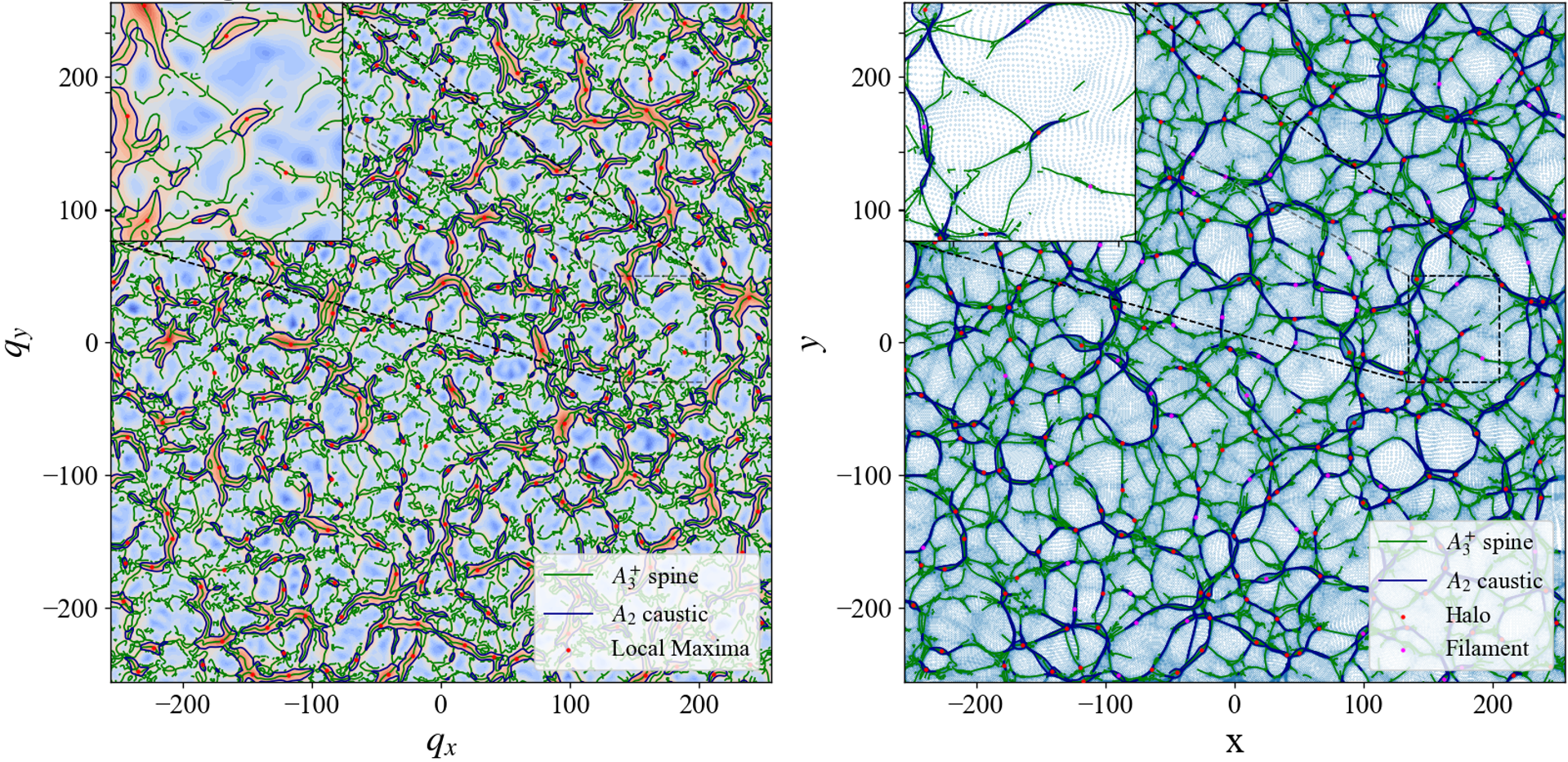}
    \caption{Snapshot from $512^2$ 1-LPT simulation corresponding to power spectrum $n = 1, R_c = 10$. The left panel shows a color-map of the higher eigenvalue $\alpha(\vec{q})$ in Lagrangian space. The right panel shows the particles in Eulerian space. Points of local maxima of the eigenvalue field are highlighted in red. The $A_3^+$ spine and $A_2$ caustics are shown as green and blue curves, respectively.}
    \label{fig:sim}
\end{figure}
\begin{figure}
    \centering
    \includegraphics[width = \linewidth]{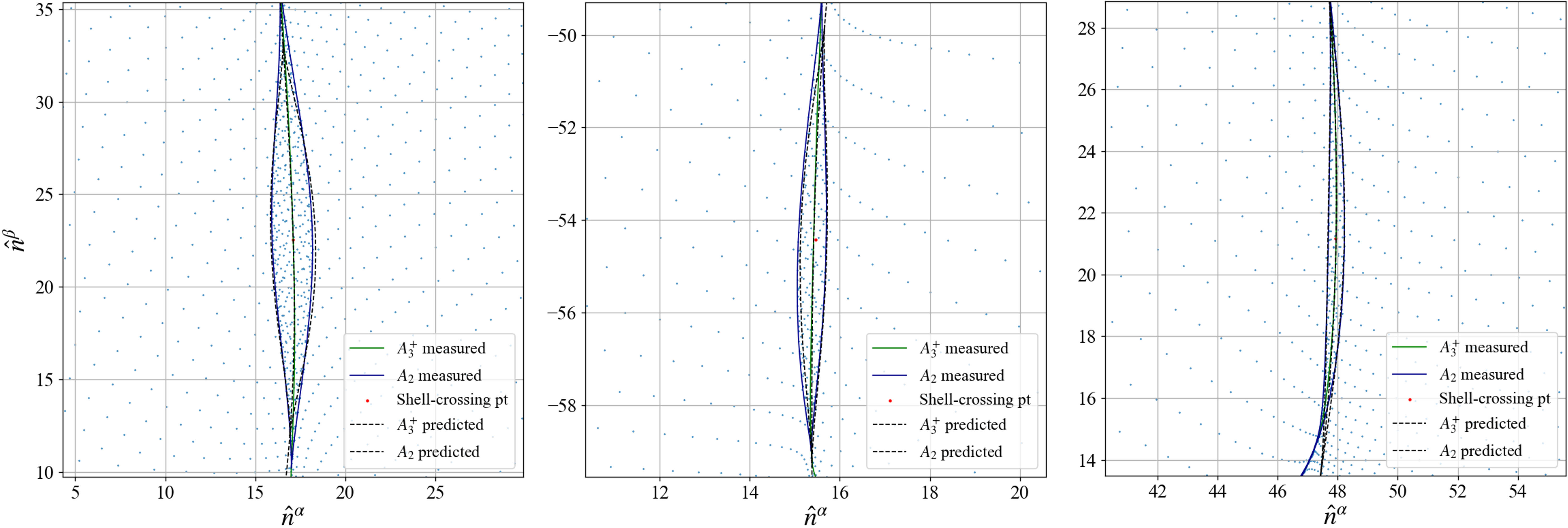}
    \caption{Three selected shell-crossing structures in a 1-LPT simulation initialized using Gaussian random field with the power-spectrum parameters $n = 1, R_c = 10$. The $A_3^+$ spine and $A_2$ caustic measured from the simulation are shown in solid lines. For comparison, the $A_3^+$ spine and $A_2$ caustic predicted from our analytical model are shown in dashed lines.}
    \label{fig:eu_shape_sim}
\end{figure}

\begin{figure}
    \centering
    \includegraphics[width=0.425\linewidth]{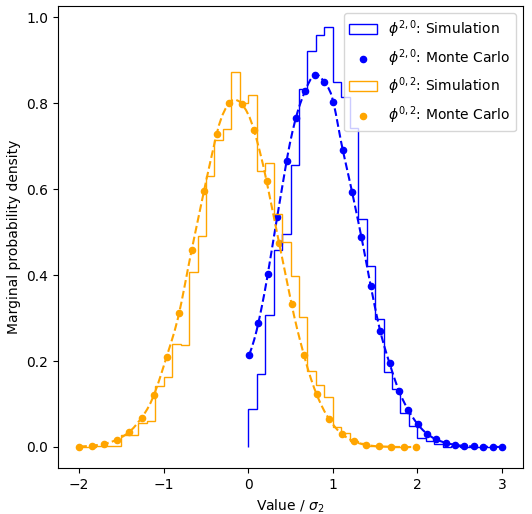}
    \includegraphics[width=0.405\linewidth]{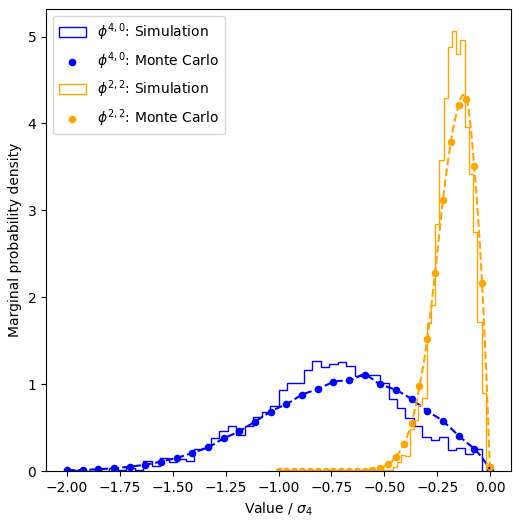}
    \includegraphics[width=0.825\linewidth]{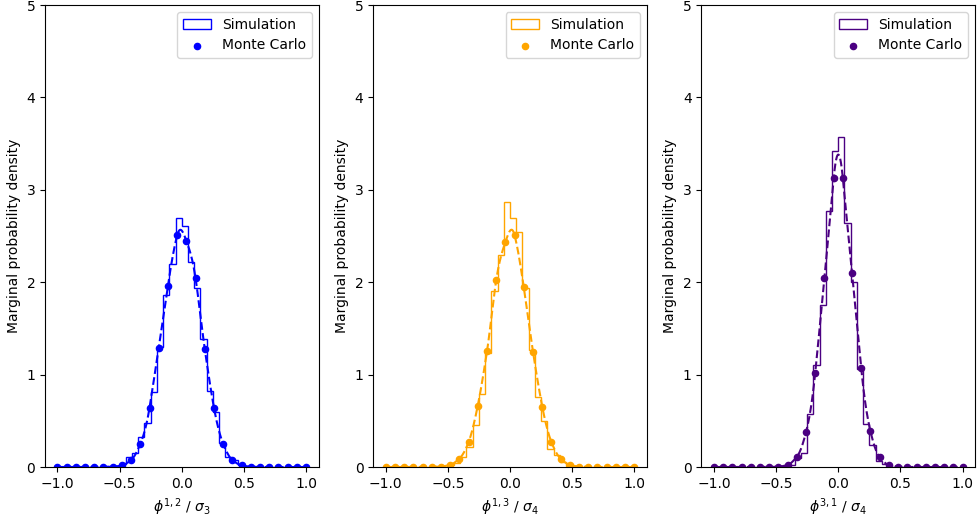}
    \includegraphics[width=0.825\linewidth]{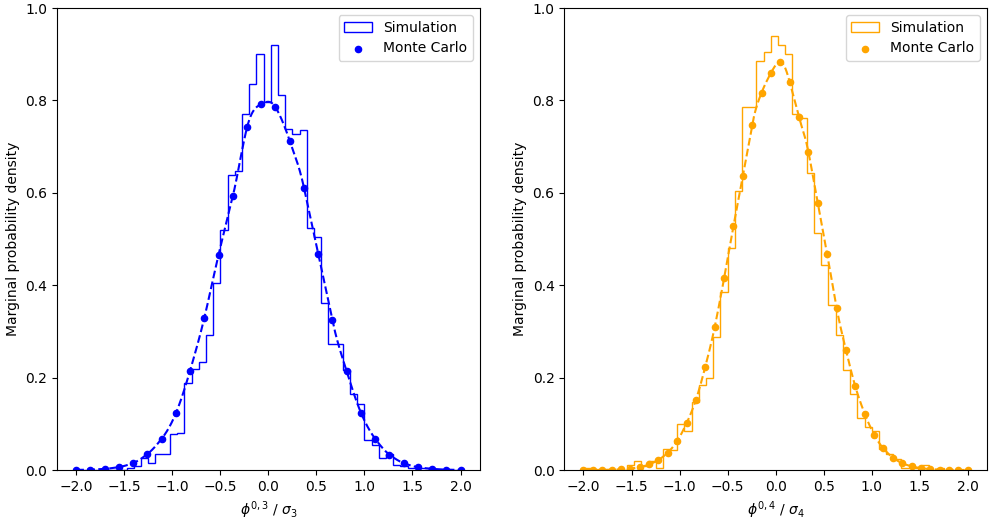}
    \caption{Comparison between marginal probability densities of derivatives of $\phi$ measured within the pancake population in 1-LPT simulations initialized using Gaussian random field with the parameters $n = 1, R_c = 10$ (shown in step histograms) and those computed using Monte Carlo integration of eq. \eqref{eq:CP_Y_9} for the same set of parameters $(n, R_c)$ (shown in interpolated curves).}
    \label{fig:hist_MP_phi}
\end{figure}

\begin{figure}
    \centering
    \includegraphics[width=0.45\linewidth]{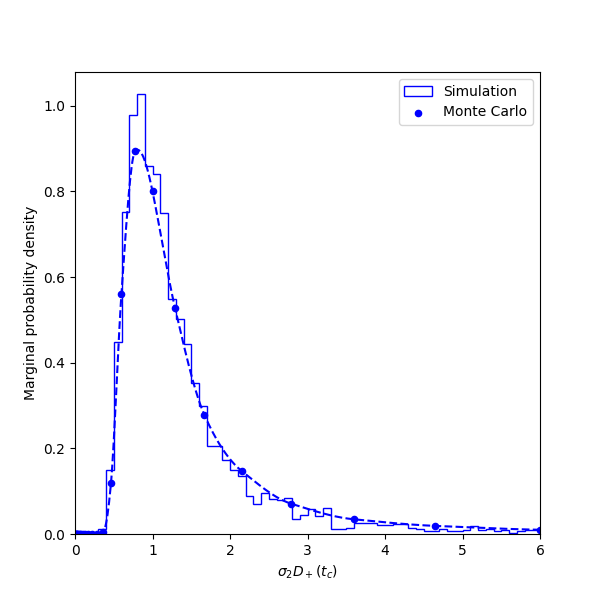}
    \includegraphics[width=0.45\linewidth]{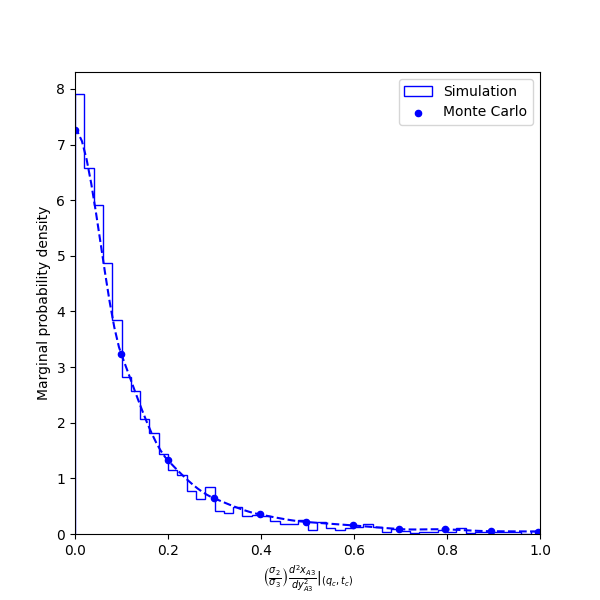}
    \includegraphics[width=0.45\linewidth]{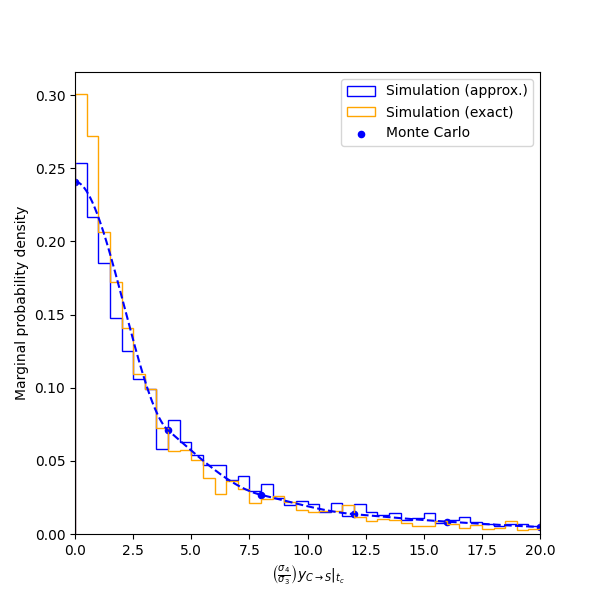}
    \includegraphics[width=0.45\linewidth]{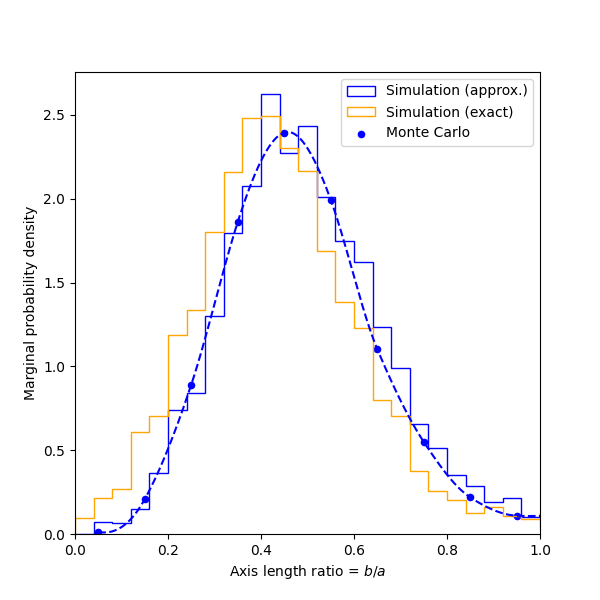}
    \caption{Comparison between marginal probability densities of shell-crossing times (top left), curvature (top right), C-S transition scale (bottom left), and axis length ratio (bottom right) measured within the pancake population in 1-LPT simulations initialized using Gaussian random field with the parameters $n = 1, R_c = 10$ (shown in step histograms) and those computed using Monte Carlo integration of eq. \eqref{eq:CP_Y_9} and transformation through eqs. \eqref{eq:P_Dtc}, \eqref{eq:P_curv}, \eqref{eq:P_CS_trans}, and \eqref{eq:P_ellip} for the same set of parameters $(n, R_c)$ (shown in interpolated curves).}
    \label{fig:hist_MP_props}
\end{figure}

We initialize a grid of $512^2$ particles using Gaussian distributed $\delta(\vec{q})$ with the matter power spectrum of the form $P_{\delta}(k) \propto k^{n} \exp{ \left( - k/R_c \right)}$, corresponding to parameters $n = 1, R_c = 10$ with exponential smoothing. The smoothing scale $R_c = 10$ is small (roughly $2 \%$) compared to the grid size $512$, so that the bias introduced by periodic boundaries is minimized. The particles are evolved according to Zeldovich flow --- the 1-LPT solution to the equations of motion \eqref{eq:lagrange_eq_motion}. Since we assume no vorticity in Lagrangian space, the motion of the particles must derive from a potential, and thus, we must restrict to 1- or 2-LPT motion in our simulations. We present comparisons against Zeldovich (1-LPT) simulations since we assumed the same in our computation of Gaussian statistics.

Fig. \ref{fig:sim} shows a snapshot from our 1-LPT simulation. To measure the $A_3^+$ and $A_2$ singularities in our simulation, we make a 2D interpolation of the eigenvalue $\alpha(\vec{q}, t)$ and the dot product $\bar{\nabla}\alpha(\vec{q},t) \cdot \bar{n}^{\alpha}(\vec{q},t)$ in Lagrangian space. We then compute their contours in Lagrangian space according to eq. \eqref{eq:A2_cond} for $A_2$ and eq. \eqref{eq:A3_cond} for $A_3^+$. The shell-crossing points are identified as local maxima of $\alpha(\vec{q}, t)$. These measurements are therefore independent of the expansion form, eq \eqref{eq:eq_motion_taylor_exp}, which forms the basis of our analytical predictions for the shape of the pancakes.

Measurements of the $A_2$ caustics and $A_3^+$ spines for three selected shell-crossings in Eulerian space are shown in fig. \ref{fig:eu_shape_sim}. Analytical predictions for the same have been overplotted for comparison. They are computed using eqs. \eqref{eq:A2_eq}, \eqref{eq:A3_eq}, and \eqref{eq:eq_motion_taylor_exp} with the values of the derivatives of $\phi$ at the shell-crossing point taken from the simulation. The $A_3^+$ spine shows great agreement close to the shell-crossing point and starts deviating only at the extremities. The same goes for the $A_2$ caustic, whose extent along the $x$-axis agrees very well, but that along the $y$-axis lags behind.

To verify our statistical results, we perform $\approx 50$ realizations of Gaussian random fields with the parameters $n = 1, R_c = 10$ and compile all the shell-crossing points ($\approx 3500$) that emerge. We record the values of derivatives of $\phi$ at the points of shell-crossing, scale them with respect to their corresponding $\sigma$ and shell-crossing time $D_+(t_c)$, eq. \eqref{eq:phi_scaling_Dtc}, and compare the histograms thus obtained with the marginal distributions (refer to figures \ref{fig:MP_q2}-\ref{fig:MP_q4}) computed using Monte Carlo integrations of eq. \eqref{eq:CP_Y_9}. Fig. \ref{fig:hist_MP_phi} shows good agreement between the distributions from simulation measurements (shown in step histograms) and those from our semi-analytic computations (shown in interpolated curves). In particular, we notice that the simulation measurement for $\phi^{4,0}$ falls short of the prediction for values close to 0. This could be because in our simulations, we fail to detect those peaks of $\alpha$ that are locally very smooth and have small heights.

We also make similar measurements for shell-crossing times, curvature, C-S transition scale, and axis length ratio according to eqs. \eqref{eq:P_Dtc}-\eqref{eq:P_ellip} within the pancake population in our simulations and compare them to the predicted curves (from figures \ref{fig:MP_Dtc}-\ref{fig:MP_ellip}) in fig. \ref{fig:hist_MP_props}. For C-S transition scale and axis length ratio, we also show the histograms (orange) corresponding to the exact eqs. \eqref{eq:A_3_C_S_trans} and \eqref{eq:A2_axis_lengths}, respectively, to cross-check our approximations in subsections \ref{sec:gaussian_stats}.5 and \ref{sec:gaussian_stats}.6 . In the case of the C-S transition scale, the additional terms contributing to the S-shape of the pancake in the exact eq. \eqref{eq:A_3_C_S_trans} expectedly reduce the extent of the C-shape compared to the approximate eq. \eqref{eq:P_CS_trans} and hence, the exact distribution of $P(y_{C \rightarrow S}(t_c))$ is narrower. In the case of the axis length ratio, the exact distribution $P(b/a)$ is shifted to lower values compared to our approximation. This indicates that the pancakes are flatter and the dynamics more anisotropic than our approximate eq. \eqref{eq:P_ellip} seems to suggest. Nevertheless, in both cases, save for the slight overshoot, our crude approximations exhibit surprisingly good agreement with the exact formulae, demonstrating that the additional contributions are indeed relatively small.

It is to be noted that measuring the distribution of pancakes' properties in our simulations using eqs. \eqref{eq:P_Dtc}-\eqref{eq:P_ellip} only verifies our semi-analytic computations of the distributions of $\phi^{i,j}$ (or their combinations) in section \ref{sec:gaussian_stats}. It does not verify if those combinations, eqs. \eqref{eq:P_Dtc}-\eqref{eq:P_ellip}, indeed map to the observable features regarding the shape of the pancake, such as the curvature and C-S transition scale. As an independent test, we could also fit $x-y$ curves in Eulerian space to the $A_3^+$ spines of pancakes identified in our simulations, compute the curvature and $y_{C\rightarrow S}$ from the fits, and match their distribution against the predicted curves, figures \ref{fig:MP_curv} and \ref{fig:MP_CS_trans}. Even simpler would be comparing the distribution of coefficients in those fits against the distribution of the coefficients in our analytical formula for the Eulerian shape of the $A_3^+$ spine, eq. \eqref{eq:A3_eq_Eul}. This is left for consideration in future works. Nevertheless, the good agreement between the shapes of the $A_2$ caustics and $A_3^+$ spines measured in our simulations and those predicted using our analytical formalism in section \ref{sec:catastrophe_theory} is a testament to the validity of eqs. \eqref{eq:P_Dtc}-\eqref{eq:P_ellip} in predicting observable features about the pancakes.

\end{document}